\newcommand{\newob}[1]{{\color{purple} #1}}

\newcommand{\newab}[1]{{\color{darkgreen} AB: #1}}

\documentclass[11pt]{article}
\usepackage[usenames,dvipsnames]{xcolor}
\definecolor{darkgreen}{rgb}{0.0, 0.5, 0.0}
\usepackage{fullpage}
\usepackage{amsmath} 
\usepackage{amsfonts}
\usepackage{amssymb}
\usepackage{authblk}
\usepackage{dsfont} 
\usepackage{graphicx}
\usepackage{pgfplots}
\usepackage{tikz}
\usepackage{subcaption}
\usepackage[colorlinks=true, linkcolor=blue, citecolor=blue, hypertexnames=false]{hyperref}  
\usepackage[authoryear,round]{natbib}
\newcommand{\journalversion}[1]{#1}
\newcommand{\blindversion}[1]{}
\usepackage{array}
\usepackage{amsthm}
\usepackage{thmtools}
\usepackage{thm-restate}

\usepackage{nameref}
\usepackage{cleveref}

\usepackage{amsmath} 
\usepackage{amssymb}
\usepackage{mathtools}

\usepackage{latexsym}
\usepackage{amsmath,amssymb}
\usepackage{epsfig}
\usepackage[right=1.0in, top=1in, bottom=1.0in, left=1.0in]{geometry}
\usepackage{setspace}
\usepackage{cleveref}
\usepackage{amssymb}
\usepackage{amsthm}
\usepackage{thmtools}
\usepackage{color}
\newtheorem{theorem}{Theorem}

\newtheorem{lemma}{Lemma}

\newtheorem{proposition}{Proposition}

\usepackage{amsmath}

\usepackage{enumerate}

\renewcommand{\theenumii}{\theenumi.\arabic{enumii}.}
\usepackage{calrsfs}

\usepackage{setspace}
\usepackage[utf8]{inputenc} 
\usepackage{nameref}
\usepackage{cleveref}     
\usepackage{url}            
\usepackage{booktabs}       
\usepackage{amsfonts}       
\usepackage{nicefrac}       
\usepackage{microtype}      
\usepackage{amsmath,mathtools}
\usepackage{cleveref}
 \usepackage{float}
\usepackage{dsfont}
\usepackage{enumerate}

\renewcommand{\theenumii}{\theenumi.\arabic{enumii}.}
\usepackage{amsmath}
\usepackage{enumerate}
\numberwithin{equation}{section}
\usepackage[all]{nowidow}
\usepackage[utf8]{inputenc}
\usepackage[inline]{enumitem}

\usepackage{bbm}

\DeclareMathAlphabet{\pazocal}{OMS}{zplm}{m}{n}

\newcommand{\PF}{$\pazocal{MP}$}
\newcommand{\PFLP}{$\pazocal{LP}$}
\newcommand{\PFLPI}{$\pazocal{LP}$-int}
\newcommand{\PFLPIU}{$\pazocal{LP}$-int-up}

\newcommand{\IN}{\mathbb{A}}

\usepackage{calrsfs}
\usepackage{graphicx}

\usepackage{xcolor}

\newcommand{\newaa}[1]{ { \color{orange} AA:{#1}}  }

\newcommand{\qm}[1][s]{q_{\frac{r_F}{2}}}
\newcommand{\qtss}[1][s]{q_{t_s}}
\newcommand{\qs}[1][s]{q^*}
\newcommand{\qw}[1][s]{q_w}
\newcommand{\qmin}[1][s]{\alpha^{\frac{1}{\alpha-1}}}
\newcommand{\rft}[1][s]{\frac{r_F}{2}}
\newcommand{\qsht}[1][s]{q_{\hat{s}}}

\newcommand{\Gad}[1]{\Gamma_{\alpha}\left(#1 \right)}
\newcommand{\Gainv}[1]{\Gamma_{\alpha}^{-1}\left(#1 \right)}

\newcommand{\qga}[1][s]{q_{\gamma r_F}}

\newcommand{\qg}[1][s]{q_{\frac{r_F}{\gamma}}}
\newcommand{\qgg}[1][s]{q_{\frac{r_F}{\gamma^2}}}
\newcommand{\qss}[1][s]{q_{s}}

\newcommand{\bF}{\overline{F}}
\newcommand{\bH}{\overline{H}}
\newcommand{\bG}{\overline{G}}
\newcommand {\bear}{\begin{eqnarray}}
\newcommand {\eear}{\end{eqnarray}}
\newcommand {\bearn}{\begin{eqnarray*}}
\newcommand {\eearn}{\end{eqnarray*}}
\newcommand {\opt}{\mbox{\normalfont opt}}
\newcommand{\Gb}{\pazocal{G}}
\newcommand{\Ex}{\mathbb{E}}

\newcommand{\Fb}{\pazocal{F}}

\newcommand{\Rb}{\pazocal{R}}
\newcommand{\Lb}{\pazocal{L}}

\newcommand{\Expect}{\mathbb{E}}

\newcommand{\Fa}{\Fb_{\alpha}}

\newcommand{\vvc}{\rho_{\alpha}}

\newcommand{\ql}{q_l}
\newcommand{\qh}{q_h}

\newcommand{\pin}{w}

\newcommand{\rl}[1][q]{\underline{r}_{\alpha}(\pin,#1)}
\newcommand{\rln}[1][q]{\underline{r}_{\alpha}(1,#1)}

\newcommand{\rh}[1][q]{\overline{r}_{\alpha}(\pin,#1)}

\newcommand{\rlone}[1][q]{\underline{r}_{\alpha}(1,#1)}
\newcommand{\rhone}[1][q]{\overline{r}_{\alpha}(1,#1)}

\newcommand{\distSet}{\pazocal{D}}
\newcommand{\MechSet}{\pazocal{P}}

\newcommand{\dMechSet}{\pazocal{P}_d}

\newcommand{\aDistSet}[2][\alpha]{\pazocal{F}_{#1}(\pin, #2)}
\newcommand{\aDistSetq}[2][\alpha]{\pazocal{F}_{#1}(\pin, \{#2\})}
\newcommand{\aDistSetqa}[2][\alpha]{\pazocal{F}_{#1}(\pin, #2)}

\newcommand{\aDistSetqprime}[2][\alpha]{\pazocal{S}_{#1, \pin, #2}}

\usepackage{tikz}
\usepackage{pgfpages}
\usepackage{pgfplots}
\usetikzlibrary{arrows,decorations,patterns,trees,matrix,calc,shapes}

\usepackage{graphicx}

\usepackage{filecontents}
\usetikzlibrary{calc,trees,positioning,arrows,chains,shapes.geometric,%
    decorations.pathreplacing,decorations.pathmorphing,shapes,%
    matrix,shapes.symbols}
\usetikzlibrary{matrix}
\usepgfplotslibrary{fillbetween}

\pgfdeclareplotmark{bluestar}{
    \node[star,star point ratio=2.25,minimum size=6pt,
          inner sep=0pt,draw=black,solid,fill=blue] {};
}
\pgfdeclareplotmark{redstar}{
    \node[star,star point ratio=2.25,minimum size=6pt,
          inner sep=0pt,draw=black,solid,fill=red] {};
}

\makeatletter
\DeclareRobustCommand*\cal{\@fontswitch\relax\mathcal}
\makeatother

\pgfplotsset{compat=1.5}
\usepackage{ifthen}

\usepackage{csvsimple}
\usepackage{pgfplotstable}
\usepackage{booktabs}

\begin{document}

\setstretch{1.5}
	
	\title{\sffamily \vspace*{-0.0cm}   Optimal Pricing with a Single Point}

	\journalversion{
		\author[1]{Amine Allouah}
		\author[2]{Achraf Bahamou}
		\author[3]{Omar Besbes}
		\affil[1]{Facebook Core Data Science, \texttt{mallouah19@gsb.columbia.edu}}
		\affil[2]{Columbia University, IEOR department, \texttt{achraf.bahamou@columbia.edu}}
		\affil[3]{Columbia University, Graduate School of Business, \texttt{obesbes@columbia.edu}}
		\date{initial version: February 26, 2021, this version: \today}
		\maketitle
	}
	
	\blindversion{
		\author{}
		\date{}
		\maketitle
		\vspace{0cm}
	}

\begin{abstract}
We study the following  fundamental data-driven pricing problem. How can/should a decision-maker price its product based on data at a single  historical price? How valuable is such data?   We consider a decision-maker who optimizes over (potentially randomized) pricing policies to maximize the worst-case ratio of the revenue she can garner compared to an oracle with full knowledge of the distribution of values, when  the latter is  only assumed to belong to a broad non-parametric set. In particular, our framework applies to the widely used regular and  monotone non-decreasing hazard rate (mhr) classes of distributions.  For settings where the seller knows the exact probability of sale associated with one historical price or only a confidence interval for it, we fully characterize optimal performance and near-optimal  pricing algorithms that adjust to the information at hand. The framework we develop  is general and allows to characterize optimal performance for deterministic or more general randomized mechanisms, and  leads to fundamental novel insights on the value of data for pricing. As examples,  against mhr distributions, we show that it is possible to guarantee  $85\%$ of oracle performance if one knows that half of the customers have bought at the historical price,  and if only $1\%$ of the customers  bought, it still possible to guarantee $51\%$ of oracle performance.
\medskip
		
\end{abstract}

\newpage

\section{Introduction}

Pricing is a central concept across a large spectrum of industries, ranging from e-commerce to transportation. A key informational dimension faced by decision-makers is the level of knowledge of customers' values. In classical settings in the literature, monopoly pricing problems are studied under the assumption that sellers have an accurate knowledge of consumer preferences through the value distribution (or the prior on values). In those cases, the seller may optimize pricing to maximize the expected revenues.

In practice, however, such information is rarely if ever available, and pricing must be conducted not based on the value distribution, but based on historical data. Typical historical data structures in the context of pricing include the prices posted and the responses of consumers observed at those prices: either a customer purchases or not. As a motivating example, consider an e-commerce firm that has been offering a product at an incumbent price $\pin$ over the past quarter to a set of heterogeneous consumers, all with values drawn from a  value distribution $F$ . The firm observes the fraction of customers who have bought the offered product at the price $\pin$; in other words the firm has an estimate of the probability of sale or conversion rate, the fraction of customers whose values are greater than or equal to $\pin$, i.e. , an estimate of $\bF(\pin) = q$ in $[0,1]$. How should the seller decide on the pricing policy in the following quarter? Can the seller take advantage of the partial demand information extracted (conversion rate at $\pin$) to refine her pricing policy? Such historical data structures  are commonplace in practice, and  typically introduce different challenges. The number of past prices that were posted is often very limited and if one only accounts for recent data, can be as low as one, as in the example above. In other words, many historical data structures have very limited price dispersion. This renders elasticity-based price optimization very challenging if not impossible in practice (without further experimentation) when trying to move from data to pricing decisions. A natural question is then if, in the absence of price dispersion,  historical data is useful in any way in order to refine pricing  decisions. The present paper offers a resounding ``yes" to this question and develops a framework to optimize prices given such limited data, and quantify the value of such data.

In more detail, we focus on a seller optimizing her pricing mechanism based on historical data with limited price dispersion. The seller does not know the value distribution of the buyer. She only knows that it belongs to some broad non-parametric class. In terms of the historical data, we anchor the paper around the setting in which the seller  has only access to  the conversion rate $q = \bF(\pin)$  at one historical price $\pin$, or potentially an interval $I$ to which $\bF(\pin)$ belongs. The questions the seller faces are then: what is an optimal  pricing mechanism given the information at hand? And how valuable is the information/data at hand? 

 To answer these question, we adopt a maximin ratio formulation in which performance is measured in comparison to the highest revenue the seller could have obtained with full information on the value distribution. The seller optimizes over  general  pricing mechanisms (we study both deterministic and randomized mechanisms). And nature may select any distribution in the class of interest to counter a pricing strategy. We are interested in characterizing the value of the maximin ratio as well as understanding the structure of optimal mechanisms. The latter quantifies the value of the collected information and the former offers concrete prescriptions. 
 
  This fundamental problem can be viewed as a fundamental building block of offline data-driven pricing and the framework we will propose will be fairly general, enabling one to add information at other points in the future. Mathematically speaking, this leads to a problem in which  the set of possible underlying value distributions is infinite dimensional, and so can be the set  of possible pricing strategies  (for randomized mechanisms). Hence, evaluating such an object is not possible without further understanding of structural properties of the problem.

Our main contributions lie in developing a general tractable characterization of deterministic and randomized  \textit{optimal} performances against any distribution in two  widely used classes of distributions: the class of regular distributions (distributions with increasing virtual values) and the  class of monotone non-decreasing hazard rate (mhr) distributions. The latter is a subclass of the former and contains a wide variety of distributions (e.g., uniform, truncated normal, logistic, extreme value, exponential, subsets of Weibull, Gamma and Beta,...);  \cite{bagnoli2005log} provides  a review of the broad set of known subclasses of mhr distributions. The class of regular distributions further incorporates additional distributions (e.g.,  subsets of Pareto,  log-normal, log-logistic,...); \cite{ewerhart2013regular} provides an overview of such classes. Our analysis is general and the exact same analysis applies to regular and mhr distributions as special cases.

From a methodological perspective, our main contributions lie in a set of problem reductions that lead to \textit{a closed form characterization of the maximin ratio for deterministic mechanisms},  and associated price prescriptions, and a sequence of \textit{finite dimensional linear programs}  that can approximate arbitrarily closely the maximin ratio for randomized mechanisms, leading to both optimal performance and associated near-optimal randomized mechanisms.  

 A first set of reductions lies in simplifying nature's optimization problem. As stated earlier, for any fixed mechanism chosen by the seller, nature's problem is an infinite dimensional problem over the class of regular (or mhr) distributions, which is non-convex. As such, evaluating the performance of a particular mechanism cannot be simply ``brute-forced" numerically.  As a first key  reduction, we establish in \Cref{thm:LB_1D} that against any mechanism, nature's worst-case optimization problem can be reduced from an infinite dimensional problem over a non-convex space to a one dimensional minimization problem over an interval. This reduction relies on exploiting the regularity (or mhr) structure to narrow down the set of candidate worst-cases to a ``small" set.
 
 Leveraging nature's problem reduction, we are able to derive a closed form (\Cref{thm:deterministic_perf_close}) for the maximin ratio for deterministic mechanisms against the classes of regular and mhr distributions. The results that one obtains  through these closed forms highlight three different ``regimes" of historical probability of sale values, and are quite illuminating with respect to the value associated with exact conversion rate information. \Cref{tab:summary} provides examples of such results.  An example of a striking theoretical result is that, with knowledge of the median, it is possible to use a simple deterministic pricing mechanism  to guarantee a substantial fraction, $85.23\%$, of the oracle revenue when the value distribution is mhr;  when the value distribution is regular, $66.62\%$ of the oracle revenue  can be guaranteed if we observe the 3rd quartile. Another highly notable theoretical result is associated with the value of low conversion rates that we uncover. We show that even if one only knows that $1\%$ of customers purchase at a particular price, then a deterministic pricing mechanism  guarantees more than 47\% of oracle performance against mhr distribution and 18\% against regular distributions. As a matter of fact, our closed form formulas indicate that, while the  the maximin ratio converges to zero as the known conversion rate converges to zero, it does so at a supra-linear, very slow, rate:  $\sqrt{q}$ for regular distributions, and $1/\log(q^{-1})$ for mhr distributions.

\begin{table}[!ht]
	\begin{center}
		\begin{tabular}{ cc|c|c }
			  & & \multicolumn{2}{c}{Maximin ratio}  \\ \hline \hline
			Distribution &  Conversion  & Randomized   &  Deterministic      \\
	Class & Rate	                     & mechanisms  & mechanisms   \\
			                     \hline
			& &   &   \\
			Regular	&$\bF(\pin) = 0.01$&31.12\%&18.18\%\\
			&$\bF(\pin) = 0.25$&67.75\%&66.62\%\\
			&$\bF(\pin) = 0.50$&55.99\%&\:\:50.00\%$^*$\\
			&$\bF(\pin) = 0.75$&41.35\%&25.00\%\\
			&&\\
	mhr		&$\bF(\pin) = 0.01$&51.17\%&47.55\%\\
			&$\bF(\pin) = 0.25$ &74.71\%&74.35\%\\
			&$\bF(\pin) = 0.50$&85.30\%&85.23\%\\
			&$\bF(\pin) = 0.75$&64.14\%&58.65\%\\
			\hline
		\end{tabular}
		\caption{\textbf{Maximin Performance.} The table provides examples of the results obtained regarding the optimal performance one may achieve as a function of the admissible set of distributions and the class of pricing mechanisms one considers. The maximin ratio is characterized exactly for deterministic mechanisms and up to at most 1\% error for randomized mechanisms. $^*$ indicates the only known result to date \citep{azar2012opt} .}
		\label{tab:summary}
	\end{center}
\end{table}

 In a second step, we study the performance of general randomized mechanisms.  To characterize such performance, we first leverage the reduction above of Nature's problem, but also establish that one can focus on mechanisms with finite support, while controlling the potential losses in performance (\Cref{thm:mechanismreduction}). In turn, we are able to derive a sequence of linear programs with order $N$ variables and order $N$ constraints (\Cref{thm:LB_LP}) that yields: $i.)$ an approximation to the maximin ratio within $O(1/\sqrt{N})$ and $ii.)$ provides a candidate randomized mechanism with near-optimal performance and support over order $N$ points. Given these, one can evaluate for every history $(\pin,\{q\})$ the performance that the seller can achieve.
 
The above characterizes the theoretical developments needed to obtain an exact characterization of the maximin ratio for randomized mechanisms against regular or mhr distributions. The results we obtain through this analysis offer novel insights on the value of information and the additional value stemming from the expanded set of  randomized pricing strategies, compared to deterministic ones. In particular, the value stemming from randomization is most prominent for values of the conversion rate close to 0 and 1. Intuitively, with historical prices providing less information, the seller can use randomization to counter uncertainty. For example, against regular distributions with a conversion rate of $1\%$, the seller can increase its guaranteed performance from about $18\%$  with a deterministic price to $31\%$ with a randomized mechanism, and with a conversion rate $75\%$, it can increase performance from $25\%$ to about $41\%$. 

\begin{figure}[!ht]
\begin{center}
\begin{tikzpicture}
\begin{axis}[
            title={},
            xmin=0,xmax=1,
	        ymin=0.0,ymax=100, 
	        width=10cm,
	        height=8cm,
	        xlabel = conversion rate $q$, 
	        ylabel = Maxmin ratio in \%, 
	        grid=both, 
	        legend pos=south west]
	        
\addplot[color=blue, mark=*, mark size=2.9pt] table[x=q,y expr=(\thisrow{ylw}+\thisrow{yup})/2] {Reg_opt_data.dat}; \addlegendentry{regular}
\addplot [name path=upper,draw=none,forget plot] table[x=q,y=yup] {Reg_opt_data.dat};
\addplot [name path=lower,draw=none,forget plot] table[x=q,y=ylw] {Reg_opt_data.dat}; 
\addplot [fill=blue!50,forget plot] fill between[of=upper and lower];

\addplot[color=red, mark=square*,  mark size=2.9pt] table[x=q,y expr=(\thisrow{ylw}+\thisrow{yup})/2] {mhr_opt_data.dat}; \addlegendentry{mhr}
\addplot [name path=upper,draw=none,forget plot] table[x=q,y=yup] {mhr_opt_data.dat};
\addplot [name path=lower,draw=none,forget plot] table[x=q,y=ylw] {mhr_opt_data.dat};
\addplot [fill=red!50, forget plot] fill between[of=upper and lower];
\end{axis}
\end{tikzpicture}
\end{center}
\caption{\textbf{Maximin ratio for randomized mechanisms against regular and mhr distribution.}}
\label{fig:maxminratios}
\end{figure}
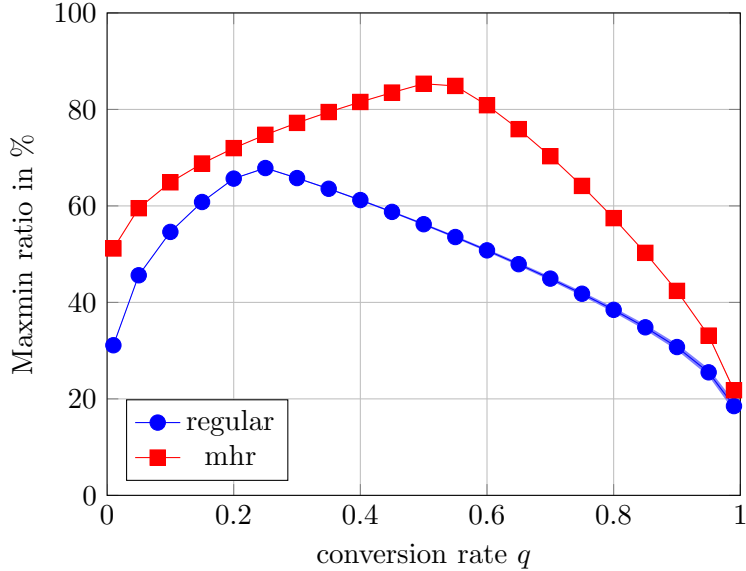

\Cref{tab:summary} presents examples of the results obtained, but the framework developed is not specific to any probability of sale value and applies to any historical price and associated probability of sale. Figure \ref{fig:maxminratios} depicts the maximin ratio for randomized mechanisms for various values of the conversion rate  ranging from 0.01 to 0.99. 

As a matter of fact, we can establish that randomization drastically affects the value that one can extract from information, leading to a fundamentally different rate of convergence of performance as $q$ goes to zero or 1. In that former regime, we establish that the rate convergence improves from  order $\sqrt{q}$ to  order $1/\log(1/q)$ (\Cref{lem:qzerbounds}), and in the latter case, we show that it goes from a linear rate to order $1/\log(1/(1-q))$ (\Cref{lem:qonebounds}). We also show that for mhr distributions, while the rate of convergence is not affected by randomization around $0$ (being order $1/\log(1/q)$ for both classes of mechanisms), it is significantly affected for values of $q$ that are close to 1, going from a linear rate to  $1/\log(1/(1-q))$ (\Cref{lem:qzerbounds}, \Cref{lem:qonebounds}).

In addition, the framework we develop is general and allows to incorporate uncertainty (or noise), in the probability of sale estimate. We establish a parallel characterization for the maximin performance  for randomized mechanisms in \Cref{thm:LB_LP_robust}. We develop  a sequence of finite dimensional linear programs that can approximate with arbitrary accuracy the maximin ratio.

Stepping back, the present paper and associated framework enables to understand the value of one measurement for pricing purposes. While such a measurement provides very limited information about the entire value distribution, we establish that it provides  significant value for pricing purposes. As such, this  leads to an important building block for future research  to better understand the value associated with an arbitrary number of measurements, or the best way to experiment and collect such responses to prices.

\subsection{Literature review}

In this section, we position our work in the landscape of related past research efforts. Our work relates and contributes to the literature on data-driven pricing with limited value distribution knowledge. A setting that has been studied is one in which the seller only knows the support of the underlying distribution. Early studies are \cite{Bergemann08} and \cite{Eren2009}, in which  the authors characterize the optimal pricing policy as well as the worst-case demand distribution with respect a min-max regret objective in the former and a competitive ratio in the latter. 
 \cite{cohen2021simple}  studies a case in which the seller has access to the maximum price at which she would still expect non-zero demand, and the authors propose to use a simple deflation mechanism and characterize its performance against some subsets of  parametric families.  \cite{Caldentey2017} characterize optimal pricing strategies in a dynamic setting where myopic or strategic customers arrive over time and only the support of their value distribution is known to the seller. 
 
 In contrast to this stream of work, we study the setting in which the seller has access to some information about the conversion rate at an incumbent price $\pin$, a typical data structure, and can adjust its decisions based on such information. We also allow value distributions with arbitrary support within  central non-parametric classes (regular or mhr). \cite{Eren2009}  study a related setting and analyzes randomized mechanisms against general discrete distributions, but with known  support bounds  information. \cite{Chehrazi2010} study a general robust decision problem while specifying a shape-preserving set of univariate functions using a constrained B-spline approximation. The framework developed in \cite{Chehrazi2010} can be applied to pricing in  environments with limited measurements and the authors illustrate their ideas using an optimal debt-settlement example. 
  In \cite{will2016demand}, the authors studied the related problem of reconstructing demand curves when only a single point has been historically observed and showed how a second point can be extracted from the sales of  discounted bundles and use it to estimate linear demand curve parameters. \cite{BesElm2020} document a setting at a large OEM where the problem is exactly one with no price dispersion in the historical data. There, the authors  applied some parametrization approach in conjunction with a robustification of prices. Recently, \cite{chen2021assortment} study model-free assortment pricing decisions from transaction data, by leveraging incentive compatibility constraints. 
 
 In a related setting, \cite{azar2012opt} and \cite{azar2013para}  assume that the seller has only access to limited statistical information about the valuation distributions (such as the median, mean and variance). In \cite{azar2012opt} for the single-bidder pricing problem, which is related to our problem, the authors analyze the case with median information and  provide a tight upper bound ($50\%$) on the best achievable  competitive ratio for regular distribution using deterministic mechanisms. This can be seen as a special case of the general framework we develop.  Our results establish the exact performance of deterministic for \textit{any} probability of sale for both regular and mhr distributions, establishing two phase transitions (small, moderate, and high probability of sale). For these settings, we also characterize the optimal performance of  randomized mechanisms. While studying a different set of questions (the performance of a Vickrey auction with duplicate bidders),  \cite{Fuetal2019EC}  also considers a setting in which the information available to the seller consists of a percentile of the value distribution, akin to the information structure in the present paper.

A number of studies look at how to collect and incorporate data on the fly for pricing purposes, in which case an exploration-exploitation trade-off emerges. See \cite{Kleinberg2003}, \cite{Besbes2009}, \cite{Broder2012}, \cite{kerkinzeevi}, \cite{bu2020online}.  The present study establishes that when initial offline data is available (such as in \cite{kerkinzeevi} or \cite{bu2020online}), it can be possible to  exploit such information, even with no price variability in the data. As such, the ideas presented here might also have implications for dynamic learning algorithms.

An alternative data structure that has received attention is one based on samples of the value distribution, as opposed to buy/no buy feedback.  \cite{Huang2018} studies the sample complexity needed to achieve near-optimal performance; see also, e.g., \cite{cole2014sample}, \cite{GolrLP21}, in the context of auctions.  The setting where the seller has access to a limited number of observed samples  has also been studied; see \cite{Huang2018},  \cite{Fu2015RandomizationBS}, \cite{Babaioff2018}, \cite{OptPricingAllouahBesbes},    \cite{Daskalakis:FactorRevealingSDPs:2020},  and \cite{ABBSamples}. These studies demonstrate that a few samples can be very informative for pricing purposes. Relatedly, the present work characterizes the value that a different type of information/data, a single percentile, has for pricing purposes.

\section{Problem formulation and approach overview}\label{sec:pbf}

We consider a seller trying to sell one indivisible good to one buyer. We assume that the buyer's value $v$ is drawn from some distribution $F$ with support included in $[0,\infty)$. The seller does not know $F$ and only knows some class information as well as partial information associated with it based on the  historical conversion rate observed at a price $\pin$. More specifically, we study the setting in which  the seller knows that the probability of sale belongs to some interval, i.e., $\bF(\pin)$ belong s to $I$ with $I \subseteq [0,1]$. An important building block is when the decision-maker knows the exact probability of sale $q$ at $\pin$, i.e., the seller knows that $\bF(\pin) = 1-F(\pin) = q$. In what follows, we will use the notation $\bF := 1- F$ to denote the complementary cumulative distribution function (ccdf).  

The problem we are interested in is the following:  how can the seller leverage the information observed at the price $\pin$ to maximize her revenue. More formally, we model the problem as a game between nature and the seller, in which the seller selects a selling mechanism and nature may choose any admissible distribution $F$ that is consistent with the observed information. We denote by $\distSet$ the set of cumulative distribution functions (cdf) on $[0,+\infty)$, i.e., the set of non-decreasing right continuous with left limits functions from $[0,+\infty)$ into $[0, 1]$ such that the limit at infinity is one. 

\paragraph{Pricing  and performance.} A (potentially) randomized pricing strategy will be characterized  by the cdf of prices the seller posts. We let $\MechSet = \{\Psi \mbox{ in } \distSet \}$ to be the set of randomized prices  that a mechanism can choose from, given the conversion rate information.

The expected revenue of the seller using a price distribution $\Psi \mbox{ in }\MechSet$, if nature is selecting a distribution $F$, is given by
\bearn
\int_{0}^{\infty}\left[\int_{0}^{\infty} p \mathbf{1}\{v \geq p\} d F(v)\right]  d \Psi(p) = \int_{0}^{\infty} p \bF(p) d \Psi(p) = \int_{0}^{\infty} Rev\left(p|\bF\right) d \Psi(p),
\eearn
where  we introduce the notation 
$$Rev\left(p|\bF\right) = p \bF(p). $$
We define $\opt(F)$ to be the maximal performance one could achieve \textit{with knowledge} of the exact distribution of buyer's values. It is known that it is a posted price \citep{Riley1983}, and is given by 
\begin{alignat}{3} 
\opt(F) \: :=\: & \sup_{p \ge 0} \: Rev\left(p|\bF\right). \label{eq:opt}
\end{alignat}

For an arbitrary distribution $\Psi \mbox{ in }\MechSet$, we define its performance against a distribution $F$ such that $\opt(F) > 0$ as follows
\bearn
R(\Psi, F)=\frac{\int_{0}^{\infty} Rev\left(p|\bF\right) d \Psi(p) }{\operatorname{opt}(F)}.
\eearn

Let $\Gb (w, I)$ denote the set of distributions with support included in $[0, \infty)$ with finite and non-zero expectation such that $\bF(w)$ belongs to $I$ where $I$ is an interval in $[0,1]$,  i.e.,  
\begin{align} \label{eq:GwI}
\Gb (w, I) \:=\:  \left\{ F: [0,\infty) \rightarrow [0,1]: F \mbox{ is in } \distSet \mbox{ and } 0 <  \Ex_F\left[v\right] < \infty  \mbox{ and } \bF(w) \mbox{ in }I \right \}.
\end{align}

Note that $\opt(F)$ is in $(0,\infty)$ for all $F$ in $\Gb (w, I)$ and hence the ratio $R(\Psi, F)$ is well defined for any element of the class $\Gb (w, I)$. For an arbitrary price distribution $\Psi \mbox{ in }\MechSet$ and for a subclass $\Fb \subseteq \Gb (w,I)$, we define nature's problem as:
\bearn
 \inf _{F \in \Fb} R(\Psi, F).
\eearn
The objective in the present paper is to characterize the maximin ratio for sub-classes of distributions $\Fb \subseteq \Gb (w,I)$ and subclasses $\MechSet' \subseteq \MechSet$ 
\bearn
\Rb(\MechSet', \Fb) &=& \sup _{\Psi \in \MechSet'} \inf _{F \in \Fb} R(\Psi, F). 
\eearn
Note that this objective is always between $0$ and $1$ and can be interpreted as a measurement of the ``value of information" when using the subclass $\MechSet'$. 
\paragraph{Mechanisms classes.} We will be interested in the performance of general randomized mechanisms $\MechSet$ but also with the performance associated  with the subclass of deterministic pricing mechanisms $\MechSet_d \subset \MechSet$, defined as the set of dirac delta, i.e.,
\bearn
\MechSet_d = \{\delta_\gamma :  \gamma \geq 0 \}.
\eearn

\paragraph{Focal classes of distributions.} Recall the definition of the set of general distributions consistent with the data,  $\Gb (w, I)$, given  in Equation \eqref{eq:GwI}. \Cref{prop:generalclass} below formalizes that it is impossible to design any randomized mechanism with a positive competitive ratio when competing against $\Gb(w, I)$ for any non-empty interval $I$.
\begin{proposition}[maximin ratio against general distributions]\label{prop:generalclass}
For any mechanism $\Psi \mbox{ in }\MechSet$, and non-empty interval $I$ in $[0,1]$, we have
$$
 \inf _{F \in \Gb (w, I)} R(\Psi, F) = 0.
$$
\end{proposition}
 In the rest of the paper, we focus on widely studied subclasses of $\Gb(w, I)$ in the pricing context. In particular, we focus on two broad subclasses. The first sublcass we analyze is the class of monotone hazard rate (mhr) distributions, i.e., distributions that admit a density, except potentially at the maximum of their support, and that have a  non-decreasing hazard rate.  
 As mentioned in the introduction, this class contains a wide variety of distributions and typical models fitted in the literature belong to subclasses of mhr distributions. See, e.g., \cite{bagnoli2005log}.

A second notable class of distributions that generalizes mhr distributions is the class of \textit{regular} distributions; these admit a density, except potentially at the maximum of their support, and  have a  non-decreasing virtual value $v - \bF(v)/f(v)$.  
 This class of distributions contains all mhr distributions but also a host of additional distributions; see, e.g., \cite{ewerhart2013regular} for a summary of widely used regular distributions. In particular, the class of regular distributions allows for heavier tails than mhr distributions. This class  is central to the pricing  and mechanism design  literatures and can be alternatively described as the class of distributions which induce a concave revenue function in the quantity space.

While we will focus on the two classes above given their central role in the literature, our analysis will be unified. In particular, the two classes above can be seen as special cases of  $\alpha$-strongly regular distributions (see, e.g., \cite{ewerhart2013regular}, \cite{cole2014sample}, \cite{schweizer2016quantitative}). These are distributions with positive density function $f$ on its support $[a, b]$, where $0 \leq a < \infty$ and $a \leq b \leq \infty$, such that $(1-\alpha) v - \bF(v)/f(v)$ is non-increasing. When $\alpha=0$, this corresponds to regular distributions and when $\alpha=1$, this corresponds to mhr distributions. We define 
\bearn
    \aDistSet{I} &=&\{F \mbox{ in }\Gb(w,I): F \mbox{ is  $\alpha$-strongly regular}\}
\eearn
to be the set of distributions that are $\alpha$-strongly regular \textit{and} consistent with the information at hand.

 It is possible to establish that when the interval $I$  contains $0$ or $1$, no pricing mechanism can guarantee a positive fraction of revenues. We formalize this result in \Cref{prop:qzerone} (presented in  Appendix \ref{apx:qonezero}). We assume throughout that the interval $I$ does not contain $0$ or $1$, i.e., $I \cap \{0,1\} = \emptyset$.

\paragraph{Approach overview.}  
We start by analyzing the case when the seller has access to the probability of sale at one price, i.e., $I = \{q\}$ is  a singleton, with $q \mbox{ in }(0,1)$, and its associated price $\pin$. This will be the focus of Sections \ref{section:naturepbreduction}-\ref{sec:rand}. In what follows, whenever the percentile is known, we will use, with some abuse of notation, $\aDistSetqa{q}$ instead of $\aDistSetq{q}$. We return in \Cref{sec:I} to the case with interval uncertainty.

The first step in analyzing  $\Rb(\MechSet', \aDistSetqa{q})$ resides in noting that it can reformulated as an equivalent mathematical program \
\begin{align}\label{eq:general-LP}\tag{\PF}
 \sup_{\Psi(\cdot) \mbox{ in }\MechSet', c \mbox{ in }[0,1]}~ &\quad c \\
s.t.~~&\quad \int_{0}^{\infty} Rev\left(p|\bF\right) d \Psi(p)  \ge c \:\:  \opt(F)  \qquad \hbox{ for all } F \mbox{ in }\aDistSetq{q}.\nonumber 
\end{align}
The value of this problem is exactly equal to the maximin ratio $\Rb(\MechSet', \aDistSetqa{q})$ and any optimal solution to the former is also optimal for $\Rb(\MechSet, \aDistSetqa{q})$. 
When $\MechSet'= \MechSet$, this is a linear program. However, the key challenge in solving such a problem and designing optimal or near-optimal pricing mechanisms resides in the fact both $\MechSet$ and $ \aDistSetqa{q}$ are infinite dimensional spaces. In turn, this is a linear program with  an infinite number of variables and constraints. (When $\MechSet'= \MechSet_d$, the set of feasible mechanisms is not convex anymore.) 

To characterize  $\Rb(\MechSet', \aDistSetqa{q})$, we will proceed in two steps. We first establish in \Cref{section:naturepbreduction} a key reduction, that many of the constraints are ``redundant" and as a result, one can restrict attention, without loss of optimality to an alternative to (\ref{eq:general-LP}) with significantly fewer constraints.  

For deterministic mechanisms $\MechSet_d$, analyzed in \Cref{section:deterministicmechanisms},   we leverage the fundamental reduction in the space of distributions to establish that the problem can be directly reframed and solved in closed form. In turn, optimal deterministic mechanisms  and optimal performance over this subclass of mechanisms can be derived explicitly against regular and mhr distributions.

In \Cref{sec:rand}, we tackle the challenge stemming from the infinite dimensional nature of the space of mechanisms of the seller in the context of general randomized mechanisms. For that, we establish that mechanisms with bounded and discrete support, can approximate (from below) the performance of general randomized mechanisms with arbitrarily high accuracy. We combine the  reductions in both the space of distributions and mechanisms  to derive a sequence of finite dimensional linear programs whose value converges (from below) to the original quantity of interest, $\Rb(\MechSet, \aDistSetqa{q})$. Furthermore, the optimal solution of any such linear program provides a (discrete) pricing distribution with a certificate of performance given by the value of the linear program and this value approaches the optimal maximin ratio. In \Cref{sec:I}, we extend the ideas above to the case when the probability of sale is only known to belong to an interval $I$ and characterize optimal performance in this more general case. 

\textbf{Notation.} With some abuse, to avoid introducing special notation at various junctions, we will interpret any ratio of a positive quantity divided by zero  as $\infty$.  Furthermore, we will use the notation $a\vee b := \max \{a,b\}$ and $a\wedge b := \min \{a,b\}$.

\section{Reduction of Nature's problem}\label{section:naturepbreduction}

In this section, we focus on Nature's problem associated with selecting a worst-case distribution against an arbitrary mechanism. For any mechanism $\Psi \mbox{ in }\MechSet$, Nature will select a worst-case distribution in the non-convex infinite dimensional space of distributions $\aDistSetqa{q}$. 
 Our first main result establishes  a fundamental reduction:  one may restrict  attention to a ``small" set of candidate worst-case distributions. In particular, we will establish that Nature's problem can be reduced to a one-dimensional optimization problem. 

For any $\alpha \mbox{ in }[0,1]$, we introduce notation for Generalized Pareto Distributions (GPD).  This class of distributions plays a central role in pricing problems  in the context of $\alpha$-strongly regular distributions (see, e.g., \cite{cole2014sample} and  \cite{schweizer2016quantitative}). Indeed, the $\alpha$-strongly regularity condition can be interpreted as a curvature restriction captured by the fact that $\alpha$ virtual value function $(1-\alpha) v - f(v)/\bF(v)$ is non-increasing. For a given value of $\alpha$, the function  defined below, $\Gamma_{\alpha}$, can be seen to be on the ``boundary" of this space as it has constant $\alpha$ virtual value function. In particular,  for any $v\ge0$,  we define
\bearn
\Gamma_{\alpha}(v)&=&
\begin{cases}
	\left(1+ (1-\alpha) \ v\right)^{-1/(1-\alpha)} &\quad \mbox{if } \alpha \mbox{ in }[0,1),\\
	e^{-v}& \quad \mbox{if } \alpha = 1.
\end{cases}
\eearn
In addition, $\Gamma_{\alpha}^{-1}$ denotes the inverse of $\Gamma_{\alpha}$ and we set $\Gamma^{-1}_{\alpha}(0):=+\infty$ and $\Gamma_{\alpha}(+\infty):=0$.  

We next introduce some notation that will allow us to define an appropriate subclass of distributions. 

For any pair of values $(s,s')$ such that $0 \le s \le s'$ and  $1 \ge q_{s} \geq q_{s'} >0$, and for any $t \ge s'$, we define on $[0, \infty)$
\bear \label{eq:G}
\overline{G}_{\alpha,t}(v | (s,q_s),(s',q_{s'})) &=& \begin{cases}
			  \Gamma_{\alpha}\left( \Gamma_{\alpha}^{-1}\left(q_{s}\right) \frac{v}{s}   \right), &\quad \mbox{if } v \in [0,s),\\ 
			q_s\Gamma_{\alpha}\left( \Gamma_{\alpha}^{-1}\left(\frac{q_{s'}}{q_{s}}\right) \frac{v-s}{s'-s}   \right) & \quad \mbox{if } v \in [s,t],\\
			0 & \quad \mbox{if } v > t.
		\end{cases}
\eear
The function $\bG_{\alpha,t}(\cdot | (s,q_s),(s',q_{s'}))$ is a  complementary cumulative distribution function (ccdf) that has conversion rate $q_s$ at price $s$ and $q_{s'}$ at price $s'$\footnote{Note that when $s=s'$ and $q_s> q_{s'}$, the ccdf $\bG_{\alpha,t}(\cdot | (s,q_s),(s',q_{s'}))$ has a mass of 
$1-q_s$ at $s=s'$. In this case, with some abuse of terminology, we continue to say that it has a conversion rate of $q_{s'}$ at $s'$ as it can be approximated arbitrarily closely by a ccdf that has this property.} and satisfies the restriction on the curvature implied by $\alpha$-strong regularity  with equality locally, on $[0,s)$, and on $[s,t]$. Furthermore, it has support $[0,t]$.

%
%
%
%

 We next define the following family of distributions, through their complementary cumulative distribution function
\bear \label{eq:F_al}
\bF_{\alpha}( v | r, (\pin, q)) := 
\begin{cases}
	\overline{G}_{\alpha,\pin}(v | (r,1), (\pin,q) ) &\quad \mbox{if } r \mbox{ in }[0,\pin), \\
	 \overline{G}_{\alpha,r}(v |  (0,1), (w,q) ) &\quad \mbox{if }  r \ge \pin.
\end{cases} 
\eear 

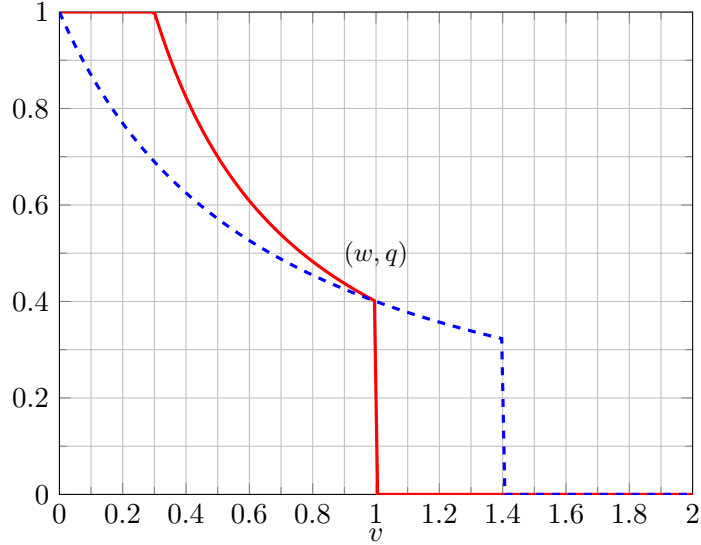
\begin{figure}[ht!]
	\centering
	\begin{tikzpicture}[scale=1, 
	declare function={Fl(\alpha,\q,\r,\x)=1*(\x<\r)+and(\x>=\r,\x<1)*((1+(\q^(\alpha-1)-1)*(\x-\r)/(1-\r))^(1/(\alpha-1)))+(\x>=1)*(0);},
	declare function={Fr(\alpha,\q,\r,\x)= (\x<\r)*((1+(\q^(\alpha-1)-1)*\x)^(1/(\alpha-1))) + (\x>=\r)*(0);}
	]
	\begin{axis}[ xmin=0, xmax=2,
	ymin=0, ymax=1,
	restrict y to domain=0:1.0,
	grid=both,
	minor tick num=1,
	axis line style={->},
	x label style={at={(axis description cs:0.5,-0.05)},anchor=north},
	y label style={at={(axis description cs:-0.05,.5)},rotate=0,anchor=south},
	xlabel={ $v$},
	width=10cm,
	height=8cm]
	]
	\addplot [domain=0:2, red, very thick, samples = 200] {Fl(0,0.4,0.3,x)};
    \addplot [domain=0:2, blue, dashed, very thick, samples = 200] {Fr(0,0.4,1.4,x)};
	\node [above] at (100, 45) { \small $(w,q)$};
	\end{axis}
	\end{tikzpicture}
	\caption{\textbf {Examples of distributions in $\aDistSetqprime{q}$.} The figure depicts examples of functions $\bF_{\alpha}( \cdot | r, (\pin, q))$: $\bF_{0}( \cdot | 0.3, (1, 0.4))$  in red  and  $\bF_{0}( \cdot | 1.5, (1, 0.4))$ in dashed blue.} \label{fig-tails-bounds}
\end{figure}

The associated cdf $F_{\alpha}( v | r, (\pin, q))$  has the upper end of its support at  $r\vee w$ and the lower end of its support at either $0$ if $r > w$ or $r$ if $r \le w$, and has a conversion rate of $q$  at $w$.  \Cref{fig-tails-bounds} depicts this distribution for two sets of parameters. This corresponds to a family of translated and truncated GPD distributions. 
 Let 
\bear \label{eq:r_bar}
\rl = \frac{\pin}{\Gainv{q}+1}, \quad \rh= \frac{\pin}{\alpha \Gainv{q}}, \mbox{ with }\rh= +\infty \mbox{ for }\alpha = 0.
\eear
We are now in a position to define the following subset of  distributions, which  is parametrized by a single parameter $r$: 
\bear \label{eq:subset}
\aDistSetqprime{q} &=& \left\{ F_{\alpha}( \cdot | r, (\pin, q)): r \mbox{ in }  [\rl,\pin) \cup [\pin,\rh ] \right\}, 
\eear
where we use the convention that whenever $\rh<\pin$, $[\pin,\rh ]:= \emptyset$. 

It is possible to establish that $\aDistSetqprime{q} \subset \aDistSetqa{q} $ as every element of this set has constant $\alpha$-virtual value on the interior of its support (see \Cref{lem:const}).

We next  state our first main result.

\begin{theorem}[Fundamental Reduction]\label{thm:LB_1D} 
Fix $\alpha \mbox{ in }[0,1]$. For any $q \mbox{ in }(0, 1)$,  for any subset of mechanisms $\MechSet' \subseteq \MechSet$, 
\bearn
\Rb(\MechSet', \aDistSetqa{q}) = \Rb(\MechSet', \aDistSetqprime{q}).
\eearn
\end{theorem}

 This result provides a central structural  property for this class of problems. When analyzing the possible response of nature to a particular mechanism, it is sufficient to only consider translated and truncated GPD families with a special structure. In particular, the candidate worst-cases are parametrized by a single parameter $r$ as outlined in the definition of $\aDistSetqprime{q}$. The only candidate worst-cases to consider are either: i.) distributions whose lower bound starts at some $r \mbox{ in }[0,\pin]$, decreases according to a GPD piece up to $\pin$ and admit a mass at $\pin$;  for those, the optimal oracle price is $r$ (cf. \Cref{lem:rr}); or ii.) distributions that  have a support starting at zero, decrease according to a GPD, and admit a mass at $ r\ge \pin$; for those distributions, the optimal  oracle price is again at $r$, but exceeds $\pin$. Intuitively, these distributions capture exactly the difficulty of not knowing the distribution of values. Indeed, when fixing any mechanism, the result above implies that one ``can think" of nature as selecting an optimal oracle price as opposed to a distribution, as conditional on the former, one can now compute the worst-case distribution. As we will see later, this structural result will be central to characterize optimal performance and derive near-optimal mechanisms for both the classes of deterministic and randomized mechanisms.

 Based on the fundamental reduction in \Cref{thm:LB_1D}, the set of constraints in (\ref{eq:general-LP})  can be significantly reduced and the problem can be equivalently stated as follows
\bearn
\sup_{\Psi \in \MechSet', c \in [0,1]}&&  c \\
s.t. &&  \int_{0}^{w} \frac{Rev\left(u| \bG_{\alpha,w}(\cdot | (r,1), (w,q) )\right)}{Rev\left(r| \bG_{\alpha,w}(\cdot | (r,1), (w,q) )\right)} d \Psi(u)  \ge c,  \mbox {for  all } r \mbox{ in }\left[\rl, w\right)   \\
     &&  \int_{0}^{r}  \frac{Rev\left(u| \bG_{\alpha,r}(\cdot | (0,1), (\pin,q) )\right)}{Rev\left(r| \bG_{\alpha,r}(\cdot | (0,1), (\pin,q) )\right)} d \Psi (u)  \ge c, \mbox{for all } r \mbox{ in }\left[w, \rh \right].
\eearn

We note that \Cref{thm:LB_1D} admits a generalization to the case in which $q$ is only known to belong to an interval (cf.  \Cref{thm:LB_1D_uncertainty_1}).

Next, we present the proof of \Cref{thm:LB_1D} together with the intuition associated with the various steps that enable this fundamental reduction.
 
 
 \subsection{Key ideas and proof of \texorpdfstring{\Cref{thm:LB_1D}}{Theorem 1} }\label{sec:proof-reduction}

We fix a  mechanism $\Psi$ throughout. The proof is organized around two main steps. In a first step, we show that  nature's optimization problem can be reduced to a  two-dimensional optimization problem, one of selecting the location the optimal oracle price  $r_F$ and the corresponding ``quantity" $q_F$. In other words, these two quantities can be seen as ``sufficient statistics" from the perspective of nature, given the limited knowledge of the seller. In a second step,  we establish that for a given value of $r_F$, the worst-case $q_F$ can be characterized explicitly, and in turn, one can further reduce the problem to a one-dimensional optimization problem, over the set of possible oracle optimal prices $r_F$. 

\textbf{Step 1.} Fix $q \mbox{ in }(0,1)$.  In this first step, we  develop a reduction of nature's problem to a two-dimensional optimization problem parametrized by the set of possible values that the optimal oracle price $r_F$ and quantity $q_F$ can take. We first define the set of feasible values for $r_F$ and  $q_F$ given the information at hand. To that end, let $ {\cal B}_{\alpha}(\pin,q) $ denote the set of feasible pairs, i.e., 
\bearn
 {\cal B}_{\alpha}(\pin,q) &:=&  \left\{(r^*,q^*) \mbox{ in }\mathbb{R}_+\times[0,1]  :  \mbox{there exists } \: F \mbox{ in } \aDistSetqa{q} \mbox{ with } r_F= r^*, q_F = q^* \right\}.
\eearn
Given such a definition, we have

\bear
 \inf_{F \in \aDistSetqa{q}} R(\Psi, F)  &= & \inf_{F \in \aDistSetqa{q}} \: \frac{\Expect_{\Psi}[  Rev\left(p|\bF\right)]} {\opt(F)} \nonumber \\
 &\stackrel{}{=}&\inf_{(r^*,q^*) \in {\cal B}_{\alpha}(\pin,q)}  \inf_{\substack{F \in \aDistSetqa{q}: \\ r_F = r^*, q_F = q^*}} \: \frac{\Expect_{\Psi}[  Rev\left(p|\bF\right)]} {\opt(F)} \nonumber\\
 &=&  \inf_{(r^*,q^*) \in {\cal B}_{\alpha}(\pin,q)} \frac{1}{r^*  q^* } \inf_{\substack{F \in \aDistSetqa{q}: \\ r_F = r^*, q_F = q^*}} \: \Expect_{\Psi}[  Rev\left(p|\bF\right)]. \label{eq:numerator}
\eear
The reduction above has allowed to ``decouple" the problem, where the denominator is fully controlled and the numerator can be minimized in the inner minimization, independently of the denominator.

Fix $(r^*,q^*) \mbox{ in }{\cal B}_{\alpha}(\pin,q)$.  Next, we characterize $\inf_{\substack{F \in \aDistSetqa{q}: \\ r_F = r^*, q_F = q^*}} \: \Expect_{\Psi}[  Rev\left(p|\bF\right)]$. We first derive a lower bound.

\bearn
 \inf_{\substack{F \in \aDistSetqa{q}: \\ r_F = r^*, q_F = q^*}} \: \Expect_{\Psi}[  Rev\left(p|\bF\right)] 
&=& \inf_{\substack{F \in \aDistSetqa{q}:\\ r_F = r^*, q_F = q^*}} \:   \int_0^{\infty} Rev\left(p|\bF\right) d\Psi(p) \\
 &\ge&  \inf_{\substack{F \in \aDistSetqa{q}: \\ r_F = r^*, q_F = q^*}}\: \left[ \int_0^{r^*\wedge \pin} Rev\left(p|\bF\right) d\Psi(p)  + \int_{r^*\wedge\pin}^{r^*\vee\pin} Rev\left(p|\bF\right) d\Psi(p)\right].
 \eearn
 
Next, we leverage the following single crossing property result from \cite[Lemma 2]{{ABBSamples}}. 
\begin{lemma}[Single Crossing Property]\label{lemma:singlecross}
	Fix $\alpha \mbox{ in }[0,1]$, $F$ in $\Fb_{\alpha}$ and  a pair of values $(s,s')$ such that $0 \le s \le s'$ and $q_{s'} = \bF(s')>0$. Then
	\bearn
	\overline{F}(v) &\ge& q_{s} \Gamma_{\alpha}\left( \Gamma_{\alpha}^{-1}\left(\frac{q_{s'}}{q_{s}}\right) \frac{v-s}{s'-s}   \right)  
	 \qquad \mbox{ if } \:  v \mbox{ in }[s, s']. \label{ineq:lb_lambda_gen_pair}\\
	\overline{F}(v) &\le& q_{s} \Gamma_{\alpha}\left( \Gamma_{\alpha}^{-1}\left(\frac{q_{s'}}{q_{s}}\right) \frac{v-s}{s'-s}   \right) 
	 \qquad \mbox{ if } \:  v \mbox{ in }(s', +\infty).
	\eearn	
\end{lemma}
\Cref{lemma:singlecross} provides a systematic way to obtain local lower and upper bounds on the ccdf of any $\alpha$-regular distribution as a function of $\bH_{\alpha}(\cdot)$. The bound coincides  with the original function at the extreme points of the interval $[s,s']$, and provides a lower bound on the interval $[s,s']$ and an upper bound  on $[s',+\infty)$ that coincides with the function at $s'$. Furthermore, the bounds are only parameterized by $\alpha$ and the quantiles at the interval extremes. For further intuition about this lemma, we refer the reader to \cite{ABBSamples}. Applying \Cref{lemma:singlecross} to the pairs $(s,s')=(0,r^* \wedge w) $ and $(s,s')=(r^*\wedge w, r^* \vee w) $, we obtain the following lower bound
\bearn
 \inf_{\substack{F \in \aDistSetqa{q}: \\ r_F = r^*, q_F = q^*}} \: \Expect_{\Psi}[  Rev\left(p|\bF\right)]  
 &\overset{}{\geq}&   \: \int_0^{r^*\wedge\pin} p  \Gamma_{\alpha}\left( \Gamma_{\alpha}^{-1}\left(q^*\vee q \right) \frac{v}{r^* \wedge w}   \right)   d\Psi(p) \\
 && + \mathbbm{1}_{r^* \neq w } \int_{r^*\wedge\pin}^{r^*\vee\pin} (q^*\vee q) \Gamma_{\alpha}\left( \Gamma_{\alpha}^{-1}\left(\frac{q^*\wedge q}{q^*\vee q}\right) \frac{v-r^* \wedge w}{r^* \vee w-r^* \wedge w}   \right) d\Psi(p)\\ 
 &\overset{}{=}& \Expect_{\Psi}\left[  p\: \overline{G}_{\alpha,r^*\vee\pin} (p| (r^*\wedge\pin,q^*\vee q), (r^*\vee\pin,q^*\wedge q ))\right] \\
 &=& \Expect_{\Psi}[  Rev\left(p|\overline{G}_{\alpha,r^*\vee\pin} \left(\cdot| (r^*\wedge\pin,q^*\vee q), (r^*\vee\pin,q^*\wedge q )\right)\right)],
\eearn
where we used the definition of $\overline{G}_{\alpha,r^*\vee\pin} (p| (r^*\wedge\pin,q^*\vee q), (r^*\vee\pin,q^*\wedge q ))$ given in \eqref{eq:G}. 

 In  \Cref{lem:G}, stated and proved in \Cref{apx:naturepbreduction}, we establish  that the distributions $\overline{G}_{\alpha,r^*\vee\pin} (p| (r^*\wedge\pin,q^*\vee q), (r^*\vee\pin,q^*\wedge q ))$ always belong to  $\{F \mbox{ in }\aDistSetqa{q}: r_{F} = r^*, q_{F} = q^*\}$.  In turn,  this implies that the inequality above is tight.  
Returning to \eqref{eq:numerator}, we have established that the problem of nature can be written as
\bear
\inf_{F \in \aDistSetqa{q}} R(\Psi, F) 
 =  \inf_{(r^*,q^*) \in {\cal B}_{\alpha}(\pin,q)}     \int_0^{\infty} \frac{Rev\left(p|\overline{G}_{\alpha,r^*\vee\pin} \left(\cdot| (r^*\wedge\pin,q^*\vee q), (r^*\vee\pin,q^*\wedge q )\right)\right)}{Rev\left(r^*|\overline{G}_{\alpha,r^*\vee\pin} \left(\cdot| (r^*\wedge\pin,q^*\vee q), (r^*\vee\pin,q^*\wedge q )\right)\right) }  d\Psi(p) . \label{eq:LB_2D}
\eear

\begin{figure}[!ht]
\centering
\begin{minipage}{\dimexpr.5\textwidth-1em}
  
\begin{center}
\begin{tikzpicture}[scale=1, 
	declare function={
	H(\r,\qs,\w,\q,\x)=min(1,\qs*1/(1+(\qs/\q-1)*(\x-\r)/(\w-\r)));
	LB(\r,\qs,\w,\q,\x)=\x*(H(0,1,\r,\qs,\x)*(\x<\r)+and(\x>=\r, \x<\w)*H(\r,\qs,\w,\q,\x)+(\x>=\w)*(0));
	UB(\r,\qs,\w,\q,\x)=(min(\r*\qs, \x*H(\r,\qs,\w,\q,\x)*(\x<\r)+and(\x>=\r, \x<\w)*min(\x*H(0,1,\r,\qs,\x),\r*\qs) +(\x>=\w)*(\x*H(\r,\qs,\w,\q,\x)));
	},
    ]
	\begin{axis}[xmin=0, xmax=2,
	ymin=0, ymax=0.7,
	restrict y to domain=0:0.7,
	grid=both,
	minor tick num=1,
	axis line style={->},
	x label style={at={(axis description cs:0.5,-0.08)},anchor=north},
	y label style={at={(axis description cs:-0.08,.5)},rotate=0,anchor=south},
	xlabel={Value $v$},
	ylabel={Expected revenue},
	width=9cm,
	height=7cm, legend pos = south west, 
	legend style={nodes={scale=0.7, transform shape}}]
	]
	\draw [fill] (100,500) circle [radius=1.5pt];
	\draw [fill] (70,595) circle [radius=1.5pt];
	\addplot[name path=L, domain=0:2, red, very thick, samples = 400] {LB(0.7,0.85,1,0.5,x)};
	\addlegendentry{Lower bounds on $Rev(v|\bF)$};

	\path (axis cs:0.6,0.18) node(x) {\color{black} \scalebox{0.6}{$Rev\left(p|\overline{G}_{\alpha,\pin} \left(\cdot| (0,1), (r^*,q^*)\right)\right)$}}
    (axis cs:0.4,0.35) node(y) {};
    \draw[->,red,thick] (x) -- (y);
    
    \path (axis cs:1.5,0.6) node(x) {\color{black} \scalebox{0.6}{$Rev\left(p|\overline{G}_{\alpha,\pin} \left(\cdot| (r^*,q^*),(\pin,q)\right)\right)$}}
    (axis cs:0.8,0.55) node(y) {};
    \draw[->,red,thick] (x) -- (y);

	\node [right] at (100, 500) { \small $(w,wq)$};
	\node [above] at (70, 620) { \small $(r^*,r^*q^*)$};
	\end{axis}
	\end{tikzpicture}
\end{center}

\end{minipage} \hspace{0.5cm}
\begin{minipage}{\dimexpr.5\textwidth-1em}
  \begin{center}
\begin{tikzpicture}[scale=1, 
	declare function={
	H(\r,\qs,\w,\q,\x)=min(1,\qs*1/(1+(\qs/\q-1)*(\x-\r)/(\w-\r)));
	LB1(\r,\qs,\w,\q,\x)=\x*(H(0,1,\w,\q,\x)*(\x<\w)+and(\x>=\w, \x<\r)*H(\w,\q, \r,\qs,\x)+(\x>=\r)*(0));
	UB1(\r,\qs,\w,\q,\x)=1*(\x*H(\w,\q,\r,\qs,\x)*(\x<\w)+and(\x>=\w, \x<\r)*min(\x*H(0,1,\w,\q,\x),\r*\qs) +(\x>=\r)*(\x*H(\w,\q,\r,\qs,\x)));
	},
    ]
	\begin{axis}[xmin=0, xmax=2,
	ymin=0, ymax=0.7,
	restrict y to domain=0:0.7,
	grid=both,
	minor tick num=1,
	axis line style={->},
	x label style={at={(axis description cs:0.5,-0.08)},anchor=north},
	xlabel={Value $v$},
    yticklabels={,,},
	width=9cm,
	height=7cm, legend pos = south west, 
	legend style={nodes={scale=0.7, transform shape}}]
	]
	\draw [fill] (100,500) circle [radius=1.5pt];
	\draw [fill] (140,510) circle [radius=1.5pt];
	\addplot[name path=L, domain=0:2, red, very thick, samples = 400] {LB1(1.4,0.37,1,0.5,x)};
	\addlegendentry{Lower bounds on $Rev(v|\bF)$};
	
	\path (axis cs:0.9,0.25) node(x) {\color{black} \scalebox{0.8}{$Rev\left(p|\overline{G}_{\alpha,r^*} \left(\cdot| (0,1),(\pin,q)\right)\right)$}}
    (axis cs:0.6,0.4) node(y) {};
    \draw[->,red,thick] (x) -- (y);
    
    \path (axis cs:1.2,0.64) node(x) {\color{black} \scalebox{0.8}{$Rev\left(p|\overline{G}_{\alpha,r^*} \left(\cdot| (\pin,q),(r^*,q^*)\right)\right)$}}
    (axis cs:1.2,0.51) node(y) {};
    \draw[->,red,thick] (x) -- (y);

	\node [above] at (100, 525) { \small $(w,wq)$};
	\node [right] at (140, 520) { \small $(r^*,r^*q^*)$};
	\end{axis}
	\end{tikzpicture}
\end{center}

\end{minipage}
\caption{\textbf{Parametrized worst-case revenue curves.} The figure depicts, conditional on the optimal oracle price $r^*$ and revenue $r^*q^*$,  the worst-case revenue functions  obtained in the proof using the single crossing property \Cref{lemma:singlecross}. The left panel corresponds to a case where $r^* < \pin$ and the right one to a case $r^* > \pin$. For these figures, $\alpha$ is set to zero.} \label{fig:illustration-bounds}
\end{figure}
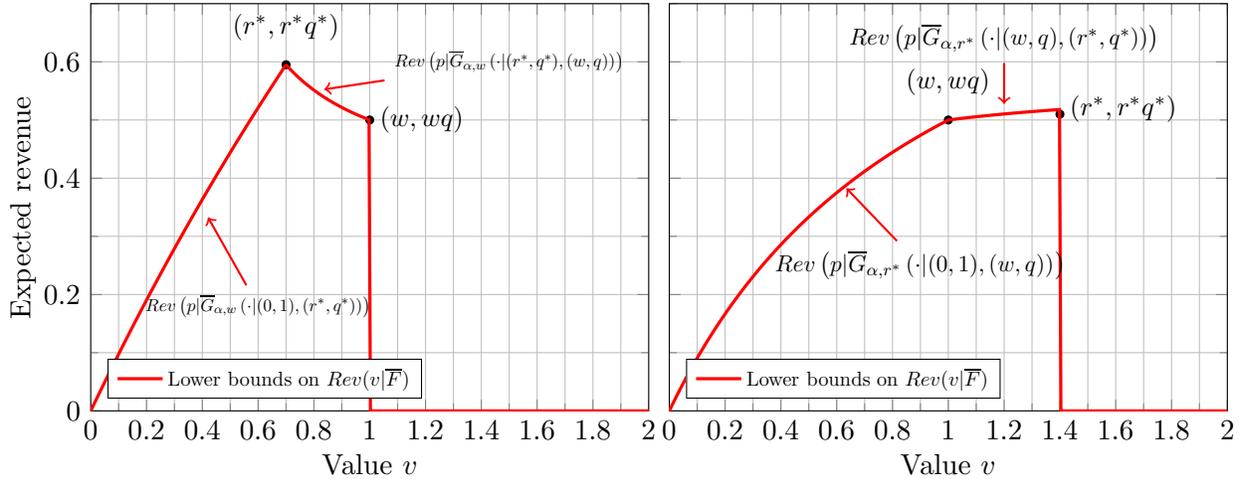

In other words, conditional on $r_F= r^*$ and $q_F = q^*$, one can pin down the worst-case distribution, and associated revenue curve. In \Cref{fig:illustration-bounds}, we illustrate the construction  of the worst-case revenue curves.

\noindent \textbf{Step 2.} In a second step, we will further reduce the minimization problem stated in Equation \eqref{eq:LB_2D} to a one dimensional minimization problem by solving exactly for the worst-case $q^*$ across instances.   To that end, we first develop an alternative characterization of  the set ${\cal B}_{\alpha}(\pin,q)$. In particular,  in \Cref{lem:opt_feas},  stated and proved in \Cref{apx:naturepbreduction}, we establish that ${\cal B}_{\alpha}(\pin,q) = {\cal B}_l \cup {\cal B}_h$, where
\bearn
{\cal B}_l &=& \Bigg\{ (r^*,q^*) \mbox{ in }[0,w) \times[0,1]: \:    q^* \geq \max \left \{\Gad{\Gainv{q} \frac{r^*}{ \pin}} , \Gad{\frac{1}{\alpha}}, \frac{q}{\Gad{\frac{\pin}{r^*}-1}}\right\}  \Bigg\}\\
{\cal B}_h &=& \Bigg\{ (r^*,q^*) \mbox{ in }[w,+\infty) \times[0,1]: \: q^* \le  \Gad{\Gainv{q} \frac{r^*}{ \pin}} , \:   q^* \geq q\Gad{\frac{1}{\alpha+\frac{\pin}{r^*- \pin}}} \Bigg\}.
\eearn
Next, define the function for each $r^*, \pin, q,$
 \bearn
  \widetilde{R}_{r^*, \pin, q}: q^* &\mapsto& \int_0^{\infty} \frac{Rev\left(p|\overline{G}_{\alpha,r^*\vee\pin} \left(\cdot| (r^*\wedge\pin,q^*\vee q), (r^*\vee\pin,q^*\wedge q )\right)\right)}{Rev\left(r^*|\overline{G}_{\alpha,r^*\vee\pin} \left(\cdot| (r^*\wedge\pin,q^*\vee q), (r^*\vee\pin,q^*\wedge q )\right)\right) }  d\Psi(p). 
\eearn
By \eqref{eq:LB_2D}, and the definition of $\widetilde{R}_{r^*, \pin, q}(\cdot)$, we have
\bearn
\inf_{F \in \aDistSetqa{q}} R(\Psi, F) 
 &=& \min \left\{ \inf_{\substack{(r^*,q^*) \in {\cal B}_l
}} \widetilde{R}_{r^*, \pin, q}(q^*), \inf_{\substack{(r^*,q^*) \in {\cal B}_{h} }} \widetilde{R}_{r^*, \pin, q}(q^*) \right\}.
\eearn
In \Cref{lem:decreasingqs}, stated and proved in \Cref{apx:naturepbreduction},  we establish the following monotonicity result: for any $r^*$ that is consistent with a pair in  ${\cal B}_{\alpha}(\pin,q)$, i.e, that belongs to $$J_{w, q} = \{ r\mbox{ s.t. there exists } q^* \mbox{ s.t. }  (r^*,q^*) \mbox{ is in }  {\cal B}_{\alpha}(\pin,q)\},$$ the function $\widetilde{R}_{r^*, \pin, q}(\cdot)$ 
is decreasing  in the set $ \{ q^*:  (r^*,q^*) \mbox{ is in } {\cal B}_{\alpha}(\pin,q)\}$. This monotonicity, in conjunction with the explicit characterization of the sets ${\cal B}_l$ and ${\cal B}_h$,  
implies that that  if $r^* < w$, fixing a feasible $r^*$,  the worst-value of $q^*$ is $1$;  and if $r^* \ge w$, fixing a feasible $r^*$, the worst-case value of $q^*$ is $ \Gamma_{\alpha}\left( \Gamma_{\alpha}^{-1}\left(q\right) \frac{r^*}{\pin}   \right)$ 
.

In turn, the problem reduces to finding the domain of possible values of $r^*$, i.e., characterizing $J_{w, q}$.  We show in \cref{lem:decreasingqs} that, for  $r^* \leq \pin$, there exists a value $q^*$ such that $(r^*,q^*) \in {\cal B}_{l}$ if and only if $r^* \geq \rl$. Hence 
\bearn
\inf_{(r^*,q^*) \in {\cal B}_{l} } \widetilde{R}_{r^*, \pin, q}(q^*) =  \inf_{r^* \in\left[\rl, w\right)} \int_0^{\pin}  \frac{Rev\left(p|\overline{G}_{\alpha,\pin} \left(\cdot| (r^*,1), (\pin,q)\right)\right)}{Rev\left(r^*|\overline{G}_{\alpha,\pin} \left(\cdot| (r^*,1), (\pin, q )\right)\right) }  d\Psi(p). 
\eearn
In turn, we show in \cref{lem:decreasingqs} that for $r^* > \pin$, there exists a value $q^*$ such that $(r^*,q^*) \in {\cal B}_{h}$ if and only if $r^* \leq \rh$. Hence 
\bearn
\inf_{(r^*,q^*) \in {\cal B}_{h}} \widetilde{R}_{r^*, \pin, q}(q^*) = \inf_{r^* \in [w, \rh]} \int_{0}^{r^*} \frac{Rev\left(p|\overline{G}_{\alpha,r^*} \left(\cdot| (0,1), (\pin,q)\right)\right)}{Rev\left(r^*|\overline{G}_{\alpha,r^*} \left(\cdot| (0,1), (\pin, q )\right)\right) }  d\Psi(p).
\eearn

Therefore we have established
\bearn
\inf_{F \in \aDistSetqa{q}} R(\Psi, F)  &=& \min \Biggl\{ \inf_{r^* \in\left[\rl, w\right)} \int_0^{\pin}  \frac{Rev\left(p|\overline{G}_{\alpha,\pin} \left(\cdot| (r^*,1), (\pin,q)\right)\right)}{Rev\left(r^*|\overline{G}_{\alpha,\pin} \left(\cdot| (r^*,1), (\pin, q )\right)\right) }  d\Psi(p) , \\
 && \inf_{r^* \in [w, \rh]} \int_{0}^{r^*} \frac{Rev\left(p|\overline{G}_{\alpha,r^*} \left(\cdot| (0,1), (\pin,q)\right)\right)}{Rev\left(r^*|\overline{G}_{\alpha,r^*} \left(\cdot| (0,1), (\pin, q )\right)\right) }  d\Psi(p))\Biggr\} \\
 &\stackrel{(a)}{=}& \min \left\{ \inf_{r \in [\rl,\pin) \cup [\pin, \rh]}  R(\Psi, F_{\alpha}( \cdot | r, (\pin, q))) \right\}\\
 &\stackrel{(b)}{=}& \inf_{F \in \aDistSetqprime{q}} R(\Psi, F),
\eearn
where $(a)$ follows from the fact that the optimal oracle price associated with $F_{\alpha}( \cdot | r, (\pin, q))$ is $r$ for all $r \mbox{ in }[\rl,\pin) \cup [\pin,\rh]$ (a fact established in \Cref{lem:rr});  in $(b)$, we use the definition of $\aDistSetqprime{q}$  in Equation \eqref{eq:subset}. 
This concludes the proof.

\section{Optimal performance for deterministic mechanisms}\label{section:deterministicmechanisms}

We are now in a position to investigate the performance of general classes of mechanisms and their associated performance.  In this section, we investigate the optimal performance when one restricts attention to  mechanisms that post a deterministic price.  
  
  \subsection{Optimal prices and performance}
  
Using \Cref{thm:LB_1D}, it is possible to obtain the following reduction.
\bearn
\Rb (\MechSet_d, \aDistSetqa{q}) \:=\: \Rb (\MechSet_d, \aDistSetqprime{q})  
                                                   \:=\: \sup_{p \in[0,\pin]} \min_{r \in[\rl,\pin) \cup [\pin,\rh ] }\frac{Rev\left(p|F_{\alpha}( \cdot | r, (\pin, q))\right) }{ \opt(F_{\alpha}( \cdot | r, (\pin, q)))},
\eearn                                                
 where for the last equality, we use the fact that one can restrict  attention to mechanisms that post a price that is less or equal than the incumbent price $\pin$, as any deterministic mechanism that posts a price strictly above the incumbent price yields zero competitive ratio in the worst-case\footnote{Any deterministic price above $\pin$ would yield a performance of zero against a distribution that puts all the mass at $\pin$.}.  In turn, we may split the possible worst-cases into   different regions to obtain   
\bear    \label{eq:main_deterministic_red}                                                
 \Rb (\MechSet_d, \aDistSetqa{q})       &=& \sup_{p \in[0,w]} \min\left \{ \min_{r \in[\rl,w)}\frac{Rev\left(p|F_{\alpha}( \cdot | r, (\pin, q))\right) }{ \opt(F_{\alpha}( \cdot | r, (\pin, q)))},  \min_{r \in[w, \rh ) } \frac{Rev\left(p|F_{\alpha}( \cdot | r, (\pin, q))\right) }{ \opt(F_{\alpha}( \cdot | r, (\pin, q)))} \right\} \nonumber\\
                                                    &\stackrel{(a)}{=}& \sup_{p \in[0,w]} \min\left \{  \min_{r \in[\rl, p)} \frac{ p\bF_{\alpha}( p | r, (\pin, q))}{ r},   \min_{r \in[p,w)} \frac{p}{r}, \min_{r \in[w, \rh ] } \frac{ p\bF_{\alpha}( p | r, (\pin, q))}{ \opt(F_{\alpha}( \cdot | r, (\pin, q)))} \right\} \nonumber\\
                                                    &\stackrel{(b)}{=}& \sup_{p \in[0,w]} \min\left \{  \min_{r \in[\rl, p)} \frac{ p\bF_{\alpha}( p | r, (\pin, q))}{ r},  \frac{p}{w}, \min_{r \in[w, \rh ] } \frac{ p\bF_{\alpha}( p | r, (\pin, q))}{ \opt(F_{\alpha}( \cdot | r, (\pin, q)))} \right\},
\eear
where $(a)$ follows from the fact that $\opt(F_{\alpha}( \cdot | r, (\pin, q)))=r$ if $r < w$ (cf. \Cref{lem:rr}) and noting that conditional on $r$ belonging to $[p,w)$, the conversion rate is equal to 1 at $p$. $(b)$ follows from noting that the worst-case in the latter case is for nature to select $r=w$. 
The reduction above highlights three ``regimes" of worst cases that may emerge, driven by the location of the oracle optimal price.\footnote{Note that in cases when $\rh<\pin$ (which happens when $\alpha \in(0,1]$ and $q < \Gad{1/\alpha}$), there are only two regimes as the last term in the brackets does not affect the worst-case.} For regular and mhr distributions, we establish that one can actually explicitly solve the problem above and characterize the spectrum of optimal transformations from data to decisions and the associated performance.
\begin{theorem}[Maximin Ratio for deterministic mechanisms]\label{thm:deterministic_perf_close}
Fix the set of mechanisms to be $\MechSet_d$.
\begin{itemize}
\item For regular distributions ($\alpha = 0$), the optimal price is given by 
\bearn
p^*_d(\pin, q) = \pin \left( \frac{ 2  \sqrt{q}}{1+\sqrt{q}} \mathbf{1}\left\{q \in\left(0,\frac{1}{4}\right]\right\} +  \frac{ q(3-4q)}{1-q} \mathbf{1}\left\{q \in\left(\frac{1}{4},\frac{1}{2}\right]\right\} +  \mathbf{1}\left\{q \in\left(\frac{1}{2}, 1\right)\right\}\right),
\eearn
and the maximin ratio is characterized as follows
\bearn
 \Rb (\dMechSet, \Fb_0(\pin, q)) =   \frac{2 \sqrt{q}}{1+\sqrt{q}} \mathbf{1}\left\{q \in\left(0,\frac{1}{4}\right]\right\} +  \frac{3-4q}{4(1-q)} \mathbf{1}\left\{q \in\left(\frac{1}{4},\frac{1}{2}\right]\right\} +  (1-q)\mathbf{1}\left\{q \in\left(\frac{1}{2}, 1\right)\right\}.
\eearn
\item For mhr distributions ($\alpha = 1$), the optimal price is given by
\bearn
p^*_d(\pin, q) = \pin \left( \beta_q\left(\frac{e}{q}\right) \mathbf{1}\{q \in(0, \hat{q}]\} +   \beta_q\left(\frac{1}{\log(q^{-1})}\right) \mathbf{1}\{q \in(\hat{q}, e^{-e^{-1}}]\} +  \mathbf{1}\{q \in(e^{-e^{-1}}, 1)\}\right),
\eearn
and the maximin ratio is characterized as follows
\bearn
\Rb (\dMechSet, \Fb_1(\pin, q)) =  \beta_q\left(\frac{e}{q}\right) \mathbf{1}\{q \in(0, \hat{q}]\} +   \rho(q) \mathbf{1}\{q \in(\hat{q}, e^{-e^{-1}}]\} +  e q \log(q^{-1}) \mathbf{1}\{q \in(e^{-e^{-1}}, 1)\},
\eearn
where if $W$ is the Lambert function defined as the inverse of $x \to x e^x$ in $[0, +\infty)$,  $\beta_q(x) = 1-\frac{1}{\log(q^{-1})}(W(x) + \frac{1}{W(x)} -2)$, $\rho(q) = \beta_q\left(\frac{1}{\log(q^{-1})}\right) e \log(q^{-1}) e^{-\log(q^{-1})\beta_q\left(\frac{1}{\log(q^{-1})}\right)}$ and $\hat{q}$ is the unique solution in $[0,1]$ to the equation $W\left(\frac{1}{\log(q^{-1})}\right) W\left(\frac{e}{q}\right) = 1$. Numerically $\hat{q} \in[0.52, 0.53]$.
\end{itemize}
\end{theorem}

The proof is presented in Appendix \ref{apx:deterministicmechanisms}.  When using deterministic mechanisms, this result enables one, quite notably,  to obtain in closed form the exact value associated with  the conversion rate at one price when using deterministic mechanisms, but  also the optimal price to post. The structure of the result is also quite instructive. There are three ``regimes": high, intermediate and low conversion rates (where those regions depend of the focal class).   For high conversion rates, the seller's optimal price is simply to continue to post the incumbent price. Intuitively, the seller may want to explore higher prices, but the seller runs the risk of losing all customers when pricing higher. In this case, the hard cases for the seller are masses at or around $\pin$.  For intermediate conversion rates and low conversion rates, the situation is different as in those cases,  a more subtle interplay arises,  and the seller needs to carefully select a price below the incumbent price to optimize its competitive ratio. We next analyze some implications of this result.

\subsection{Performance analysis}

While \Cref{thm:deterministic_perf_close} provides  a full characterization, there are various notable observations with regard to the implications of the result.

\begin{figure}[!ht]
\begin{center}
\begin{tikzpicture}
\begin{axis}[
            title={},
            xmin=0,xmax=1,
	        ymin=0.0,ymax=1, 
	        width=10cm,
	        height=8cm,
	        xlabel =  $q$, 
	        ylabel = Normalized optimal price, 
	        grid=both, 
	        legend pos=south west]
	        
\addplot[color=blue, mark=*, mark size=2.9pt] table[x=q,y=d0] {best_d.dat}; \addlegendentry{regular}

\addplot[color=red, mark=square*,  mark size=2.9pt] table[x=q,y=d] {best_d.dat}; \addlegendentry{mhr}

\end{axis}
\end{tikzpicture}
\caption{\textbf{Optimal deterministic normalized price $p^*_d(\pin, q)/w$ as a function of the probability of sale $q$.}} \label{fig:det-p} 
\end{center}
\end{figure}
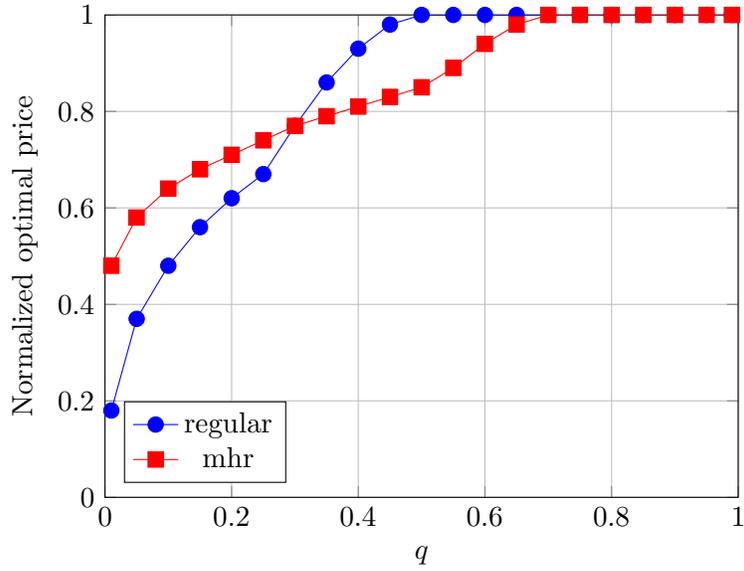

In \Cref{fig:det-p}, we plot the optimal price to post given the data at hand. 
As highlighted in \Cref{thm:deterministic_perf_close}, there are various regimes. When $q$ is ``high", then the optimal price for the seller is simply to post the incumbent price $\pin$. However, when the observed probability of sale decreases, the seller will price below the incumbent price $\pin$, potentially much below the latter. 

\begin{figure}[!ht]
\begin{center}
\begin{tikzpicture}
\begin{axis}[
            title={},
            xmin=0,xmax=1,
	        ymin=0.0,ymax=100, 
	        width=10cm,
	        height=8cm,
	        xlabel =  $q$, 
	        ylabel = Maxmin ratio in \%, 
	        grid=both, 
	        legend pos=south west]
	        
\addplot[color=blue, mark=*, mark size=2.9pt] table[x=q,y expr=(\thisrow{ylw}+\thisrow{yup})/2] {Reg_optd_data.dat}; \addlegendentry{regular}
\addplot [name path=upper,draw=none,forget plot] table[x=q,y=yup] {Reg_optd_data.dat};
\addplot [name path=lower,draw=none,forget plot] table[x=q,y=ylw] {Reg_optd_data.dat}; 
\addplot [fill=blue!50,forget plot] fill between[of=upper and lower];

\addplot[color=red, mark=square*,  mark size=2.9pt] table[x=q,y expr=(\thisrow{ylw}+\thisrow{yup})/2] {MHR_optd_data.dat}; \addlegendentry{mhr}
\addplot [name path=upper,draw=none,forget plot] table[x=q,y=yup] {MHR_optd_data.dat};
\addplot [name path=lower,draw=none,forget plot] table[x=q,y=ylw] {MHR_optd_data.dat};
\addplot [fill=red!50, forget plot] fill between[of=upper and lower];
\end{axis}
\end{tikzpicture}
\caption{\textbf{Performance of deterministic mechanisms as a function of the probability of sale $q$.}} \label{fig:det} 
\end{center}
\end{figure}
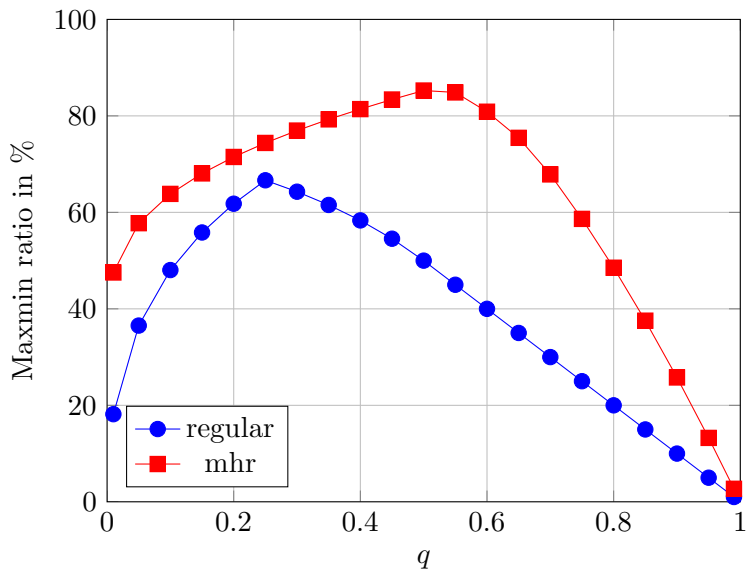

In \Cref{fig:det}, we plot the maximin ratio against regular and mhr distributions as a function of the conversion rate rate observed for the incumbent price. This value can be interpreted as measuring the value of information associated with the data when using deterministic mechanisms.

 A first striking implication is associated with the levels of performance that one can achieve with a simple deterministic mechanism, despite the very limited information available at hand. For example, when  simply knowing that the conversion rate of customers is $50\%$ at a particular price, and that the value distribution  is mhr, there exists a deterministic pricing mechanism that can guarantee more than 85\% of oracle performance!  For regular distribution, with a $25\%$ conversion rate, the seller can guarantee more than 66\% of oracle performance using an adequate deterministic mechanism. Our results provide a full mapping from historical conversion rate to achievable performance and prescription. 

 As shown in Figure \ref{fig:det}, the maximin ratio can be quite different depending on the underlying class of distributions. The  difference between the two curves highlights the ``price" of heavier tails that one could face  under regular distributions. 

  While the optimal maximin ratio converges to zero as $q \to 0$ (i.e simply knowing that no customer purchases beyond a given price does not guide pricing decisions), quite strikingly, the rate of convergence is quite \textit{slow}. Indeed, this is illustrated in Figure \ref{fig:det}. For example, with a conversion rate of $1\%$, it is still possible to guarantee more than $47\%$ against mhr distributions and more than $18\%$ against regular distributions with appropriate prices. As a matter fact, the closed form formulas in  \Cref{thm:deterministic_perf_close} show that the maximin ratio, while converging to zero as $q$ becomes small, it does so only at rate 
  $\Theta(1/\log(q^{-1}))$ for mhr and rate $\Theta(\sqrt{q})$ for regular distributions.  In other words, this  highlights that even knowing only  that a ``small" fraction of customers purchases at a given price  is very informative. We will further see that significant  more value can be captured through randomized mechanisms.

  The limiting behavior at 0  has also other implications with regard to the earlier literature. The fact that the performance converges to zero as $q$ tends to zero shows that the results in  \cite{cohen2021simple}, that a simple pricing rule can guarantee a fraction of revenues with knowledge  of the exact value of the upper bound  of the support for particular parametric families (corresponding to $q=0^+$ in the present paper), do not extend to non-parametric classes such as mhr or regular. At the same time, as highlighted above, against such distributions, the convergence of performance to zero is very slow.

\section{Optimal performance for randomized mechanisms}\label{sec:rand}

In this section, we now turn to the analysis of general randomized mechanisms. These will allow to measure the full value of information associated with percentile data.

  \subsection{Near optimal mechanisms and performance} \label{sec:rand-opt}

\paragraph{Reduction to bounded discrete mechanisms.}  As mentioned following the statement of Problem \ref{eq:general-LP}, one of the  challenges in analyzing randomized mechanisms stems from the infinite dimensional nature of the space of the seller's strategies. Next, we establish that bounded discrete mechanisms can approximate arbitrary closely general randomized mechanisms. 

More specifically, consider an increasing sequence of $N$ positive reals $\IN = \{a_i\}_{i=1}^{N}$ for $N \mbox{ in }\mathbb{N}^*$ and  define the set of discrete mechanisms on $\IN$ as 

\bearn
\MechSet_{\IN} = \left\{ \Psi \in\MechSet \: : \: 
\Psi(x) = \sum_{j = 1}^N p_j \mathbf{1} \{ x \ge a_{i}\} , \mbox{ for some } 0 \leq p_i \leq 1, \: \sum_{j = 1}^{N} p_i = 1 \right\}.
\eearn

\begin{proposition}\label{thm:mechanismreduction} 
Fix $\Psi$ in  $\MechSet$, $q \mbox{ in }(0,1)$, $N > 1$,  and any finite sequence of increasing reals $\IN= \{a_i\}_{i=1}^{N}$ such that $0 < a_1 \leq \pin \leq a_N$ . Then there exists $\Psi_{\IN} \mbox{ in }\MechSet_{\IN}$ such that 
    \bearn
    \inf_{F \in \aDistSetqa{q}} R(\Psi_{\IN}, F) \geq   \inf_{F \in \aDistSetqa{q}} R(\Psi, F) - \frac{\Delta(\IN)}{a_1}  -  \frac{1}{q(1+(q^{-1}-1)a_N)} \mathbf{1}\{a_N < \rh\},
    \eearn    
    where $\Delta(\IN) = \sup_i \{a_i - a_{i-1}\}$.

\end{proposition}
The proof of \Cref{thm:mechanismreduction} is presented in  \Cref{apx:mechanismreduction}.  This result implies two main points. First, it is possible to approximate arbitrarily closely the performance of general randomized mechanisms through \textit{discrete and bounded} mechanisms. Second, the optimality gap between the two classes can be quantified and is driven by two terms: a discretization term ($\Delta(\IN)/a_1$)associated with how fine the grid is,  and a truncation term. For mhr distributions, since $\rh$ is finite, the truncation term can be eliminated by selecting $a_N$ high enough. 

The key ideas underlying the result revolve around, first, quantifying how much the seller looses by restricting the support of the mechanism to a bounded interval $[0, b]$ with $b > \pin$. We quantify this error by leveraging the concavity of the revenue function in the quantity space as well as the upper-bound on the tail of the distribution obtained from the regularity assumption. The second step consists of quantifying how much the seller looses by restricting to the class of mechanisms that randomize over a finite set of prices using the single crossing property of regular distributions applied to local intervals (\Cref{lemma:singlecross}). It is important to note that here, because there is no exogenously imposed uniform positive lower bound on $\opt(F)$, to appropriately control losses, it is key to perform an analysis that maintains the coupling between the achieved revenues and the oracle revenues.

\paragraph{A family of factor revealing finite dimensional linear programs.} We are now ready to present the sequence of finite dimensional linear programs that will be central to our analysis. For any $q \mbox{ in }(0,1)$, and $\alpha \mbox{ in }[0,1]$, we will define a linear program parametrized by  a finite sequence of increasing positive reals $\IN = \{a_i\}_{i=0}^{2N}$, where $N >1$, such that $a_N < \pin, a_{N+1} = \pin$. In particular,  we define the following linear program.
\begin{align}
\underline{\Lb}_{\alpha, q, \IN} \:\:=\:\:  \max_{ \mathbf{p}, c} &  ~~c  \label{eq:small-LP}\tag{\PFLP}\\
s.t.    ~~   & \frac{1}{\opt(F_{\alpha}( \cdot | a_{i+1}^{-}, (\pin, q)))} \sum_{j=0}^{2N}   a_j \bF_{\alpha}( a_j | a_{i}, (\pin, q)) p_j \ge c \quad i=0,...2N, \nonumber\\
& \sum_{j=0}^{2N} p_j \le 1, \quad p_i \ge 0 \quad i=0,...2N,\nonumber
\end{align}
with $\opt(F_{\alpha}( \cdot | a_{i+1}^{-}, (\pin, q))) = \lim_{x \to a_{i+1}^-} \opt(F_{\alpha}( \cdot | x, (\pin, q)))$ for any $i = 0, \cdots, 2N$.

\begin{theorem}[Maximin Ratio for Randomized Mechanisms] \label{thm:LB_LP}
Fix $q \mbox{ in }(0,1)$, and $\alpha \mbox{ in }[0,1]$. 
\begin{enumerate}[label=(\roman*)]
\item  For any sequence  of increasing positive reals $\IN =  \{a_i\}_{i=0}^{2N}$, where $N >1$, such that $a_{N+1} = \pin$, the solution to Problem \eqref{eq:small-LP} provides a feasible distribution of prices and its performance is lower bounded by $\underline{\Lb}_{\alpha, q, \IN}$, implying that
    \bearn
    \Rb (\MechSet, \aDistSetqa{q}) \geq \underline{\Lb}_{\alpha, q, \IN}.
    \eearn
\item   Furthermore, there exists a sequence  of increasing prices $\mbox{ in }= \{a_i\}_{i=0}^{N}$ such that:
    \bearn
    \Rb (\MechSet, \aDistSetqa{q}) \leq \underline{\Lb}_{\alpha, q, \IN} +  \mathcal{O}\left(\frac{1}{\sqrt{N}}\right),
    \eearn
 where the   $\mathcal{O}$ notation includes constants that depend only on $\alpha$ and $q$. 
\end{enumerate}
\end{theorem}

The proof of \Cref{thm:LB_LP} is presented in  Appendix \ref{apx:LP_q}. \Cref{thm:LB_LP} is notable in two respects. First, it provides a systematic procedure to obtain a lower bound on performance through a linear program but also an associated pricing distribution that guarantees such performance. The second notable point is that, by judiciously constructing a discrete grid,  the values associated with a  sequence of linear programs constructed converge to the maximin ratio of interest as the grid becomes finer. The proof  is constructive and provides such a sequence.  In addition, the result implies that it suffices to solve a linear program with order $N$ variables and order $N$ constraints to yield an approximation within order $1/\sqrt{N}$ of the maximin ratio and an associated near-optimal prescription.

\paragraph{Remark (alternative feasible price sets and constraints).} \label{ref1:discrete} We note that above, we assumed that the set of feasible prices to post was any non-negative number. In practice, often, there are are constraints on the set of feasible prices to use. Such constraints can be  encoded in the framework.  Indeed, one can still apply \Cref{thm:LB_1D} as it applies to any subclass of mechanisms $\MechSet' \subseteq \MechSet$. For example, if the set of feasible prices is a discrete set $\left\{\phi_{1}, \phi_{2}, \cdots, \phi_{K}\right\}$,  the result applies when the subset $\MechSet'$ is the set of mechanisms that can only put mass over a subset of these prices. With this result in hand, one may then develop a special case of the Linear Program  \ref{eq:small-LP} in which one replaces the sequence $\IN= \{a_i\}_{i=0}^{2N}$ by the sequence $\left\{\phi_{1}, \phi_{2}, \cdots, \phi_{K}\right\}$ (and partition the latter between values below $\pin$ and above $\pin$ to partition the constraints).  In turn, the result would follow from \Cref{thm:LB_LP}. The main difference is that if we start with a discrete set of prices, one does not need to call on \Cref{thm:mechanismreduction} as the sequence on which one randomizes is pre-determined, and the discretization error would only stem from the discretization of the set of constraints. In general, the framework is flexible and could allow other constraints such as, e.g., to never put too much weight on prices above $\pin$ (as these ``riskier''). Such a constraint could be easily added to Problem \ref{eq:small-LP}.

\subsection{Performance analysis} \label{sec:rand-perf}

\begin{figure}[!ht]
\centering
\begin{minipage}{.5\textwidth}
  
\begin{center}
\begin{tikzpicture}
\begin{axis}[
            title={},
            xmin=0,xmax=1,
	        ymin=0.0,ymax=100, 
	        width=7cm,
	        height=7cm,
	        xlabel =  $q$, 
	        ylabel = maximin ratio in \%, 
	        grid=both, 
	        legend pos=south west, legend style={nodes={scale=0.8, transform shape}}]
\addplot[color=blue, mark=*, mark size=2.9pt] table[x=q,y expr=(\thisrow{ylw}+\thisrow{yup})/2] {Reg_opt_data.dat}; \addlegendentry{regular-Randomized}
\addplot [name path=upper,draw=none,forget plot] table[x=q,y=yup] {Reg_opt_data.dat};
\addplot [name path=lower,draw=none,forget plot] table[x=q,y=ylw] {Reg_opt_data.dat}; 
\addplot [fill=blue!50,forget plot] fill between[of=upper and lower];

\addplot[color=blue, mark=diamond*,  mark size=2.9pt] table[x=q,y expr=(\thisrow{ylw}+\thisrow{yup})/2] {Reg_optd_data.dat}; \addlegendentry{regular-Deterministic}
\addplot [name path=upper,draw=none,forget plot] table[x=q,y=yup] {Reg_optd_data.dat};
\addplot [name path=lower,draw=none,forget plot] table[x=q,y=ylw] {Reg_optd_data.dat};
\addplot [fill=blue!50, forget plot] fill between[of=upper and lower];

\node[anchor=east, align=center] (source) at (axis cs:0.2,30){$\Theta(\sqrt{q})$};
\path (axis cs:0.1,27) node(x) {} (axis cs:0.02,17) node(y) {};
\draw[->,black,thick] (x) -- (y);

\node[anchor=east, align=center] (source) at (axis cs:0.69,30){$\Theta(1-q)$};
\path (axis cs:0.6,25) node(x) {} (axis cs:0.99,0.3) node(y) {};
\draw[->,black,thick] (x) -- (y);

\node[anchor=east, align=center] (source) at (axis cs:0.3,75){$\Theta(\frac{1}{\log(\frac{1}{q})})$};
\path (axis cs:0.05,70) node(x) {} (axis cs:0.01,37) node(y) {};
\draw[->,black,thick] (x) -- (y);

\node[anchor=east, align=center] (source) at (axis cs:1,60){$\Theta(\frac{1}{\log(\frac{1}{1-q})})$};
\path (axis cs:0.9,52) node(x) {} (axis cs:0.99,29) node(y) {};
\draw[->,black,thick] (x) -- (y);

\end{axis}
\end{tikzpicture}
\end{center}

\end{minipage}%
\begin{minipage}{.5\textwidth}
  \begin{center}
\begin{tikzpicture}
\begin{axis}[
            title={},
            xmin=0,xmax=1,
	        ymin=0.0,ymax=100, 
	        width=7cm,
	        height=7cm,
	        xlabel =  $q$, 
	        ylabel = maximin ratio in \%, 
	        grid=both, 
	        legend pos=south west, legend pos=south west, legend style={nodes={scale=0.8, transform shape}}]
	        
\addplot[color=red, mark=*, mark size=2.9pt] table[x=q,y expr=(\thisrow{ylw}+\thisrow{yup})/2] {MHR_opt_data.dat}; \addlegendentry{mhr-Randomized}
\addplot [name path=upper,draw=none,forget plot] table[x=q,y=yup] {MHR_opt_data.dat};
\addplot [name path=lower,draw=none,forget plot] table[x=q,y=ylw] {MHR_opt_data.dat}; 
\addplot [fill=red!50,forget plot] fill between[of=upper and lower];

\addplot[color=red, mark=diamond*,  mark size=2.9pt] table[x=q,y expr=(\thisrow{ylw}+\thisrow{yup})/2] {MHR_optd_data.dat}; \addlegendentry{mhr-Deterministic}
\addplot [name path=upper,draw=none,forget plot] table[x=q,y=yup] {MHR_opt_data.dat};
\addplot [name path=lower,draw=none,forget plot] table[x=q,y=ylw] {MHR_opt_data.dat};
\addplot [fill=red!50, forget plot] fill between[of=upper and lower];

\node[anchor=east, align=center] (source) at (axis cs:0.35,40){$\Theta(\frac{1}{\log(\frac{1}{q})})$};
\path (axis cs:0.05,42) node(x) {} (axis cs:0.01,49) node(y) {};
\draw[->,black,thick] (x) -- (y);

\node[anchor=east, align=center] (source) at (axis cs:0.9,30){$\Theta(1-q)$};
\path (axis cs:0.75,25) node(x) {} (axis cs:0.99,0.3) node(y) {};
\draw[->,black,thick] (x) -- (y);

\node[anchor=east, align=center] (source) at (axis cs:0.35,85){$\Theta(\frac{1}{\log(\frac{1}{q})})$};
\path (axis cs:0.07,83) node(x) {} (axis cs:0.01,60) node(y) {};
\draw[->,black,thick] (x) -- (y);

\node[anchor=east, align=center] (source) at (axis cs:1,90){$\Theta(\frac{1}{\log(\frac{1}{1-q})})$};
\path (axis cs:0.8,83) node(x) {} (axis cs:0.98,39) node(y) {};
\draw[->,black,thick] (x) -- (y);

\end{axis}
\end{tikzpicture}
\end{center}

\end{minipage}
\caption{\textbf{Maximin ratio as a function of the probability of sale.} The figure depicts the performance of optimal randomized and deterministic mechanisms, as well as the rates of convergence to zero when $q$ approaches $0$ or 1. The left panel corresponds to regular distributions and the right one to mhr distributions.} \label{fig:rand}
\end{figure}
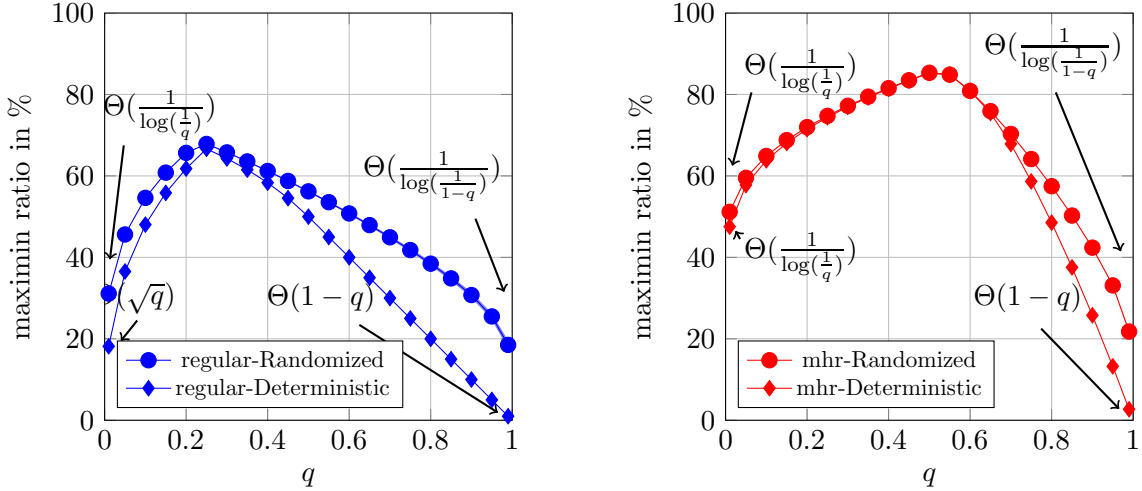

We next discuss the implications of \Cref{thm:LB_LP} in terms of performance. In  \Cref{fig:rand}, we plot the maximin ratio for randomized mechanisms superimposed with that for deterministic mechanisms.  For randomized mechanisms, we note that we plot a lower bound that is obtained by selecting the sequence used in \Cref{thm:LB_LP} (we provide further details in \Cref{apx:upb_params}). Furthermore, all the lower bounds depicted can be shown to be within 1\% of the maximin ratio by solving an alternative, but related,  linear program that can be shown to yield an upper bound (such an LP is presented in Appendix \ref{apx:upb_params}). 

The figure highlights various points. First the  value of randomization is limited for a historical price with a ``moderate" probability of sale and is more critical against distributions with heavier tails (regular) versus mhr. At the same time, the value of randomization can be quite significant for low and high conversion rates. For example, with access to a price with a probability of sale of $1\%$ against regular distributions, the performance improves from 18\% for deterministic mechanisms to 31\% for randomized ones. For a probability of sale of $75\%$, the performance improves from 25\% to 41\%. Against mhr distributions, for the previous probabilities of sale of 1\% and 75\% the performance improves from 47\% to 51\% and from 58\% to 64\%, respectively.

In \Cref{fig:rand-mech}, we illustrate the structure of the cdf associated with near-optimal mechanisms obtained by solving Problem \eqref{eq:small-LP} for two values of the probability of sale,  one in which a very small fraction of customers purchases ($q=0.01$)  and another in which  a  large fraction of customers purchase ($q=0.75$).  Some  examples for other values of $q$ are presented in Appendix \ref{apx:optmechq}. For these, without loss of generality, we fix $w=1$. 
\begin{figure}[!ht]
\centering
\begin{minipage}{.5\textwidth}
\begin{center}
\begin{tikzpicture}
	\begin{axis}[ xmin=0.001, xmax=250,
	ymin=0, ymax=1,
	restrict y to domain=0:1.0,
	grid=both,
	xmode=log,
	minor tick num=1,
	axis line style={->},,
	xlabel={ $p$},
	ylabel={ $\Psi(p)$},
	width=7cm, legend pos=north west,
	height=7cm, cycle list name=black white]
	]
	\addplot [black!60!green, very thick,dashed ]  table [ col sep=comma ] {optmech_001.csv};
	\addlegendentry{$q=0.01$};
	\addplot [black!100, very thick ]  table [ col sep=comma ] {optmech_075.csv};
	\addlegendentry{$q=0.75$};
	\end{axis}
	\end{tikzpicture}
	\caption*{\textbf{regular distributions}}
\end{center}
\end{minipage}%
\begin{minipage}{.5\textwidth}
  \begin{center}
\begin{tikzpicture}
	\begin{axis}[ xmin=0, xmax=4,
	ymin=0, ymax=1,
	restrict y to domain=0:1.0,
	grid=both,
	minor tick num=1,
	axis line style={->},
	xlabel={ $p$},
	ylabel={ $\Psi(p)$},
	width=7cm, legend pos=south east,
	height=7cm, cycle list name=black white]
	]
	\addplot [black!60!green, very thick,dashed ]  table [ col sep=comma ] {optmech_101.csv};
	\addlegendentry{$q=0.01$};
	\addplot [black!100, very thick ]  table [ col sep=comma ] {optmech_175.csv};
	\addlegendentry{$q=0.75$};
	\end{axis}
	\end{tikzpicture}
	\caption*{\textbf{mhr distributions}}
\end{center}
\end{minipage}
\caption{\textbf{Illustration of near optimal mechanisms.} The figure depicts near optimal pricing distributions for $\pin = 1$, $q = 0.01$ and $q = 0.75$. The left panel corresponds to regular distributions (plotted using a log scale) and the right panel to  mhr distributions (on a regular scale).} \label{fig:rand-mech}
\end{figure}
Recall from \Cref{fig:det-p} that for $q=0.01$, against regular distributions, the optimal deterministic price was about $0.18$. Indeed, intuitively, with so few customers purchasing at the incumbent price, the seller should consider decreasing her price.  When randomization is allowed, the (near) optimal mechanism puts mass over values between $0.05$ and $1$. This careful randomization yields an improvement in performance from $18\%$ to more than $31\%$. 
When $q=0.75$, the optimal deterministic price was simply given by $\pin =1$. Indeed, with so many customers already purchasing, it seems natural that the seller would not want to consider a decrease in price. For deterministic mechanisms, she cannot increase the price as nature could counter such a price to yield zero performance (with as mass at $\pin$).  A (near) optimal randomized mechanism puts significant mass right around 1 (about 50\% of the mass), but also inflates the current price and puts the remaining mass between $1$ and $\infty$. Here, the benefits of randomization are substantial (from a ratio of 25\% for deterministic prices to about 41.35\% for randomized prices). 
Against mhr distributions, the structure of (near) optimal mechanisms is equally rich, with mass spread below $1$, right around $1$ or above $1$ depending on the cases. 

 \paragraph{On the values  of small and large probabilities of sale.} We next explore in more detail the value of randomization for low and high values of $q$.  Our next result provides theoretical lower and upper bounds on the optimal performance $\Rb(\MechSet, \aDistSetqa{q})$ as $q \to 0$.  

\begin{proposition}\label{lem:qzerbounds}
For any $q \mbox{ in }(0, 0.4)$, and $\alpha \mbox{ in }[0,1]$ there exist $c_1, c_2 > 0$ such that 
\bearn
\frac{c_1}{\log(q^{-1})} \leq \Rb(\MechSet, \aDistSetqa{q}) \leq 
\frac{c_2}{\log(q^{-1})}.
\eearn
\end{proposition}

The proof of \Cref{lem:qzerbounds} is presented in Appendix \ref{apx:qzerbounds}.   While the optimal maximin ratio converges to zero as $q \to 0$, quite strikingly, the rate of convergence is \textit{extremely slow}, $\Theta(\log(q^{-1}))$ for both mhr and regular distributions.  In other words, this shows that even very low conversion rates are quite informative for pricing purposes. Furthermore, recalling the result for deterministic mechanisms, we see that, for regular distributions,  randomization allows to fundamentally alter the rate of convergence as $q$ approaches zero, from $\Theta(\sqrt{q})$  to $\Theta(\log(q^{-1}))$, altering the  value that can be extracted from the data. Randomization is extremely valuable with very low conversion rates. Such an effect is less pronounced for mhr distributions as the rate of convergence to zero was already extremely slow for deterministic mechanisms.

 We now explore the value of randomization for high values of $q$. Our next result provides theoretical lower and upper bounds on the optimal performance $\Rb(\MechSet, \aDistSetqa{q})$ as $q \to 1$.  

\begin{proposition}\label{lem:qonebounds}
For any $q \mbox{ in }(0.5, 1)$, and $\alpha \mbox{ in }[0,1]$, there exist $c_3, c_4 > 0$ such that
\bearn
\frac{c_3}{\log((1-q)^{-1})} \leq \Rb(\MechSet, \aDistSetqa{q}) \leq 
\frac{c_4}{\log((1-q)^{-1})}.
\eearn
\end{proposition}
The proof of \Cref{lem:qonebounds} is presented in \cref{apx:qzerbounds}. This result further highlights the significant value of randomization against both regular and mhr  distributions. Indeed, with deterministic mechanisms, the performance decreased linearly with $q$ as $q$ approached one. Now, the performance only decreases at the significantly slower rate of  $\Theta(1/\log((1-q)^{-1}))$. The value of randomization is again extremely high, as illustrated in \Cref{fig:rand}.

\section{Optimal pricing with uncertainty on the probability of sale}\label{sec:I}

In this section, we show how the ideas established in the previous sections can be generalized when the seller does not know the exact value of the probability of sale but only an interval to which it belongs $[\ql, \qh]$. In particular, we focus on  general randomized mechanisms and the object of interest is now
\bearn
\Rb (\MechSet, \aDistSet{[\ql, \qh]}) &:=&  \sup_{\Psi \in \MechSet} \inf_{F \in \aDistSet{[\ql, \qh]}} R(\Psi, F).
\eearn

 For any $\ql<\qh \mbox{ in }(0,1)^2$, $\alpha \mbox{ in }[0,1]$, $N > 1$ and any finite sequence of increasing prices $\IN= \{a_i\}_{i=1}^{2N}$, such that $a_N < \pin, a_{N+1} = \pin$, define a generalized version of \ref{eq:small-LP} given by
\begin{align}
\underline{\Lb}_{\alpha, \ql, \qh, \IN} \:\:=\:\:  \max_{ \mathbf{p}, c} &  ~~c  \label{eq:small-LP-2}\tag{\PFLPI}\\
s.t.    ~~   & \frac{1}{\opt(F_{\alpha}( \cdot | a_{i+1}^{-}, (\pin, q_l)))} \sum_{j=1}^{2N}   a_j \bF_{\alpha}( a_j | a_{i}, (\pin, q_l)) p_j \ge c \quad i=1,...N, \nonumber\\
& \frac{1}{\opt(F_{\alpha}( \cdot | a_{i+1}, (\pin, q_h)))} \sum_{j=1}^{2N}   a_j \bF_{\alpha}( a_j | a_{i}, (\pin, q_h)) p_j \ge c \quad i=N+1,...2N, \nonumber\\
& \sum_{j=1}^{2N} p_j \le 1, \quad p_i \ge 0 \quad i=1,...2N,\nonumber
\end{align}
with $\opt(F_{\alpha}( \cdot | a_{i+1}^{-}, (\pin, q_l))) = \lim_{x \to a_{i+1}^-} \opt(F_{\alpha}( \cdot | x, (\pin, q_l)))$ for any $i = 0, \cdots, N$.

\begin{theorem}\label{thm:LB_LP_robust}
Fix $\ql < \qh $ in  $(0,1)^2$, and $\alpha \mbox{ in }[0,1]$. 
\begin{enumerate}
\item  For any sequence  of increasing positive reals $\IN= \{a_i\}_{i=1}^{2N}$, where $N >1$, such that $a_{N+1} = \pin$, the solution to Problem \eqref{eq:small-LP-2} provides a feasible distribution of prices and its performance   is lower bounded by $\underline{\Lb}_{\alpha, \ql, \qh, \IN}$, yielding that
    \bearn
    \Rb (\MechSet, \aDistSet{[\ql, \qh]}) \geq \underline{\Lb}_{\alpha, \ql, \qh, \IN}.
    \eearn
\item   Furthermore, there exists a sequence  of increasing prices $\IN= \{a_i\}_{i=0}^{2N+1}$  with $a_0 = \rl[\ql] , a_{N+1}= \pin,$ such that:
    \bearn
    \Rb (\MechSet, \aDistSet{[\ql, \qh]}) \leq \underline{\Lb}_{\alpha, \ql, \qh, \IN} +  \mathcal{O}\left(\frac{1}{\sqrt{N}}\right),
    \eearn
 where the   $\mathcal{O}$ notation includes constants that depend only on $\alpha$ and $\ql, \qh$. 
\end{enumerate}
\end{theorem}

The full Proof of \Cref{thm:LB_LP_robust} is presented in Appendix \ref{apx:LP_q}. The result highlights that uncertainty in the probability of sale can be incorporated and optimal performance can again be approximated with arbitrary accuracy by solving a finite dimensional linear program.

\begin{figure}[!ht]
\centering
\begin{minipage}{.5\textwidth}
  
\begin{center}
\begin{tikzpicture}
\begin{axis}[
            title={},
            xmin=0,xmax=1,
	        ymin=20.0,ymax=90, 
	        width=7cm,
	        height=7cm,
	        xlabel = mid-point of interval $\hat{q}$, 
	        ylabel = maximin ratio in \%, 
	        grid=both, 
	        legend cell align={left},
	        legend pos=north east, legend style={nodes={scale=0.8, transform shape}}]
\addplot[color=blue!25, mark=*, mark size=2.pt] table[x=q,y=r100] {Reg_robust_data.dat}; \addlegendentry{regular, $N = 100$}
\addplot[color=blue!50, mark=*, mark size=2.pt] table[x=q,y=r500] {Reg_robust_data.dat}; \addlegendentry{regular, $N = 500$}
\addplot[color=blue!75, mark=*, mark size=2.pt] table[x=q,y=r1000] {Reg_robust_data.dat}; \addlegendentry{regular, $N = 1000$}
\addplot[color=blue!100, mark=bluestar, mark size=2.pt] table[x=q,y=rinf] {Reg_robust_data.dat}; \addlegendentry{regular, $N = \infty$}
\end{axis}
\end{tikzpicture}
\end{center}
\end{minipage}%
\begin{minipage}{.5\textwidth}
  \begin{center}
\begin{tikzpicture}
\begin{axis}[
            title={},
            xmin=0,xmax=1,
	        ymin=20.0,ymax=90, 
	        width=7cm,
	        height=7cm,
	        xlabel = mid-point of interval $\hat{q}$, 
	        ylabel = maximin ratio in \%, 
	        grid=both, 
	        legend cell align={left},
	        legend pos=south west, legend style={nodes={scale=0.8, transform shape}}]
\addplot[color=red!25, mark=*, mark size=2.pt] table[x=q,y=r100] {MHR_robust_data.dat}; \addlegendentry{mhr, $N = 100$}
\addplot[color=red!50, mark=*, mark size=2.pt] table[x=q,y=r500] {MHR_robust_data.dat}; \addlegendentry{mhr, $N = 500$}
\addplot[color=red!75, mark=*, mark size=2.pt] table[x=q,y=r1000] {MHR_robust_data.dat}; \addlegendentry{mhr, $N = 1000$}
\addplot[color=red!100, mark=redstar, mark size=2.pt] table[x=q,y=rinf] {MHR_robust_data.dat}; \addlegendentry{mhr, $N = \infty$}
\end{axis}
\end{tikzpicture}
\end{center}
\end{minipage}
\caption{\textbf{Maximin ratio as a function of the uncertainty.} The figure depicts the performance of optimal randomized  mechanisms in face of uncertainty in the probability of sale. The left panel corresponds to regular distributions and the right one to mhr distributions.} \label{fig:I}
\end{figure}
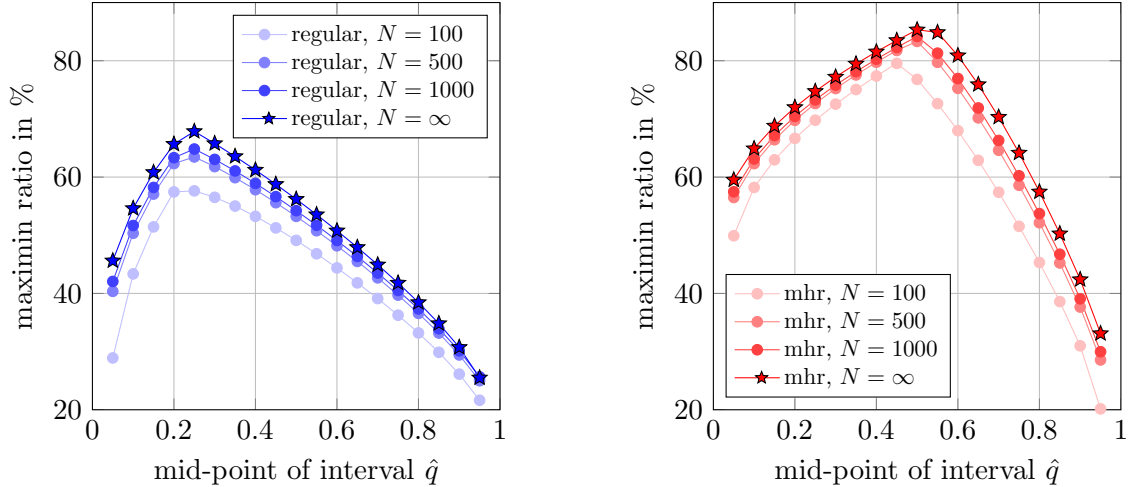

\textbf{Sensitivity Analysis.} We next illustrate the impact of uncertainty on the maximin performance. While the framework above applies to any interval, to anchor ideas we focus on the following experiment to parametrize the interval with a more ``physical" quantity.  Consider a setting in which the seller has access to $N$ buy/no-buy decisions of customers at the fixed price $\pin$. The seller can then use this data to estimate the conversion rate $\bF(p)$ through the following estimator
$$
\hat{q} = \frac{\#\mbox{buy decisions}}{N}, 
$$
We will fix an uncertainty interval through 
$$[\ql,\qh] = \left[\hat{q} - 1.96 \frac{\sqrt{\hat{q}(1-\hat{q})}}{\sqrt{N}}, \hat{q} + 1.96 \frac{\sqrt{\hat{q}(1-\hat{q})}}{\sqrt{N}}\right],$$
inspired by a 95\% confidence interval. In  \Cref{fig:I}, we present the maximin ratio as a function of the mid-point $\hat{q}$ and $N$ for various values of  $N \mbox{ in }[100, 500, 1000, \infty]$, for both mhr and regular distributions.

The figure quantifies the impact of uncertainty in the probability of sale on performance degradation. The case $N=\infty$ corresponds the case of known probability of sale. We observe the seller can still leverage the noisy information to achieve high levels of performance despite the uncertainty in the conversion rate.

\section{Conclusion}

In the present paper, we present a systematic analysis of pricing in the presence of a common data structure,  historical data at a single price. We propose a novel and general framework that allows to obtain how such data should be optimally used and the best performance one can achieve. The novel framework is powerful but also yields novel insights on the value of such information, the power of simple deterministic mechanisms, but also the incremental value of randomized mechanisms and the regimes in which it is most significant.

There are many avenues of future research that this work opens up. This framework offers a framework to quantify in a robust manner the value of a single measurement. As such, it offers a foundation for future work that can tackle how to leverage  measurements at multiple prices. There, a key question would be how to find a parallel reduction to \Cref{thm:LB_1D}.   More broadly, a promising direction is also to leverage such analyses to inform the the design of static and dynamic price  experiments.

\setstretch{1.0}
\bibliographystyle{agsm}
\bibliography{Main_MS_paper_R1}

\newgeometry{right=0.8in, top=1in, bottom=1.2in, left=0.8in}

\newpage

\begin{center}
 {\Large \textbf{Electronic Companion: 
 Appendix for \\
Optimal Pricing with a Single Point\\}}
\medskip
Amine Allouah, Facebook Core Data Science, \texttt{mallouah19@gsb.columbia.edu}\\
Achraf Bahamou, Columbia University, IEOR department, \texttt{achraf.bahamou@columbia.edu}\\
Omar Besbes, Columbia University, Graduate School of Business, \texttt{obesbes@columbia.edu}

\end{center}

\pagenumbering{arabic}
\renewcommand{\thepage}{App-\arabic{page}}
\setcounter{page}{1}
\setcounter{section}{0}
\setcounter{proposition}{0}
\setcounter{lemma}{0}

\renewcommand{\theequation}{\thesection-\arabic{equation}}
\renewcommand{\thetheorem}{\thesection-\arabic{equation}}
\renewcommand{\theproposition}{\thesection-\arabic{proposition}}
\renewcommand{\thelemma}{\thesection-\arabic{lemma}}

\appendix 
\setcounter{equation}{0}
\setcounter{proposition}{0}
\setcounter{lemma}{0}

\section{Preliminaries and properties of Generalized Pareto Distributions} 

Throughout the paper, whenever a distribution $F$ is defined, and when clear from context, we use $q_w$ to denote $\bF(w)$ to lighten the notation. We also use the generalized inverse of a distribution $F$ in $\distSet$, defined by  $F^{-1}(1-q):=\inf \{ v \mbox{ in } \mathbb{R}^+ \mbox{ s.t. } F(v) \ge 1-q  \}$ for all $q$ in $[0,1]$.

For a distribution with positive density function $f$ on its support $[a, b]$, where $0 \leq a < \infty$ and $a \leq b \leq \infty$, we denote the $\alpha$-virtual value function for $v \in [a,b]$ by
\bearn 
\phi_{F}^{\alpha}(v) := (1-\alpha) v - \frac{\bF(v)}{f(v)}.
\eearn

\begin{lemma}\label{lem:const} 
	Fix two scalars $\beta > 0$ and $s\ge0$.	The cumulative distribution  function $ \Gamma_{\alpha}\left(\beta\left(v-s\right)\right)$ for $v \ge s$ admits a constant $\alpha$-virtual value  function given by  
\bearn
 (1-\alpha) s - \frac{1}{\beta}.
\eearn
\end{lemma}

\begin{proof}[\textbf{\underline{Proof of \Cref{lem:const}}}] 
Let us first explicitly compute the $\alpha$-virtual value function.  The  derivative of $ \Gamma_{\alpha}\left(\beta\left(v-s\right)\right)$, for any $v \ge s$, is given by
\bearn
-\beta  \left(\Gad{ \beta(v-s)}\right)^{2-\alpha}.
\eearn
Therefore, the $\alpha$-virtual value function evaluated at $v \ge s$ is given by
\bearn
 (1-\alpha) v - \frac{\Gamma_{\alpha}\left( \beta (v-s) \right)}{\beta  \left(\Gad{ \beta(v-s)}\right)^{2-\alpha}} 
&=& (1-\alpha) v - \frac{1}{\beta} \left(\Gad{ \beta(v-s)}\right)^{\alpha-1} \\
&=& (1-\alpha) v - \frac{1}{\beta} \left( 1 + (1-\alpha) \Gainv{\beta(v-s)}\right) \\
&=& (1-\alpha) v - \frac{1}{\beta} - (1-\alpha) (v-s) \\
&=& (1-\alpha) s - \frac{1}{\beta}. 
\eearn
This completes the proof.
\end{proof}

We will also need the following result derived in \citep[Lemma E-1]{ABBSamples}.

\begin{lemma}\label{lemma:reserve_price_gamma_lambda}
	Fix $\alpha \in [0,1]$, two scalars $\beta \ge 0$ and $w'\ge0$.	The revenue function $v \Gamma_{\alpha}\left(\beta\left(v-w'\right)\right)$ for $v \ge w'-\frac{1}{(1-\alpha)\beta}$ is unimodal and attains its maximum at 
	\bearn
	r = \max\left\{\frac{1-(1-\alpha) \beta w'}{\beta \alpha }, w'-\frac{1}{(1-\alpha)\beta}\right\}.
	\eearn	
	With the following conventions: 	$\max\{+\infty, v\}=+\infty$, and  $\max\{-\infty, v\}=v$  for any real number  $v$.

\end{lemma}

\setcounter{equation}{0}
\setcounter{theorem}{0}
\setcounter{lemma}{0}
\setcounter{proposition}{0}

\setcounter{equation}{0}
\setcounter{proposition}{0}
\setcounter{lemma}{0}

\setcounter{equation}{0}
\setcounter{theorem}{0}
\setcounter{lemma}{0}
\setcounter{proposition}{0}

\setcounter{equation}{0}
\setcounter{proposition}{0}
\setcounter{lemma}{0}

\section{Proofs and auxiliary results for \texorpdfstring{\Cref{sec:pbf}}{Section 2} }\label{apx:generalDistribution}

\begin{proof}[\underline{\textbf{Proof of \Cref{prop:generalclass}}}]
If $q \mbox{ in }\{0,1\}$, then the result follows from \Cref{prop:qzerone}. For $q \mbox{ in }(0,1)$, let $\Psi$ a mechanism in $\MechSet$. We know that 
\bearn
\lim_{u \to \infty} \bar{\Psi}(u) := 1- \Psi(u) = 0.
\eearn
Fix $\epsilon>0$. By definition of the limit,  there exists $M \ge w$ such that for any $u \geq M $, we have:
\bear
\bar{\Psi}(u) \leq \frac{\epsilon}{2}. \label{ineq:limitCCDFGeneral}
\eear
For any integer $N$, consider the following distribution $F_{\Psi, N}$ defined through its Complementary Cumulative Distribution Function  $\bF_{\Psi, N}$:
\bearn
\bF_{\Psi, N}(v) &=& \begin{cases}
             1 &\quad \mbox{if } v < 0,\\
             q &\quad \mbox{if } v \mbox{ in } [0, w), \\
             \frac{q}{N} & \quad \mbox{if } v \mbox{ in }[w, N^2 M), \\
             0 & \quad \mbox{if } v \mbox{ in }[N^2 M, +\infty).
        \end{cases}
\eearn
Note that $F_{\Psi, N} \mbox{ in }\Gb (w, I)$  and that $F_{\Psi, N}$ represents a three point mass distribution with mass at points $0$, $w$ and $N^2 M$. 

Note also that $\opt(F_{\Psi, N}) = \max \left\{q w, q N M \right\}.$
 Since $M \ge w$ and $N \ge 1$, we have $\opt(F_{\Psi, N}) =  q N M$. Thus the performance of mechanism $\Psi$ is given by
\bearn
R(\Psi, F_{\Psi, N}) &=& \frac{1}{{q N M}} \left( {\int_{[0,w)} u \bF_{\Psi, N} (u) d \Psi(u) + \int_{[w,N^2M)}u \bF_{\Psi, N} (u) d \Psi(u) + \int_{[N^2M,+\infty)} u \bF_{\Psi, N} (u) d \Psi(u)} \right) \\
&=&  \frac{1}{N M}   \left({\int_{[0,w)} u d \Psi(u) + \frac{1}{N} \int_{[w,N^2M)} u d \Psi(u)} \right)\\
&=& \frac{1}{N M} \left(  {\int_{[0,w)}u d \Psi(u) + \frac{1}{N} \int_{[w,M)} u d \Psi(u) + \frac{1}{N} \int_{[M,N^2M)} u d \Psi(u)} \right) \\
&\leq& \frac{1}{N M} \left(   {w \Psi(w)} + {\frac{1}{N} M (\Psi(M)- \Psi(0))} +{\frac{1}{N} N^2M (\Psi(N^2M)-\Psi(M))} \right) \\
&\leq&  \frac{w}{N M} + \frac{1}{N^2} + (1-\Psi(M)),
\eearn
where in the last step we use the fact that for any $u \ge 0$ $\Psi(u)$ is in $[0, 1].$

Let us now choose $N$ large enough such that $$ \frac{w}{N M} + \frac{1}{N^2} \leq \frac{\epsilon}{2},$$

Combining the latter with \eqref{ineq:limitCCDFGeneral}, we get
\bearn
R(\Psi, F_{\Psi, N}) \leq \frac{\epsilon}{2} + \frac{\epsilon}{2} = \epsilon.
\eearn
Taking $\epsilon \to 0$ concludes the proof.
\end{proof}

\subsection{Additional result for the cases \texorpdfstring{$q \mbox{ in }\{0, 1\}$}{}}\label{apx:qonezero} 

\begin{lemma}\label{prop:qzerone}
	For any mechanism $\Psi \mbox{ in }\MechSet$, and any $\alpha \mbox{ in }[0,1]$, if $q \mbox{ in }\{0,1\}$, then 
	\bearn
	\inf _{F \in \aDistSetq[\alpha]{q}} R(\Psi, F)   &=& 0.
	\eearn
\end{lemma}

\begin{proof}[\textbf{\underline{Proof of \Cref{prop:qzerone}}}] We will first show the case when $q=0$ then the case when $q=1$. For both cases, we will exhibit worst case families of distributions for which the seller cannot achieve a non-trivial guarantee.

For any $r>0$, let us introduce  the following distribution through its Complementary Cumulative Distribution Function:
\bearn
		\bF_r(v) &=&
		\begin{cases}
			1 &\quad \mbox{if } v \mbox{ in }[0,r),\\
			0 & \quad \mbox{if } v \mbox{ in }[r, +\infty).
		\end{cases}
\eearn
The latter distribution represents a point mass at $r$.

\paragraph{Case $q=0$.}  We have, for any  $r<w$, $F_r \mbox{ in }\Fa(w,\{0\})$ and $\opt(F_r) = r$. Furthermore,  for any mechanism $\Psi \mbox{ in }\MechSet$, we have:

\bearn
\inf_{F \in \Fa(w,\{0\})} R(\Psi, F) &\leq& \frac{\Expect_{\Psi}[  p\bF_r(p)]} {\opt(F_r)} = \frac{1}{r} \int_{0}^{r} u d\Psi(u) = \frac{1}{r} \int_{0}^{r} \int_{0}^{u} ds d\Psi(u) \\
    &\overset{(a)}{=}& \frac{1}{r} \int_{0}^{r} \int_{s}^{r} d\Psi(u) ds = \frac{1}{r} \int_{0}^{r} \left(\Psi(r) - \Psi(s) \right)ds \\ 
   &\le& \Psi(r) - \Psi(0). 
\eearn
Where in equality (a), we used Fubini–Tonelli theorem as $(s,u) \to 1$ is a non-negative measurable function and $([0,r], d\Psi)$ and $([0,r], dx)$ are $\sigma$-finite measure spaces.

The right hand side above converges to zero as $r \to 0^+$ since $\Psi \mbox{ in }\MechSet = \distSet$ and is therefore right continuous. We conclude the case $q = 0.$

\paragraph{Case $q=1$.} 
We have, for any  $r>w$, $F_r \mbox{ in }\Fa(w,\{1\})$ and $\opt(F_r) = r$. Furthermore, for any mechanism $\Psi$, we have:
\bearn
    \inf_{F \in \Fa(w,\{1\})} R(\Psi, F)  &\leq& \frac{\Expect_{\Psi}[  p\bF_r(p)]} {\opt(F_r)} = \frac{1}{r} \int_{0}^{r} u d\Psi(u) = \frac{1}{r} \int_{0}^{r} (\Psi(r)-\Psi(u)) du.
\eearn
Since $\Psi \mbox{ in }\MechSet$, we have $\lim_{r \rightarrow +\infty} \Psi(r) = 1$, therefore, for any $\epsilon > 0$, there exists $A >0$ such that:
\bear
1-\Psi(r) = |\Psi(r)-1| < \epsilon \quad \mbox{if } r  \in [A,+\infty). \label{ineq:limitCDF}
\eear
Let $r \mbox{ in }(A,+\infty)$, we have:
\bearn
    \inf_{F \in \Fa(w,\{1\})} R(\Psi, F)   &\leq& \frac{1}{r} \int_{0}^{r} (\Psi(r)-\Psi(u)) du \\
    &=&  \frac{1}{r} \left( \int_{[0,A)} \left(\Psi(r)-\Psi(u) \right)du + \int_{[A,r]} \left(\Psi(r)-\Psi(u) \right)du \right) \\
        &\overset{(a)}{ \le }&  \frac{1}{r} \left( _{[0,A)}  du + \int_{[A,r]} \left(1-\Psi(u) \right)du \right) \\
      &=&  \frac{A}{r} + \frac{1}{r}  \int_{A}^{r} \left( 1-\Psi(u) \right) du  \\
      &\overset{(b)}{ \le }& \frac{A}{r} + \frac{r-A}{r} \epsilon \overset{(c)}{ \le } \frac{A}{r} + \epsilon,
\eearn
where in $(a)$ we use the fact that for any $u \ge 0$, we have $0 \le \Psi(u) \le 1.$ And in $(b)$ we use  \eqref{ineq:limitCDF}. In (c), we used the fact that $r-A \le r.$

Hence we conclude that for any $r \ge A/\epsilon$, we get that
\bearn
   \inf_{F \in \Fa(w,\{1\})} R(\Psi, F) \leq 2 \epsilon.
\eearn
Since $\epsilon$ was arbitrary, this completes the proof for the case $q=1$.
\end{proof}

\setcounter{equation}{0}
\setcounter{theorem}{0}
\setcounter{lemma}{0}
\setcounter{proposition}{0}

\setcounter{equation}{0}
\setcounter{proposition}{0}
\setcounter{lemma}{0}

\section{Proofs and auxiliary results  for \texorpdfstring{\Cref{section:naturepbreduction}}{Section 3}}\label{apx:naturepbreduction}

\begin{lemma}\label{lem:alpharegG}
For any $(s, q_s), (s', q_{s'}) \mbox{ in }([0,+\infty) \times [0,1])^2$ such that $s \le s'$ and $q_s \ge q_{s'}>0$, the distribution $G_{\alpha,t}(\cdot | (s, q_s),(s', q_{s'}))$ with $t \ge s'$, defined in \eqref{eq:G}, belongs to $\Fa(s, q_s) \cap \Fa(s', q_{s'})$ if and only if
\bearn
q_s\ge \Gad{\Gainv{q_{s'}}\frac{s}{s'}}.
\eearn
\end{lemma} 

\begin{proof}[\underline{\textbf{Proof of \Cref{lem:alpharegG}}}]

Let us show each direction.

$\Longrightarrow$) If the distribution  $G_{\alpha,t}(\cdot | (s, q_s),(s', q_{s'}))$ belongs to $\Fa(s, q_s) \cap \Fa(s', q_{s'})$ then $G_{\alpha,t}(\cdot | (s, q_s),(s', q_{s'}))$ is $\alpha-$regular, therefore by \Cref{lemma:singlecross} applied to the interval $[0, s']$, we have that 
\bearn
q_s =  \bG_{\alpha,t}(s | (s, q_s),(s', q_{s'})) \ge 
  \Gad{\Gainv{q_{s'}}\frac{s}{s'}},
\eearn
and hence the first direction is established.  Let us now show the other direction.

$\Longleftarrow$) Suppose now that $q_s\ge \Gad{\Gainv{q_{s'}}\frac{s}{s'}}$. By definition of $G_{\alpha,t}(\cdot | (s, q_s),(s', q_{s'}))$, we have:
\bearn
\bG_{\alpha,t}(s | (s, q_s),(s', q_{s'})) &=& q_s \\
\bG_{\alpha,t}(s' | (s, q_s),(s', q_{s'})) &=& q_{s'}.
\eearn
Therefore, we only have to show that the distribution $G_{\alpha,t}(\cdot | (s, q_s),(s', q_{s'}))$ is $\alpha$-regular. Using \Cref{lem:const}, the associated $\alpha$-virtual value function  is given by
\bearn
\phi_{G_{\alpha,t}(\cdot | (s, q_s),(s', q_{s'}))}^{\alpha}(v) &=& \begin{cases}
			-\frac{s}{\Gainv{q_s}} &\quad \mbox{if } v \mbox{ in }[0,s],\\
			(1-\alpha) s - \frac{s'-s}{\Gainv{\frac{q_{s'}}{q_s}}} &\quad \mbox{if } v \mbox{ in }(s, s'].	\end{cases}
\eearn

Thus the $\alpha$-virtual value function is piece-wise constant. Now we need to show that $\phi_{G_{\alpha,t}(\cdot | (s, q_s),(s', q_{s'}))}^{\alpha}(v) $ is non-decreasing. Next, we evaluate the difference between the two constant values that the virtual value function is taking. 
\bearn
&&(1-\alpha) s - \frac{s'-s}{\Gainv{\frac{q_{s'}}{q_s}}} - (-\frac{s}{\Gainv{q_s}}) \\
&=& s \frac{(1-\alpha)\Gainv{q_s}\Gainv{\frac{q_{s'}}{q_s}} + \Gainv{q_s} + \Gainv{\frac{q_{s'}}{q_s}} - \frac{s'}{s}\Gainv{q_s}}{\Gainv{q_s}\Gainv{\frac{q_{s'}}{q_s}}} \\
 &\overset{(a)}{=}& s \frac{\Gainv{q_s\frac{q_{s'}}{q_s}} - \frac{s'}{s}\Gainv{q_s}}{\Gainv{q_s}\Gainv{\frac{q_{s'}}{q_s}}} \\
  &=& s \frac{\Gainv{q_{s'}} - \frac{s'}{s}\Gainv{q_s}}{\Gainv{q_s}\Gainv{\frac{q_{s'}}{q_s}}}\overset{(b)}{\geq}  0, 
\eearn
where (a) stems from the fact that $\Gainv{uv} =  \Gainv{u} +  \Gainv{v} + (1-\alpha) \Gainv{u} \Gainv{v}$ and (b) is due to the fact that by assumption $q_s \geq \Gad{\frac{s}{s'} \Gainv{q_{s'}}}$ and that the function $\Gad{\cdot}$ is non increasing. This shows that the $\alpha$-virtual value function of $G_{\alpha,t}(\cdot |  (s, q_s),(s', q_{s'}))$ is non decreasing. This concludes the proof.
\end{proof}

\begin{lemma}\label{lem:rr} 
Let $\alpha \mbox{ in }[0,1]$, $\pin >0$,  $q \mbox{ in }(0,1)$, and $r \mbox{ in }[\rl,\pin) \cup[\pin,\rh]$. Then the optimal price associated with $F_{\alpha}( \cdot | r, (\pin, q))$ is given by  $r$.
\end{lemma}

\begin{proof}[\textbf{\underline{Proof of \Cref{lem:rr}}}]
We compute the virtual value function  for the function $F_{\alpha}( \cdot | r, (\pin, q))$. Since the definition of $F_{\alpha}( \cdot | r, (\pin, q))$ depends on whether $r < w$ or $r \ge w$, we treat each case separately.

\paragraph{Case 1: $r \in [\rl,\pin)$:}
By applying \Cref{lem:const} for the pair $((r,1), (\pin,q))$, we get the virtual value function at $v \ge r$ satisfies
\bearn 
\phi_{F_{\alpha}( \cdot | r, (\pin, q))}^{0}(v)
&=& \alpha v + (1-\alpha) r - \frac{w-r}{\Gainv{q}}\\
&\ge& r \left(1+ \frac{1}{\Gainv{q}}\right) -  \frac{w}{\Gainv{q}}\\
&=& \left(1+ \frac{1}{\Gainv{q}}\right) \left(r  -  \rl \right),
\eearn
since $r \ge \rl$, then we conclude that $ \phi_{F_{\alpha}( \cdot | r, (\pin, q))}^{0}(v) \ge 0$. Now, since $r$ is the lower support of the distribution $F_{\alpha}( \cdot | r, (\pin, q))$ in the case $r < w$ and $F_{\alpha}( \cdot | r, (\pin, q))$ is regular,  we conclude that necessarily the optimal price is  at $r$. 

\paragraph{Case 2: $r \in [\pin,\rh]$:} In this case, we assume $\rh \ge \pin$, otherwise the set is empty. 
Similarly, by applying \Cref{lem:const} for the pair $((0,1), (w,q)),$ we get that the virtual value function at $v < r$ satisfies
\bearn 
\phi_{F_{\alpha}( \cdot | r, (\pin, q))}^{0}(v) 
&=& \alpha v + \left(0 - \frac{w}{\Gainv{q}} \right) 
\:=\: \alpha \left(v - \rh \right).
\eearn
Since $v < r \le \rh$ , we conclude that $\phi_{F_{\alpha}( \cdot | r, (\pin, q))}^{0}(v) \le 0$. Now, since $r$ is the upper support of the distribution $F_{\alpha}( \cdot | r, (\pin, q))$ in the case $r \ge w$ and $F_{\alpha}( \cdot | r, (\pin, q))$ is regular, we conclude that necessarily the optimal price is given by $r$. 

\end{proof}

\begin{lemma}\label{lem:G}
The distribution $G_{\alpha,r^*\vee\pin} (\cdot| (r^*\wedge\pin,q^*\vee q), (r^*\vee\pin,q^*\wedge q ))$, defined in Eq. \eqref{eq:G}, belongs to $\{F \mbox{ in } \aDistSetqa{q}: r_{F} = r^*, q_{F} = q^*\}$ if and only if $(r^*, q^*)$ belongs to ${\cal B}_{\alpha}(\pin,q)$.
\end{lemma} 

\begin{proof}[\underline{\textbf{Proof of \Cref{lem:G}}}] One direction of the proof is direct. In particular, 
if the distribution $G_{\alpha,r^*\vee\pin} (\cdot| (r^*\wedge\pin,q^*\vee q), (r^*\vee\pin,q^*\wedge q ))$ belongs to $\{F \mbox{ in }\aDistSetqa{q}: r_{F} = r^*, q_{F} = q^*\}$ then by definition we have $(r^*, q^*) \mbox{ in }{\cal B}_{\alpha}(\pin,q)$. 

Let us now show the other direction, and suppose that $(r^*, q^*)$ belongs to ${\cal B}_{\alpha}(\pin,q)$ and let $F \mbox{ in }\aDistSetqa{q}$ be a corresponding distribution with $r_F=r^*$ and $q_F=q^*$. 
We will first  show that $G_{\alpha,r^*\vee\pin} (\cdot| (r^*\wedge\pin,q^*\vee q), (r^*\vee\pin,q^*\wedge q ))$ belongs to $\Fa$ and that the revenue curve of $F$ is lower bounded by the revenue curve of $G_{\alpha,r^*\vee\pin} (\cdot| (r^*\wedge\pin,q^*\vee q), (r^*\vee\pin,q^*\wedge q ))$. In a second step, we will show that the optimal revenue of $G_{\alpha,r^*\vee\pin} (\cdot| (r^*\wedge\pin,q^*\vee q), (r^*\vee\pin,q^*\wedge q ))$ is achieved at $r^*$.

\paragraph{Step 1:} We  separate the cases  $r^* < \pin$ and $r^* \geq \pin$.

\underline{\textbf{Case 1:}} $r^* < \pin$. By \Cref{lemma:singlecross}, note that we have that $q^* = \bF(r^*) \ge \Gad{\Gainv{q} \frac{r^*}{\pin}}$. By \Cref{lem:alpharegG},  applied to the following parameters $(s, q_s)=(r^*, q^*)$ and  $(s', q_{s'})=(w, q)$, $G_{\alpha,\pin} (\cdot| (r^*,q^*), (\pin,q ))$ belongs to $\Fa$. Furthermore, by \Cref{lemma:singlecross} again,  we have that
\bearn
v \bF(v) &\geq& \begin{cases}
			 v \bG_{\alpha,\pin} (v| (r^*,q^*), (\pin,q ))  &\quad \mbox{if } v \in [0,r^*], \\
			 v \bG_{\alpha,\pin} (v| (r^*,q^*), (\pin,q )) & \quad \mbox{if } v \in (r^*,\pin].
		\end{cases}
\eearn

\underline{\textbf{Case 2:}} $r^* \geq \pin$. By \Cref{lemma:singlecross}, we have that $q = \bF(\pin) \ge  \Gad{\Gainv{q^*} \frac{\pin}{r^*}}$ and hence, by \Cref{lem:alpharegG} applied to the following parameters $(s, q_s)=(w, q)$ and  $(s', q_{s'})=(r^*, q^*)$, $G_{\alpha,r^*} (\cdot|  (\pin,q ),(r^*,q^*))$ belongs to $\Fa$. Furthermore, by \Cref{lemma:singlecross} again, we have that
\bearn
v \bF(v) &\geq& \begin{cases}
			 v \bG_{\alpha,r^*} (v| (\pin,q ), (r^*,q^*))  &\quad \mbox{if } v \mbox{ in }[0,\pin], \\
			 v \bG_{\alpha,r^*} (v| (\pin,q ), (r^*,q^*)) & \quad \mbox{if } v \mbox{ in }[\pin, r^*].
		\end{cases}
\eearn

Therefore in both cases, we have that $v \bF(v) \geq v \bG_{\alpha,r^*\vee\pin} (v| (r^*\wedge\pin,q^*\vee q), (r^*\vee\pin,q^*\wedge q ))$ for all $v \mbox{ in }[0, +\infty)$ and $G_{\alpha,r^*\vee\pin} (\cdot| (r^*\wedge\pin,q^*\vee q), (r^*\vee\pin,q^*\wedge q ))$ is $\alpha$-regular. 

\paragraph{Step 2:}
To conclude the proof we will show that the optimal revenue associated with the distribution $G_{\alpha,r^*\vee\pin} (\cdot| (r^*\wedge\pin,q^*\vee q), (r^*\vee\pin,q^*\wedge q ))$ is achieved at $r^*.$ We will show that by contradiction. 

Since $G_{\alpha,r^*\vee\pin} (\cdot| (r^*\wedge\pin,q^*\vee q), (r^*\vee\pin,q^*\wedge q ))$ is $\alpha$-regular, then  the associated revenue function is unimodal and achieves its maximum at some point $r_{G}$ in $[0, +\infty)$.  Suppose for a moment  $r_G \bG_{\alpha,r^*\vee\pin} (r_G| (r^*\wedge\pin,q^*\vee q), (r^*\vee\pin,q^*\wedge q )) > r^* \bG_{\alpha,r^*\vee\pin} (r^*| (r^*\wedge\pin,q^*\vee q), (r^*\vee\pin,q^*\wedge q )) $. Then, using the above lower-bounds,  one would have \bearn 
r_G\bF(r_G) &\geq& r_G \bG_{\alpha,r^*\vee\pin} (r_G| (r^*\wedge\pin,q^*\vee q), (r^*\vee\pin,q^*\wedge q )) \\
&>& r^* \bG_{\alpha,r^*\vee\pin} (r^*| (r^*\wedge\pin,q^*\vee q), (r^*\vee\pin,q^*\wedge q ))\\
&=& r^* q^* = r_F\bF(r_F),
\eearn 
which would contradict the optimality of $r_F$. 
 Hence $G_{\alpha,r^*\vee\pin} (\cdot| (r^*\wedge\pin,q^*\vee q), (r^*\vee\pin,q^*\wedge q ))$ belongs to $\{F \mbox{ in }\aDistSetqa{q}: r_{F} = r^*, q_{F} = q^*\}$.

\end{proof}


\begin{lemma}\label{lem:opt_feas}
A pair $(r^*,q^*)$ in $\mathbb{R}_{+} \times[0,1]$ belongs to  ${\cal B}_{\alpha}(\pin,q)$ if and only if it belongs to ${\cal B}_l \cup {\cal B}_h $, where
\bearn
{\cal B}_l &=& \Bigg\{ (r^*,q^*) \mbox{ in }[0,w) \times[0,1]: \:    q^* \geq \max \left \{\Gad{\Gainv{q} \frac{r^*}{ \pin}} , \Gad{\frac{1}{\alpha}}, \frac{q}{\Gad{\frac{\pin}{r^*}-1}}\right\}  \Bigg\},\\
{\cal B}_h &=& \Bigg\{ (r^*,q^*) \mbox{ in }[w,+\infty) \times[0,1]: \: q^* \le  \Gad{\Gainv{q} \frac{r^*}{ \pin}} , \:   q^* \geq q\Gad{\frac{1}{\alpha+\frac{\pin}{r^*- \pin}}} \Bigg\}.
\eearn
\end{lemma}
\begin{proof}[\underline{\textbf{Proof of \Cref{lem:opt_feas}}}]

By \Cref{lem:G}, we have that $(r^*, q^*) \mbox{ in }{\cal B}_{\alpha}(\pin,q)$ if and only if the distribution $G_{\alpha,r^*\vee\pin} (\cdot| (r^*\wedge\pin,q^*\vee q), (r^*\vee\pin,q^*\wedge q ))$ belongs to $\{F \mbox{ in }\aDistSetqa{q}: r_{F} = r^*, q_{F} = q^*\}$.  


\paragraph{Case 1:} Suppose $r^* < \pin$. By definition, we have that
\bearn
v \bG_{\alpha,\pin} (v| (r^*,q^*), (\pin, q )) &=& \begin{cases}
			v \Gad{\Gainv{q^*} \frac{v}{ r^*}} &\quad \mbox{if } v  \in [0,r^*], \\
			v q^*\Gamma_{\alpha}\left( \Gamma_{\alpha}^{-1}\left(\frac{q}{q^*}\right) \frac{v-r^*}{\pin-r^*}   \right)
			& \quad \mbox{if } v \in (r^*,\pin],\\
			0  &\quad \mbox{if } v \in  [\pin, \infty).
		\end{cases}
\eearn



 By applying \Cref{lem:alpharegG} to the following parameters $(s, q_s):=(r^*, q^*)$ and  $(s', q_{s'}):=(w, q)$, we have that $G_{\alpha,\pin} (\cdot| (r^*,q^*), (\pin, q ))$  belongs to $\aDistSetqa{q}$ and $\Fa(r^*, q^*)$ if and only if $q^* = \bF(r^*) \ge \Gad{\Gainv{q} \frac{r^*}{ \pin}}$.

Furthermore,  using \Cref{lemma:reserve_price_gamma_lambda},  the revenue function $v \mapsto v \Gad{\Gainv{q} \frac{v}{ \pin}}$  is maximized  at $r_1 = r^*/(\alpha \Gainv{q^*})$ and  the revenue function $v \mapsto v q^*\Gamma_{\alpha}\left( \Gamma_{\alpha}^{-1}\left(\frac{q}{q^*}\right) \frac{v-r^*}{\pin-r^*}   \right)$ is maximized at 
\bearn
r_2 = \frac{1}{\alpha} \left(\frac{\pin-r^*}{\Gainv{\frac{q}{q^*}}}-(1-\alpha)r^* \right).
\eearn

Thus,  when $G_{\alpha,\pin} (\cdot| (r^*,q^*), (\pin, q ))$ belongs to $\aDistSetqa{q}$, the optimal revenue associated with $G_{\alpha,\pin} (\cdot| (r^*,q^*), (\pin, q ))$ is achieved at $r^*$ if and only if $r_2 \le r^* \le r_1$. We have $r_2 \le  r^*$ if and only if:
\bearn
\frac{1}{\alpha} \left(\frac{\pin-r^*}{\Gainv{\frac{q}{q^*}}}-(1-\alpha)r^* \right) \le r^* &\mbox{iff}& \frac{\pin-r^*}{\Gainv{\frac{q}{q^*}}} \le r^* \quad \mbox{iff} \quad \frac{w}{r^*}-1 \le \Gainv{\frac{q}{q^*}} \quad \mbox{iff} \quad q^* \ge \frac{q}{\Gad{\frac{\pin}{r^*}-1}},
\eearn
and $r^* \le  r_1$ if and only if:
\bearn
r^* \le  \frac{r^*}{\alpha \Gainv{q^*}}  &\mbox{iff}&  \Gainv{q} \le \frac{1}{\alpha} \quad \mbox{iff} \quad q^* \ge \Gad{\frac{1}{\alpha}}.
\eearn
Therefore $r_2 \le r^* \le r_1$ is equivalent to 
\bearn
q^* \geq \max \left \{ \Gad{\frac{1}{\alpha}}, \frac{q}{\Gad{\frac{\pin}{r^*}-1}}\right\}.
\eearn

We have established that when $r^* < \pin$, $G_{\alpha,\pin} (\cdot| (r^*,q^*), (\pin, q ))$ belongs to $\{F \mbox{ in }\aDistSetqa{q}: r_{F} = r^*, q_{F} = q^*\}$ if and only if
\bearn
q^* \geq \max \left \{\Gad{\Gainv{q} \frac{r^*}{ \pin}} ,   \Gad{\frac{1}{\alpha}}, \frac{q}{\Gad{\frac{\pin}{r^*}-1}}\right\}.
\eearn
We have hence established that when $r^* \leq \pin$,  $(r^*, q^*)$ belongs to ${\cal B}_{\alpha}(\pin,q)$ if and only if $(r^*,q^*)$ belongs to ${\cal B}_l$.

\paragraph{Case 2:} Suppose now $r^* \geq \pin$. By definition, we have that
\bearn
v \bG_{\alpha,r^*} (v|  (\pin, q ), (r^*,q^*))
 &=& \begin{cases}
			v \Gad{\Gainv{q} \frac{v}{ \pin}}
			&\quad \mbox{if } v \in [0,\pin],\\
			v q\Gamma_{\alpha}\left( \Gamma_{\alpha}^{-1}\left(\frac{q^*}{q}\right) \frac{v-\pin}{r^*-\pin}   \right) 
			&\quad \mbox{if } v \in (w,r^*],\\
			0  &\quad \mbox{if } v \mbox{ in } (r^*, \infty).
		\end{cases}
\eearn

 By applying \Cref{lem:alpharegG} to the following parameters $(s, q_s):=(w, q)$ and  $(s', q_{s'}):=(r^*, q^*)$, we have that $G_{\alpha,r^*} (\cdot|  (\pin, q ), (r^*,q^*))$  belongs to $\aDistSetqa{q}$ and $\Fa(w^*, \{q^*\})$ if and only if $q = \bF(\pin) \ge \Gad{\Gainv{q^*} \frac{\pin}{ r^*}}$ 
 which can be rewritten as $q^* \leq \Gad{\Gainv{q} \frac{r^*}{ \pin}}$. 

 
 Using \Cref{lemma:reserve_price_gamma_lambda}, the oracle price for the revenue function $v \mapsto v q\Gamma_{\alpha}\left( \Gamma_{\alpha}^{-1}\left(\frac{q^*}{q}\right) \frac{v-\pin}{r^*-\pin}   \right)$ 
  is achieved at \bearn
r' = \frac{1}{\alpha} \left(\frac{r^*- \pin}{\Gainv{\frac{q^*}{q}}}-(1-\alpha)w \right).
\eearn 

Given that $r^*$ is at the end of its support and that the revenue curve is unimodal, the optimal revenue associated $G_{\alpha,r^*} (\cdot|  (\pin, q ), (r^*,q^*))$ is achieved at $r^*$ if and only if $ r^* \leq r'$, which, in turn, is true if and only if:
\bearn
 r^* \leq \frac{1}{\alpha} \left(\frac{r^*- \pin}{\Gainv{\frac{q^*}{q}}}-(1-\alpha)w \right) &\mbox{iff}& \alpha r^* + (1-\alpha)w \leq  \frac{r^*- \pin}{\Gainv{\frac{q^*}{q}}}  \\
&\mbox{iff}&   \Gainv{\frac{q^*}{q}} \leq  \frac{r^*- \pin}{w + \alpha (r^*-w)} \\
&\mbox{iff}& \Gad{\frac{1}{\alpha + \frac{w}{r^*-w}}} \le \frac{q^*}{q} \quad \mbox{iff} \quad q\Gad{\frac{1}{\alpha+\frac{\pin}{r^*- \pin}}} \leq q^* .
\eearn
We have established that when $r^* \geq  \pin$, $G_{\alpha,r^*} (\cdot|  (\pin, q ), (r^*,q^*))$ belongs to $\{F \mbox{ in }\aDistSetqa{q}: r_{F} = r^*, q_{F} = q^*\}$ if and only if 
\bearn
q\Gad{\frac{1}{\alpha+\frac{\pin}{r^*- \pin}}} \leq q^* \leq \Gad{\Gainv{q} \frac{r^*}{ \pin}}.
\eearn
In turn, we have hence established that when $r^* \geq \pin$,  $(r^*, q^*) \mbox{ in }{\cal B}_{\alpha}(\pin,q)$ if and only if $(r^*,q^*) \mbox{ in }{\cal B}_h$.
 This concludes the proof.
\end{proof}


\begin{lemma}\label{lem:decreasingqs}
Let $J_{w, q} = \{ r: \mbox{ there exists }  q^* \mbox{ s.t. }  (r^*,q^*) \mbox{ in } {\cal B}_{\alpha}(\pin,q)\}$. We have $J_{w, q} = [\rl, w) \cup [w, \rh]$ and for any $r^* \mbox{ in }J_{w, q}$, the function $\widetilde{R}_{r^*, \pin, q}(\cdot)$ 
is decreasing  in the set $ \{ q^*:  (r^*,q^*) \mbox{ in } {\cal B}_{\alpha}(\pin,q)\}$.
\end{lemma}

\begin{proof}[\underline{\textbf{Proof of \Cref{lem:decreasingqs}}}]

We will show that $J_{w, q} =  [\rl, w) \cup [w, \rh]$ by analyzing  the cases when $r^* < \pin$ and $r^* \geq \pin$ separately.

Suppose first that $r^* < \pin$. In this case, we show that there exists a value $q^*$ such that $(r^*,q^*) \in {\cal B}_{l}$ if and only if $r^* \geq \rl$. 

 If there exists $(r^*,q^*) \in {\cal B}_{l}$ then we have y \Cref{lem:opt_feas} that $\frac{q}{\Gad{\frac{\pin}{r^*}-1}} \leq q^*$ and since $q^* \leq 1$, we have $\frac{q}{\Gad{\frac{\pin}{r^*}-1}} \leq 1$, which implies that $r^* \ge \frac{w}{1 + \Gainv{q}} = \rl$. 
Now if $r^* \in [\rl, \pin)$, note that $(r^*, 1) \in {\cal B}_{l}$, as we have seen above that $1 \geq \frac{q}{\Gad{\frac{\pin}{r^*}-1}}$. Furthermore, we have
\bearn
1 \geq \Gad{\Gainv{q} \frac{r^*}{ \pin}}  \quad \mbox{and} \quad 1 \geq \Gad{\frac{1}{\alpha}}
\eearn
since $\Gad{x} \leq 1$ for all $x \geq 0$. 

Now suppose that $r^* \geq \pin$. In this case, we show that there exists a value $q^*$ such that $(r^*,q^*) \in {\cal B}_{h}$ if and only if $\rh \ge \pin$ (which is equivalent to $q \geq \Gad{\frac{1}{\alpha}}$) and $r^* \leq \rh$. 

If there exists $(r^*,q^*) \in {\cal B}_{h}$ then, by \Cref{lem:opt_feas},  we have $q\Gad{\frac{1}{\alpha+\frac{\pin}{r^*- \pin}}} \leq q^* \leq \Gad{\Gainv{q} \frac{r^*}{ \pin}}$. Note that we have
\bearn
q\Gad{\frac{1}{\alpha+\frac{\pin}{r^*- \pin}}} \leq  \Gad{\Gainv{q} \frac{r^*}{\pin}} &\mbox{iff}& q^{\alpha-1}\left(\Gad{\frac{1}{\alpha+\frac{\pin}{r^*- \pin}}}\right)^{\alpha-1} \geq \left(\Gad{\Gainv{q} \frac{r^*}{\pin}}\right)^{\alpha-1} \\
&\mbox{iff}& \left(1+\left(1-\alpha\right)\Gainv{q}\right)\left(1+\frac{1-\alpha}{\alpha+\frac{\pin}{r^*- \pin}}\right) \geq 1 + (1-\alpha)\Gainv{q} \frac{r^*}{\pin} \\ 
&\mbox{iff}& \frac{1-\alpha}{\alpha+\frac{1}{\frac{r^*}{\pin}- 1}}
+ \left(1-\alpha\right)\Gainv{q}\left(1+\frac{1-\alpha}{\alpha+\frac{1}{\frac{r^*}{\pin}- 1}}\right) \geq  (1-\alpha)\Gainv{q} \frac{r^*}{\pin} \\
&\mbox{iff}& \frac{1}{\alpha+\frac{1}{\frac{r^*}{\pin}- 1}}
+ \Gainv{q}\left(1+\frac{1-\alpha}{\alpha+\frac{1}{\frac{r^*}{\pin}- 1}}\right) \geq  \Gainv{q} \frac{r^*}{\pin} \\
&\mbox{iff}& \frac{1}{\alpha+\frac{1}{\frac{r^*}{\pin}- 1}}
+ \Gainv{q}\left(\frac{1-\alpha}{\alpha+\frac{1}{\frac{r^*}{\pin}- 1}} - (\frac{r^*}{\pin}- 1)\right) \geq 0 \\
&\mbox{iff}& \frac{1 + \Gainv{q} \left(1-\alpha -\alpha \frac{r^*}{\pin} + \alpha \right)}{\alpha+\frac{1}{\frac{r^*}{\pin}- 1}} \geq 0 \\ 
&\mbox{iff}& \frac{1 - \Gainv{q} \alpha \frac{r^*}{\pin} }{\alpha+\frac{1}{\frac{r^*}{\pin}- 1}} \geq 0 \\
&\mbox{iff}&  1 - \Gainv{q} \alpha \frac{r^*}{\pin} \geq 0 \quad \mbox{iff} \quad r^* \leq \frac{\pin}{\alpha \Gainv{q}} = \rh.
\eearn
Additionally, since $r^* \geq \pin$,  the above inequality implies that $\rh \geq \pin$ (which in turns implies that $q \geq \Gad{\frac{1}{\alpha}}$). 

 Now if $q \geq \Gad{\frac{1}{\alpha}}$ then $\rh \ge \pin$  and therefore, if $r^* \in [\pin, \rh]$, we always have that $(r^*, \Gad{\Gainv{q} \frac{r^*}{\pin}}) \in {\cal B}_{h}$, as we showed above that $q\Gad{\frac{1}{\alpha+\frac{\pin}{r^*- \pin}}} \leq  \Gad{\Gainv{q} \frac{r^*}{\pin}}$.

Next, we show the $\widetilde{R}_{r^*, \pin, q}(\cdot)$ monotonicity property by analyzing  the cases when $r^* < \pin$ and $r^* \geq \pin$ separately.

Suppose first that $r^* < \pin$. In this case,  we have 
\bearn
 \widetilde{R}_{r^*, \pin, q}(q^*)
 &\stackrel{}{=}& \int_0^{r^*} \frac{p}{r^*} \frac{\Gad{\Gainv{q^*} \frac{p}{r^*}}}{q^*} d\Psi(p)  + \int_{r^*}^{\pin} \frac{p}{r^*}  \Gad{\Gainv{\frac{q}{q^*}} \frac{p-r^*}{\pin-r^*}} d\Psi(p) \\
  &=& \int_0^{r^*} \frac{p}{r^*} \frac{\Gad{\Gainv{q^*} \frac{p}{r^*}}}{\Gad{\Gainv{q^*}}} d\Psi(p)  + \int_{r^*}^{\pin} \frac{p}{r^*}  \Gad{\Gainv{\frac{q}{q^*}} \frac{p-r^*}{\pin-r^*}} d\Psi(p) \\
  &\stackrel{}{=}& \int_0^{r^*} \frac{p}{r^*} \left( \Gad{\frac{1- p/r^*}{1/\Gainv{q^*} + (1-\alpha)} } \right)^{-1}d\Psi(p)  + \int_{r^*}^{\pin} \frac{p}{r^*}  \Gad{\Gainv{\frac{q}{q^*}} \frac{p-r^*}{\pin-r^*}} d\Psi(p),
\eearn
where the last equality follows from the fact that  for $u \ge v$, $\Gad{u}/\Gad{v} = \Gad{(u-v)/(1+(1-\alpha)v)}$.
Each term on the RHS  above  is decreasing in $q^*$, since $\Gad{\cdot}$ is decreasing. Hence  $\widetilde{R}_{r^*, \pin, q}(q^*)$ is decreasing in this case.

Suppose now that $r^* \geq \pin$ and $q \geq \Gad{\frac{1}{\alpha}}$ so that the interval $[w, \rh]$ is non-empty. In this case,  we have 
\bearn
  \widetilde{R}_{r^*, \pin, q}(q^*) 
&\stackrel{}{=}& \int_0^{\pin} \frac{p}{r^*} \frac{ \Gad{\Gainv{q} \frac{p}{ \pin}}}{q^*} d\Psi(p)  + \int_{\pin}^{r^*} \frac{p}{r^*} \frac{q \: \Gad{\Gainv{\frac{q^*}{q}} \frac{p-\pin}{r^*- \pin}}}{q^*} d\Psi(p) \\
&=& \int_0^{\pin} \frac{p}{r^*}\frac{\Gad{\Gainv{q} \frac{p}{ \pin}}}{q^*} d\Psi(p)  + \int_{\pin}^{r^*} \frac{p}{r^*} \frac{ \: \Gad{\Gainv{\frac{q^*}{q}} \frac{p-\pin}{r^*- \pin}}}{\Gad{\Gainv{\frac{q^*}{q}}}} d\Psi(p) \\
&\stackrel{}{=}& \int_0^{\pin} \frac{p}{r^*}\frac{\Gad{\Gainv{q} \frac{p}{ \pin}}}{q^*} d\Psi(p)  + \int_{\pin}^{r^*} \frac{p}{r^*} \left( \: \Gad{ \frac{\frac{r^*- p}{r^*- \pin}} {1/\Gainv{\frac{q^*}{q}} + (1-\alpha)\frac{p-\pin}{r^*- \pin}}} \right)^{-1} d\Psi(p) ,
\eearn 
where the last equality follows from the fact that  for $u \ge v$, $\Gad{u}/\Gad{v} = \Gad{(u-v)/(1+(1-\alpha)v)}$.
Each term in the above equality is decreasing in $q^*$, therefore the result also holds for this case.
This concludes the proof.
\end{proof}

\setcounter{equation}{0}
\setcounter{proposition}{0}
\setcounter{lemma}{0}

\section{Proofs and auxiliary results  for \Cref{section:deterministicmechanisms}}\label{apx:deterministicmechanisms}
We define the following useful notation used throughout this section:
\bear
\Tilde{v}_\alpha &:=& \begin{cases}
            1 &\quad \mbox{if } \alpha = 0,\\
			\alpha^{\frac{\alpha}{1-\alpha}} &\quad \mbox{if } \alpha \text{ in } (0, 1),\\
			e^{-1} &\quad \mbox{if } \alpha = 1,
		\end{cases} \label{eq:vtilde}\\
\underline{q}_\alpha &:=& \begin{cases}
            0 &\quad \mbox{if } \alpha = 0,\\
			\Gad{\frac{1}{\alpha}} &\quad \mbox{if } \alpha \text{ in } (0, 1].
		\end{cases} \label{eq:qlow}
\eear
One can easily check that, for any $\alpha$ in $[0,1]$, $\underline{q}_\alpha < \Gad{\Tilde{v}_\alpha}$.

\begin{proof}[\underline{\textbf{Proof of \Cref{thm:deterministic_perf_close}}}]
The proof is divided into three separate parts. 

In a first step, we simplify the problem given in \eqref{eq:main_deterministic_red} in      \Cref{section:deterministicmechanisms}. We show in \Cref{thm:deterministic_perf_pb} that the seller's posted price has to counter at most  3 worst-case distributions, where two of these are fixed, and the third one is a function of the price selected.  In a second step,  we  analyze the case of  regular distributions, and in a third step,  we analyze the case of mhr distributions.

Recall the definition  of $\underline{q}_\alpha$ introduced in \eqref{eq:qlow}.  
\begin{proposition}[Worst-case Distributions against Deterministic Mechanisms]\label{thm:deterministic_perf_pb}
For any $\alpha \mbox{ in }\left[0,1\right]$, we have \
\bearn \label{eq:deterministic_reduction}
& & \Rb \left(\MechSet_d, \aDistSetqa{q}\right)\\ &=& 
\begin{cases}
\sup_{p \in \left[0,1\right]} \min\left \{\frac{Rev\left(p | \bF_{\alpha}\left( \cdot | \mu_{\alpha, q}\left(p\right), (1, q)\right)\right)}{\opt\left(F_{\alpha}\left( \cdot | \mu_{\alpha, q}\left(p\right), (1, q)\right)\right)}, \frac{Rev\left(p|\delta_1\right)}{Rev\left(1|\delta_1\right)}, \frac{Rev\left(p| \bF_{\alpha}\left( \cdot|  \rhone, (1, q)\right)\right)}{\opt\left(F_{\alpha}\left( \cdot |  \rhone, (1, q)\right)\right)}\right\},\quad \mbox{ if } q \in  [\underline{q}_\alpha,1), \\
\sup_{p \in \left[0,1\right]} \min\left \{\frac{Rev\left(p | \bF_{\alpha}\left( \cdot | \mu_{\alpha, q}\left(p\right), (1, q)\right)\right)}{\opt\left(F_{\alpha}\left( \cdot | \mu_{\alpha, q}\left(p\right), (1, q)\right)\right)}, \frac{Rev\left(p|\delta_1\right)}{Rev\left(1|\delta_1\right)}\right\}, \quad \mbox{ if } q \in (0,  \underline{q}_\alpha),
\end{cases}
\eearn
with 
\bearn
\mu_{\alpha, q}\left(p\right) &=& 1- \frac{\sqrt{\Delta_{\alpha,q}(p)}-\alpha \Gainv{q} \left(1-p\right)}{2q^{\alpha-1}}\\
\Delta_{\alpha,q}(p) &=& \left(\alpha \Gainv{q} \left(1-p\right)\right)^2 + 4 \Gainv{q} \left(1-p\right) q^{\alpha-1},
\eearn
%
\end{proposition}
In the above result, the initial price $\pin$ is normalized to 1 without loss of generality. The above establishes that, when restricting attention to deterministic prices, one can restrict attention to worst-case distributions consisting of a GPD distribution with support starting at $\mu_{\alpha, q}\left(p\right)$ and truncated at 1, a mass at 1, or a GPD distribution truncated at $\rhone$ (when $q \ge \underline{q}_\alpha$).  The proof is deferred to \Cref{app:aux-det}. 

We now leverage the above reduction to explicitly derive optimal deterministic mechanisms against regular and mhr distributions.

Let us introduce the following functions that represent the ratios of the worst families for $p \ge \rln$  
\bearn
R_{1, \alpha}\left(p, q\right) &:=& \frac{Rev\left(p | \bF_{\alpha}\left( \cdot | \mu_{\alpha, q}\left(p\right), (1, q)\right)\right)}{\opt\left(F_{\alpha}\left( \cdot | \mu_{\alpha, q}\left(p\right), (1, q)\right)\right)},\\
R_{2, \alpha}\left(p, q\right) &:=& \frac{Rev\left(p|\delta_1\right)}{Rev\left(1|\delta_1\right)} ,\\
R_{3, \alpha}\left(p, q\right) &:=& \frac{Rev\left(p| \bF_{\alpha}\left( \cdot| \rhone, (1, q)\right)\right)}{\opt\left(F_{\alpha}\left( \cdot | \rhone, (1, q)\right)\right)}, \quad \text{defined only for $q \geq  \underline{q}_\alpha$}. 
\eearn

We next analyze some properties of the above defined ratios. To that end, recall the definitions of $\Tilde{v}_\alpha$ introduced in \eqref{eq:vtilde} and of $\underline{q}_\alpha$ in \eqref{eq:qlow}.  

\begin{lemma} \label{ratio_alpha_properties}
We have the following properties: 
\begin{enumerate}
    \item If $q \in [\underline{q}_\alpha, \Gad{\Tilde{v}_\alpha})$, then there exists $p_{13, \alpha, q}$ in $\left[\rln,1\right]$  such that $R_{1, \alpha}\left(\cdot,q\right) \ge R_{3, \alpha}\left(\cdot,q\right)$ in $\left[\rln, p_{13, \alpha, q}\right]$ and $R_{1, \alpha}\left(\cdot,q\right) \le R_{3, \alpha}\left(\cdot,q\right)$ in $\left[p_{13, \alpha, q},1\right]$. Else, if $q \in \left[\Gad{\Tilde{v}_\alpha}, 1\right]$, then $R_{1, \alpha}\left(p,q\right) \ge R_{3, \alpha}\left(p,q\right)$ for all $p$ in $\left[\rln, 1\right]$.
    \item For any $q \in (0,1)$, there exists $p_{12, \alpha, q}$ such that $R_{1, \alpha}\left(\cdot,q\right)\ge R_{2, \alpha}\left(\cdot,q\right)$ in $\left[\rln, p_{12, \alpha, q}\right]$ and $R_{1, \alpha}\left(\cdot,q\right)\le R_{2, \alpha}\left(\cdot,q\right)$ in $\left[p_{12, \alpha, q},1\right]$.
    \item If $q \ge \underline{q}_\alpha$, then we have that $R_{3, \alpha}\left(\cdot, q\right)$ is non decreasing in $\left[\rln,1\right]$.
    \item For $\alpha \mbox{ in }\{0,1\}$, $R_{1, \alpha}\left(\cdot,q\right)$ is non increasing in $ \left[\rln,1\right]$, $R_{2, \alpha}\left(\cdot,q\right)$ is non decreasing in $ \left[\rln,1\right]$.
\end{enumerate}
\end{lemma}

\begin{lemma} \label{r3_properties}
For $\alpha \mbox{ in }\{0,1\}$, there exists a unique $\hat{q}_{\alpha}$ in $[\underline{q}_\alpha, \Gad{\Tilde{v}_\alpha}]$ solution to the equation $p_{13, \alpha, q} = p_{12, \alpha, q}$, and we have for $q$ in $[\underline{q}_\alpha, \Gad{\Tilde{v}_\alpha}]: \quad p_{13, \alpha, q} \leq p_{12, \alpha, q} \quad \text{if and only if} \quad q \leq \hat{q}_{\alpha}$. Furthermore, we have the following expressions:
\bearn
&&\hat{q}_{0} = \frac{1}{4},\quad p_{13, 0, q} = 1- \frac{\left(1-2q\right)^2}{1-q}, \quad p_{12, 0, q} = 1 - \frac{\left(1-\sqrt{q}\right)^2}{1-q}, \\
&&\hat{q}_{1}  = \hat{q},\quad p_{13, 1, q} = \mu_{1, q}^{-1}\left(W\left(\frac{1}{\log\left(q^{-1}\right)}\right)\right),\quad p_{12, 1, q} = \mu_{1, q}^{-1}\left(\frac{1}{W\left(\frac{e}{q}\right)}\right),
\eearn
Where $\hat{q}$ is the unique solution in $[0,1]$ to the equation $W\left(\frac{1}{\log(q^{-1})}\right) W\left(\frac{e}{q}\right) = 1$, $W$ is the Lambert function defined as the inverse of $x \to x e^x$ in $\left[0, +\infty\right)$. Numerically $\hat{q} \in[0.52, 0.53]$.
\end{lemma}

The proof can be found in \Cref{app:aux-det}. We proceed by analyzing three main cases $q$ in $\left(0, \hat{q}_{\alpha}\right]$, $q$ in $\left(\hat{q}_{\alpha}, \Gad{\Tilde{v}_\alpha}\right]$ and $q$ in $\left(\Gad{\Tilde{v}_\alpha}, 1\right)$.

Below, we fix $\alpha$ in $\{0,1\}$.

\noindent \underline{\textbf{Case $q$ in $\left(0, \hat{q}_{\alpha}\right]$}:} 

We analyze the two sub-cases $q$ in $(0, \underline{q}_{\alpha})$ and $q$ in $[\underline{q}_{\alpha}, \hat{q}_{\alpha}]$, which both will lead the same final result. Note that the first sub-case is empty for $\alpha = 0$.

\underline{\textbf{Sub-case:}} $\alpha = 1$ and $q$ in $(0, \underline{q}_{\alpha})$: In this case, based on \cref{thm:deterministic_perf_pb}, we have:
\bearn
\Rb \left(\dMechSet, \Fa\left(\pin, q\right)\right) &=& \max \Biggl\{ \max_{p \in \left[\rln, p_{12, \alpha, q}\right]}  \min \left(R_{1, \alpha}\left(p, q\right), R_{2, \alpha}\left(p, q\right)\right),  \\&& \max_{p \in \left[p_{12, \alpha, q}, 1\right]}  \min \left(R_{1, \alpha}\left(p, q\right), R_{2, \alpha}\left(p, q\right)\right) \Biggl\}.
\eearn
Now let us simplify each term.
\begin{itemize}
    \item Using \Cref{ratio_alpha_properties}-2, we have that, $R_{1,\alpha}\left(\cdot,q\right)$ is above $R_{2,\alpha}\left(\cdot,q\right)$ in $\left[\rln, p_{12, \alpha, q}\right]$, therefore:
    \bearn
    \max_{p \in \left[\rln, p_{12, \alpha, q}\right]}  \min \left(R_{1, \alpha}\left(p, q\right), R_{2, \alpha}\left(p, q\right)\right) &=&  \max_{p \in \left[\rln, p_{12, \alpha, q}\right]}  R_{2, \alpha}\left(p, q\right) \overset{\left(a\right)}{=} R_{2, \alpha}\left(p_{12, \alpha, q}, q\right), 
    \eearn
    where in (a), we used the result in \Cref{ratio_alpha_properties}-4 that states that $R_{2, \alpha}\left(\cdot,q\right)$ is non decreasing in $ \left[\rln, 1\right]$.
    \item Using \Cref{ratio_alpha_properties}-2, we have that, $R_{1,\alpha}\left(\cdot,q\right)$ is below $R_{2,\alpha}\left(\cdot,q\right)$ in $\left[p_{12, \alpha, q}, 1\right]$, therefore:
    \bearn
    \max_{p \in \left[p_{12, \alpha, q},1\right]}  \min \left(R_{1, \alpha}\left(p, q\right), R_{2, \alpha}\left(p, q\right)\right) &=&  \max_{p \in \left[p_{12, \alpha, q},1\right]}  R_{1, \alpha}\left(p, q\right) 
    \overset{\left(b\right)}{=} R_{1, \alpha}\left(p_{12, \alpha, q}, q\right),
    \eearn
    where in (b), we used the result in \Cref{ratio_alpha_properties}-4 that states that $R_{1, \alpha}\left(\cdot,q\right)$ is non increasing in $\left[\rln, 1\right]$.
\end{itemize}
Therefore, we have
\bearn
\Rb \left(\dMechSet, \Fa\left(\pin, q\right)\right) &=& \max \{ R_{2, \alpha}\left(p_{12, \alpha, q}, q\right), R_{1, \alpha}\left(p_{12, \alpha, q}, q\right) \} \overset{\left(c\right)}{=} p_{12, \alpha, q},
\eearn 
where in (c), we used the fact that, by definition, $R_{2, \alpha}\left(p_{12, \alpha, q}, q\right) = R_{1, \alpha}\left(p_{12, \alpha, q}, q\right)$. We also note that the value above is achieved at $p = p_{12, \alpha, q}$. 

\underline{\textbf{Sub-case:}}  $q$ in $[\underline{q}_{\alpha}, \hat{q}_{\alpha}]$: In this case, based on \Cref{r3_properties}, we have that  $p_{13, \alpha, q} \le p_{12, \alpha, q},$ thus we have
\bearn
\Rb \left(\dMechSet, \Fa\left(\pin, q\right)\right) &=& \max \Biggl\{ \max_{p \in \left[\rln, p_{13, \alpha, q}\right]}  \min \left(R_{1, \alpha}\left(p, q\right), R_{2, \alpha}\left(p, q\right),  R_{3, \alpha}\left(p, q\right)\right), \\&& \max_{p \in \left[p_{13, \alpha, q}, p_{12, \alpha, q}\right]}  \min \left(R_{1, \alpha}\left(p, q\right), R_{2, \alpha}\left(p, q\right),  R_{3, \alpha}\left(p, q\right)\right), \\&& \max_{p \in \left[p_{12, \alpha, q}, 1\right]}  \min \left(R_{1, \alpha}\left(p, q\right), R_{2, \alpha}\left(p, q\right),  R_{3, \alpha}\left(p, q\right)\right) \Biggl\}.
\eearn

Now let us simplify each term.
\begin{itemize}
    \item Using \Cref{ratio_alpha_properties}-1, we have that, for $q \in [\underline{q}_{\alpha}, \hat{q}_{\alpha}]  \subseteq [\underline{q}_{\alpha},  \Gad{\Tilde{v}_\alpha}] $,  $R_{1,\alpha}\left(\cdot,q\right)$ is above $R_{3,\alpha}\left(\cdot,q\right)$ in $\left[\rln, p_{13, \alpha, q}\right]$, therefore:
    \bearn
    \max_{p \in \left[\rln, p_{13, \alpha, q}\right]}  \min \left(R_{1, \alpha}\left(p, q\right), R_{2, \alpha}\left(p, q\right),  R_{3, \alpha}\left(p, q\right)\right) &=&  \max_{p \in \left[\rln, p_{13, \alpha, q}\right]}  \min \left(R_{2, \alpha}\left(p, q\right),  R_{3, \alpha}\left(p, q\right)\right).
    \eearn
    \item Using \Cref{ratio_alpha_properties}-1 and \Cref{ratio_alpha_properties}-2, we have that both $R_{3,\alpha}\left(p,q\right) \ge R_{1,\alpha}\left(p,q\right) \ge R_{2,\alpha}\left(p,q\right)$ in $\left[p_{13, \alpha, q}, p_{12, \alpha, q}\right]$, therefore: 
    \bearn
    \max_{p \in \left[p_{13, \alpha, q}, p_{12, \alpha, q}\right]}  \min \left(R_{1, \alpha}\left(p, q\right), R_{2, \alpha}\left(p, q\right),  R_{3, \alpha}\left(p, q\right)\right) &=&  \max_{p \in \left[p_{13, \alpha, q}, p_{12, \alpha, q}\right]}  R_{2, \alpha}\left(p, q\right).
    \eearn
    \item Using \Cref{ratio_alpha_properties}-1 and \Cref{ratio_alpha_properties}-2, we have that both $R_{2,\alpha}\left(\cdot,q\right)$ and $R_{3,\alpha}\left(\cdot,q\right)$ are above $R_{1,\alpha}\left(\cdot,q\right)$ in $\left[p_{12, \alpha, q}, 1\right]$, therefore: 
    \bearn
    \max_{p \in \left[p_{12, \alpha, q}, 1\right]}  \min \left(R_{1, \alpha}\left(p, q\right), R_{2, \alpha}\left(p, q\right),  R_{3, \alpha}\left(p, q\right)\right) &=&  \max_{p \in \left[p_{12, \alpha, q}, 1\right]}  R_{1, \alpha}\left(p, q\right).
    \eearn
\end{itemize}
Therefore, we have
\bearn
\Rb \left(\dMechSet, \Fa\left(\pin, q\right)\right) &=& \max \Biggl\{ \max_{p \in \left[\rln, p_{13, \alpha, q}\right]}  \min \left(R_{2, \alpha}\left(p, q\right),  R_{3, \alpha}\left(p, q\right)\right), \max_{p \in \left[p_{13, \alpha, q}, p_{12, \alpha, q}\right]}  R_{2, \alpha}\left(p, q\right), \\
&& \max_{p \in \left[p_{12, \alpha, q}, 1\right]}  R_{1, \alpha}\left(p, q\right) \Biggl\}\\
&\overset{\left(a\right)}{=}& \max \Biggl\{ \max_{p \in \left[p_{13, \alpha, q}, p_{12, \alpha, q}\right]}  R_{2, \alpha}\left(p, q\right),\max_{p \in \left[p_{12, \alpha, q}, 1\right]}  R_{1, \alpha}\left(p, q\right) \Biggl\},
\eearn
where in (a), we used the fact that 
\bearn
\max_{p \in \left[\rln, p_{13, \alpha, q}\right]}  \min \left(R_{2, \alpha}\left(p, q\right),  R_{3, \alpha}\left(p, q\right)\right) \leq \max_{p \in \left[\rln, p_{13, \alpha, q}\right]} R_{2, \alpha}\left(p, q\right)  \overset{\left(b\right)}{\leq} \max_{p \in \left[p_{13, \alpha, q}, p_{12, \alpha, q}\right]}  R_{2, \alpha}\left(p, q\right), 
\eearn
and in (b), we used the result in \Cref{ratio_alpha_properties}-4 that states that $R_{2, \alpha}\left(\cdot,q\right)$ is non decreasing in $ \left[\rln,1\right]$. 

Using the latter and also the fact that  $R_{1, \alpha}\left(\cdot,q\right)$ is non increasing in $ \left[\rln,1\right]$ we have
\bearn
\Rb \left(\dMechSet, \Fa\left(\pin, q\right)\right) &=& \max \{ R_{2, \alpha}\left(p_{12, \alpha, q}, q\right), R_{1, \alpha}\left(p_{12, \alpha, q}, q\right) \} \\
&\overset{\left(c\right)}{=}& p_{12, \alpha, q}, 
\eearn 
where in (c), we used the fact that, by definition, $R_{2, \alpha}\left(p_{12, \alpha, q}, q\right) = R_{1, \alpha}\left(p_{12, \alpha, q}, q\right)$. We also note that the value above is achieved at $p = p_{12, \alpha, q}$.

We conclude that, for $q$ in $\left(0, \hat{q}_{\alpha}\right]$, $\Rb \left(\dMechSet, \Fa\left(\pin, q\right)\right) = p_{12, \alpha, q}$ and is achieved at $p = p_{12, \alpha, q}$.
For $\alpha = 0$, using the expressions in \Cref{r3_properties},  we obtain $\left(0, \hat{q}_{\alpha}\right] \overset{\alpha = 0}{=}  (0, \frac{1}{4}]$ and:
\bearn
\Rb \left(\dMechSet, \Fa\left(\pin, q\right)\right) &\overset{\alpha = 0}{=}&  p_{12, 0, q} = 1 - \frac{\left(1-\sqrt{q}\right)^2}{1-q}, \quad  \mbox{which is achieved at $p = 1 - \frac{\left(1-\sqrt{q}\right)^2}{1-q}$}.
\eearn 
For $\alpha = 1$, using the expressions in \Cref{r3_properties}, we obtain $\left(0, \hat{q}_{\alpha}\right] \overset{\alpha = 1}{=}  (0, \hat{q}]$ and:
\bearn
\Rb \left(\dMechSet, \Fa\left(\pin, q\right)\right) &\overset{\alpha = 1}{=}&  p_{12, 1, q} = \mu_{1, q}^{-1}\left(\frac{1}{W\left(\frac{e}{q}\right)}\right), \quad \mbox{which is achieved at $p = \mu_{1, q}^{-1}\left(\frac{1}{W\left(\frac{e}{q}\right)}\right)$} \\
&=& 1-\frac{1}{\log\left(q^{-1}\right)}\left(W\left(\frac{e}{q}\right) +\frac{1}{W\left(\frac{e}{q}\right)} - 2\right) := \beta_q\left(\frac{e}{q}\right).
\eearn 

\noindent \underline{\textbf{Case $q$ in $\left(\hat{q}_{\alpha}, \Gad{\Tilde{v}_\alpha}\right]:$}}
In this case, based on \Cref{r3_properties}, we have that  $p_{12, \alpha, q} \le p_{13, \alpha, q},$ thus we have
\bearn
\Rb \left(\dMechSet, \Fa\left(\pin, q\right)\right) &=& \max \Biggl\{ \max_{p \in \left[\rln, p_{12, \alpha, q}\right]}  \min \left(R_{1, \alpha}\left(p, q\right), R_{2, \alpha}\left(p, q\right),  R_{3, \alpha}\left(p, q\right)\right), \\&& \max_{p \in \left[p_{12, \alpha, q}, p_{13, \alpha, q}\right]}  \min \left(R_{1, \alpha}\left(p, q\right), R_{2, \alpha}\left(p, q\right),  R_{3, \alpha}\left(p, q\right)\right), \\&& \max_{p \in \left[p_{13, \alpha, q}, 1\right]}  \min \left(R_{1, \alpha}\left(p, q\right), R_{2, \alpha}\left(p, q\right),  R_{3, \alpha}\left(p, q\right)\right) \Biggl\}.
\eearn

Now let us simplify each term.
\begin{itemize}
    \item Using \Cref{ratio_alpha_properties}-1, we have that, for $q \in \left(\hat{q}_{\alpha},  \Gad{\Tilde{v}_\alpha}\right]  \subseteq [\underline{q}_{\alpha},  \Gad{\Tilde{v}_\alpha}] $,  $R_{1,\alpha}\left(\cdot,q\right)\ge R_{3,\alpha}\left(\cdot,q\right)$ in $\left[\rln, p_{13, \alpha, q}\right]$, therefore:
    \bearn
    \max_{p \in \left[\rln, p_{12, \alpha, q}\right]}  \min \left(R_{1, \alpha}\left(p, q\right), R_{2, \alpha}\left(p, q\right),  R_{3, \alpha}\left(p, q\right)\right) &=&  \max_{p \in \left[\rln, p_{12, \alpha, q}\right]}  \min \left(R_{2, \alpha}\left(p, q\right),  R_{3, \alpha}\left(p, q\right)\right).
    \eearn
    \item Using \Cref{ratio_alpha_properties}-1 and \Cref{ratio_alpha_properties}-2, we have that both $R_{2, \alpha}\left(p,q\right) \ge R_{1, \alpha}\left(p,q\right) \ge R_{3, \alpha}\left(p,q\right)$  in $\left[p_{12, \alpha, q}, p_{13, \alpha, q}\right]$, therefore: 
    \bearn
    \max_{p \in \left[p_{12, \alpha, q}, p_{13, \alpha, q}\right]}  \min \left(R_{1, \alpha}\left(p, q\right), R_{2, \alpha}\left(p, q\right),  R_{3, \alpha}\left(p, q\right)\right) &=&  \max_{p \in \left[p_{12, \alpha, q}, p_{13, \alpha, q}\right]}  R_{3, \alpha}\left(p, q\right).
    \eearn
    \item Using \Cref{ratio_alpha_properties}-1 and \Cref{ratio_alpha_properties}-2, we have that both $R_{2,\alpha}\left(\cdot,q\right)$ and $R_{3,\alpha}\left(\cdot,q\right)$ are above $R_{1,\alpha}\left(\cdot,q\right)$ in $\left[p_{13, \alpha, q}, 1\right]$, therefore: 
    \bearn
    \max_{p \in \left[p_{13, \alpha, q}, 1\right]}  \min \left(R_{1, \alpha}\left(p, q\right), R_{2, \alpha}\left(p, q\right),  R_{3, \alpha}\left(p, q\right)\right) &=&  \max_{p \in \left[p_{13, \alpha, q}, 1\right]}  R_{1, \alpha}\left(p, q\right).
    \eearn
\end{itemize}
Therefore, we have

\bearn
\Rb \left(\dMechSet, \Fa\left(\pin, q\right)\right) &=&  \max \Biggl\{ \max_{p \in \left[\rln,p_{12, \alpha, q}\right]}  \min \left( R_{2, \alpha}\left(p, q\right),  R_{3, \alpha}\left(p, q\right)\right), \max_{p \in \left[p_{12, \alpha, q}, p_{13, \alpha, q}\right]}  R_{3, \alpha}\left(p, q\right), \\&& \max_{p \in \left[p_{13, \alpha, q}, 1\right]}  R_{1, \alpha}\left(p, q\right)  \Biggl\} \\
&\overset{\left(a\right)}{=}& \max \Biggl\{ \max_{p \in \left[p_{12, \alpha, q}, p_{13, \alpha, q}\right]}  R_{3, \alpha}\left(p, q\right), \max_{p \in \left[p_{13, \alpha, q}, 1\right]}  R_{1, \alpha}\left(p, q\right) \Biggl\},
\eearn
where in (a), we used the fact that 
\bearn
\max_{p \in \left[\rln, p_{12, \alpha, q}\right]}  \min \left(R_{2, \alpha}\left(p, q\right),  R_{3, \alpha}\left(p, q\right)\right) \leq \max_{p \in \left[\rln, p_{12, \alpha, q}\right]} R_{3, \alpha}\left(p, q\right)  \overset{\left(b\right)}{\leq} \max_{p \in \left[p_{12, \alpha, q}, p_{13, \alpha, q}\right]}  R_{3, \alpha}\left(p, q\right),
\eearn
and  in (b), we used the result in \Cref{ratio_alpha_properties}-3 that states that $R_{3, \alpha}\left(\cdot,q\right)$ is non decreasing in $ \left[\rln,1\right]$. Therefore, using the property \Cref{ratio_alpha_properties}-4, we have, additionally, that $R_{1, \alpha}\left(\cdot,q\right)$ is non increasing in $ \left[\rln,1\right]$ and therefore
\bearn
\Rb \left(\dMechSet, \Fa\left(\pin, q\right)\right) &=& \max \{ R_{3, \alpha}\left(p_{13, \alpha, q}, q\right), R_{1, \alpha}\left(p_{13, \alpha, q}, q\right) \} \\
&\overset{\left(c\right)}{=}& R_{3, \alpha}\left(p_{13, \alpha, q}, q\right), \quad  \mbox{which is achieved at $p = p_{13, \alpha, q}$}.
\eearn 
where in (c), we used the fact that $R_{3, \alpha}\left(p_{13, \alpha, q}, q\right) = R_{1, \alpha}\left(p_{13, \alpha, q}, q\right)$.

\noindent For $\alpha = 0$, using the expressions in \Cref{r3_properties}, we obtain $(\hat{q}_{\alpha}, \Gad{\Tilde{v}_\alpha}]  \overset{\alpha = 0}{=} (\frac{1}{4}, \frac{1}{2}]$ and:
\bearn
\Rb \left(\dMechSet, \Fa\left(\pin, q\right)\right) &\overset{\alpha = 0}{=}&  R_{3, 0}\left(p_{13, 0, q}, q\right) = \frac{3-4q}{4\left(1-q\right)}, \quad   \mbox{which is achieved at $p_{13, 0, q} = \frac{q\left(3-4q\right)}{1-q}$}.
\eearn 
For $\alpha = 1$, using the expressions in \Cref{r3_properties}, we obtain $(\hat{q}_{\alpha}, \Gad{\Tilde{v}_\alpha}] \overset{\alpha = 1}{=} (\hat{q}, e^{-e^{-1}}]$ and:
\bearn
\Rb \left(\dMechSet, \Fa\left(\pin, q\right)\right) &\overset{\alpha = 1}{=}&   R_{3, 1}\left(p_{13, 1, q}, q\right)  = \mu_{1, q}^{-1}\left(W\left(\frac{1}{\log\left(q^{-1}\right)}\right)\right) e \log\left(q^{-1}\right) e^{-\log\left(q^{-1}\right)\mu_{1, q}^{-1}\left(W\left(\frac{1}{\log\left(q^{-1}\right)}\right)\right)}:=  \rho\left(q\right) \\ && \mbox{which is achieved at $p=\mu_{1, q}^{-1}\left(W\left(\frac{1}{\log\left(q^{-1}\right)}\right)\right) = \beta_q\left(\frac{1}{\log\left(q^{-1}\right)}\right)$}. 
\eearn 

\noindent \underline{\textbf{Case $q$ in $(\Gad{\Tilde{v}_\alpha}, 1)$}}:  Using \Cref{ratio_alpha_properties}-1, we have that for $q$ in $(\Gad{\Tilde{v}_\alpha}, 1)$, $R_{1,\alpha}\left(\cdot,q\right)$ is above $R_{3,\alpha}\left(\cdot,q\right)$ in $\left[\rln, 1\right]$. Therefore, we have
\bearn
\Rb \left(\dMechSet, \Fa\left(\pin, q\right)\right) &=&  \max_{p \in \left[\rln, 1\right]}  \min \left(R_{1, \alpha}\left(p, q\right), R_{2, \alpha}\left(p, q\right),  R_{3, \alpha}\left(p, q\right)\right) \\
&=& \max_{p \in \left[\rln, 1\right]}  \min \left(R_{2, \alpha}\left(p, q\right),  R_{3, \alpha}\left(p, q\right)\right) \\
&\overset{\left(a\right)}{=}& \min \left(R_{2, \alpha}\left(1, q\right),  R_{3, \alpha}\left(1, q\right)\right) \\
&=& \min\left(1, R_{3, \alpha}\left(1, q\right)\right) \overset{\left(b\right)}{=} R_{3, \alpha}\left(1, q\right).
\eearn
In (a), we used the results in \Cref{ratio_alpha_properties}-3 that states that $R_{3, \alpha}\left(\cdot,q\right)$ is non decreasing in $ \left[\rln,1\right]$ and \Cref{ratio_alpha_properties}-4, that states that $R_{2, \alpha}\left(\cdot,q\right)$ is non decreasing in $ \left[\rln,1\right]$ as in this case $q \geq \Gad{\Tilde{v}_\alpha} \geq \underline{q}_\alpha$. In (b), we used the fact that $R_{3, \alpha}\left(1, q\right) \leq 1$. 

For $\alpha = 0$, using the expressions in \Cref{r3_properties}, we obtain $(\Gad{\Tilde{v}_\alpha}, 1)  \overset{\alpha = 0}{=} (\frac{1}{2}, 1)$ and:
\bearn
\Rb \left(\dMechSet, \Fa\left(\pin, \{q\}\right)\right) &\overset{\alpha = 0}{=}&  R_{3, 0}\left(1, q\right) = 1-q, \quad \mbox{which is achieved at $p = 1$}.
\eearn 
For $\alpha = 1$, using the expressions in \Cref{r3_properties}, we obtain $(\Gad{\Tilde{v}_\alpha}, 1)  \overset{\alpha = 1}{=} (e^{-e^{-1}}, 1)$ and:
\bearn
\Rb \left(\dMechSet, \Fa\left(\pin, q\right)\right) &\overset{\alpha = 1}{=}&  R_{3, 1}\left(1, q\right) = eq \log\left(q^{-1}\right), \quad \mbox{which is achieved at $p = 1$}.
\eearn 
This completes the proof of \Cref{thm:deterministic_perf_close}. 
\end{proof}

\subsection{Proofs of auxiliary results}\label{app:aux-det}

\begin{proof}[\underline{\textbf{Proof of \Cref{thm:deterministic_perf_pb}}}]

Following the reasoning in \Cref{section:deterministicmechanisms} in Eq.\eqref{eq:main_deterministic_red}, we have that:
\bearn
&& \Rb \left(\MechSet_d, \aDistSetqa{q}\right) \\ &=& \sup_{p \in[0,w]} \min\left \{\min_{r \in[\rl, p)} \frac{ p\bF_{\alpha}( p | r, (\pin, q))}{ r}, \frac{p}{w}, \min_{r \in[w, \rh ] } \frac{ p\bF_{\alpha}( p | r, (\pin, q))}{ \opt(F_{\alpha}( \cdot | r, (\pin, q)))}\right\}\\
 &=& \sup_{\frac{p}{\pin} \in \left[0,1\right]} \min \Biggl\{\inf_{\frac{r}{\pin} \in [\rlone, \frac{p}{\pin})}  \frac{\frac{p}{\pin} \bF_{\alpha}( p/\pin | r/\pin, (1, q))}{\frac{r}{\pin}}, \frac{p}{\pin}, \inf_{\frac{r}{\pin} \in\left[1, \rhone\right]} \frac{\frac{p}{\pin}\bF_{\alpha}( p/\pin | r/\pin, (1, q))}{\frac{r}{\pin} \bF_{\alpha}( r/\pin | r/\pin, (1, q))}   \Biggr\} \\
  &\overset{\left(a\right)}{=}& \sup_{\tilde{p} \in \left[0,1\right]} \min \Biggl\{\inf_{\tilde{r} \in [\rlone, \tilde{p})}  \frac{\tilde{p} \bF_{\alpha}\left( \tilde{p} |\tilde{r}, (1, q)\right)}{\tilde{r}}, \tilde{p}, \inf_{\tilde{r} \in\left[1, \rhone\right]} \frac{\tilde{p} \bF_{\alpha}\left( \tilde{p}   |\tilde{r}, (1, q)\right)}{\tilde{r}  \bF_{\alpha}\left( \tilde{r}   |\tilde{r}, (1, q)\right)}   \Biggr\}, 
\eearn 
where in $(a)$ we used two changes of variables to remove the dependency on $\pin$, namely $\tilde{p}=p/w$ and $\tilde{r}=r/w$. Note that when $q < \underline{q}_\alpha$, the last term in the brackets does not affect the worst-case. Thus we conclude that
\bear  \label{eq:det_mech_inter_1}  
 \Rb \left(\MechSet_d, \aDistSetqa{q}\right)   = \sup_{p \in \left[0,1\right]} \min \Biggl\{ \inf_{r \in [\rlone, p)}  \frac{p \bF_{\alpha}\left( p |r, (1, q)\right)}{r} , p , \inf_{r \in\left[1, \rhone\right]} \frac{p \bF_{\alpha}\left( p   |r, (1, q)\right)}{r  \bF_{\alpha}\left( r   |r, (1, q)\right)} \Biggr\}. 
\eear

For each (normalized) price $p$ in $[0,1]$, we have three  terms that determine the worst case performance. We analyze each term separately. The second term is just the identity stemming from nature selecting a point mass at $1$. We next analyze the first and third terms with the brackets in \eqref{eq:det_mech_inter_1}. 

\paragraph{Third term.} The third term is only present if $q \geq \underline{q}_\alpha$ (ensuring that $\left[1, \rhone\right] \neq \emptyset$). In this case, for any $p$ in $[0,1]$ and , the third term can be shown to be equal to
\bearn
 \inf_{r \in \left[1, \rhone\right]} \frac{p \bF_{\alpha}\left( p   |r, (1, q)\right)}{r  \bF_{\alpha}\left( r   |r, (1, q)\right)} = \frac{p \bF_{\alpha}\left( p   |\rhone, (1, q)\right)}{ \sup_{r \in\left[1, \rhone\right]} r  \bF_{\alpha}\left( r   |r, (1, q)\right)}. 
\eearn
Indeed,  fix $q \geq \underline{q}_\alpha$. For any $r \in\left[1, \rhone\right]$, $\bF_{\alpha}\left(p   |r, (1, q)\right) =  \Gad{\Gainv{q} p} = \bF_{\alpha}\left(p   |\rhone, (1, q)\right)$.  By \Cref{lemma:reserve_price_gamma_lambda} applied to $\beta:= \Gainv{q}$ and $w':=0$, we have that the function $v \to v\bF_{\alpha}\left( v   |\rhone, (1, q)\right)$ is maximized at $\rhone$ thus it achieves its maximum at $\rhone$ on the interval $[1,\rhone].$ Hence  we get that 
\bear
 \inf_{r \in \left[1, \rhone\right]} \frac{p \bF_{\alpha}\left( p   |r, (1, q)\right)}{r  \bF_{\alpha}\left( r   |r, (1, q)\right)} &=&  \inf_{r \in \left[1, \rhone\right]}  \frac{p \bF_{\alpha}\left( p   |\rhone, (1, q)\right)}{ r  \bF_{\alpha}\left(r| \rhone \left(1,q\right)\right)} =   \frac{p \bF_{\alpha}\left( p   |\rhone, (1, q)\right)}{\sup_{r \in \left[1, \rhone\right]} r  \bF_{\alpha}\left(r| \rhone \left(1,q\right)\right)} \nonumber \\
 &=& \frac{p \bF_{\alpha}\left( p   |\rhone, (1, q)\right)}{ \opt(  F_{\alpha}\left(\cdot| \rhone,\left(1,q\right)\right))} .\label{eq:det_mech_inter_2}
\eear
One can easily check that
\bearn
\opt(F_{\alpha}\left(\cdot| \rhone,\left(1,q\right)\right)) &=& \lim_{v\to \rhone} v  \bF_{\alpha}\left(v|\rhone,\left(1,q\right)\right) = \frac{\Tilde{v}_\alpha}{\Gainv{q}},
\eearn
where $\Tilde{v}_\alpha$ was defined in \eqref{eq:vtilde}.

\paragraph{First term.}  For any $p$ in $[0,1],$ the first term in \eqref{eq:det_mech_inter_1} can be rewritten as 
\bearn
\inf_{r \in [\rlone, p)}  \frac{p \bF_{\alpha}\left( p  |r, (1, q)\right)}{r} = \inf_{r \in [\rlone, p)} \Phi(r),
\eearn
where
\bearn  
\Phi(r): = \frac{p}{r} \Gad{\Gainv{q}\frac{p-r}{1-r}}.
\eearn

 We  study the function $\Phi(\cdot)$ and by analyzing its derivative,  determine exactly where its minimum is achieved.  In particular, we will establish the following claim.
On $[\rlone, p)$, the function $\Phi(\cdot)$ is minimized at 
\bearn
\mu_{\alpha, q}\left(p\right) &=& 1- \frac{\sqrt{\left(\alpha \Gainv{q} \left(1-p\right)\right)^2 + 4 \Gainv{q} \left(1-p\right) q^{\alpha-1}}-\alpha \Gainv{q} \left(1-p\right)}{2q^{\alpha-1}}. 
\eearn

For any $p$ in $[0,1],$ at  any $r$ in $[\rlone, p)$, $\Phi(\cdot)$ is differentiable  with derivative given by 

\bearn
\frac{d\Phi}{dr}(r)  &=& - \frac{p}{r^2} \Gad{\Gainv{q}\frac{p-r}{1-r}} - \frac{p}{r} \left( \Gainv{q}\frac{-(1-r)+(p-r)}{(1-r)^2} \right)  \left(\Gad{\Gainv{q}\frac{p-r}{1-r}}\right)^{2-\alpha}\\
&=& - \frac{p}{r^2} \Gad{\Gainv{q}\frac{p-r}{1-r}} - \frac{p}{r} \left( \Gainv{q}\frac{-(1-p)}{(1-r)^2} \right)  \left(\Gad{\Gainv{q}\frac{p-r}{1-r}}\right)^{2-\alpha}\\
&=& - \frac{p}{r^2(1-r)^2} \left(\Gad{\Gainv{q}\frac{p-r}{1-r}}\right)^{2-\alpha} \left [ \left(\Gad{\Gainv{q}\frac{p-r}{1-r}}\right)^{-(1-\alpha)} (1-r)^2 -  \Gainv{q} r(1-p)   \right] \\
&=& - \frac{p}{r^2(1-r)^2} \left(\Gad{\Gainv{q}\frac{p-r}{1-r}}\right)^{2-\alpha} \left [ \left(1+ (1-\alpha) \Gainv{q}\frac{p-r}{1-r}\right) (1-r)^2  -  \Gainv{q} r(1-p)   \right] \\
&=& - \frac{p}{r^2(1-r)^2} \left(\Gad{\Gainv{q}\frac{p-r}{1-r}}\right)^{2-\alpha} \left [ (1-r)^2 + (1-\alpha) \Gainv{q} (p-r)(1-r)   -  \Gainv{q} r(1-p)   \right].
\eearn

Note that the sign of the derivative of $\Phi$ is determined  by that of the quadratic
\bearn
 \varphi(r)&:=& - \left[ (1-r)^2 + (1-\alpha) \Gainv{q} (p-r)(1-r)   -  \Gainv{q} r(1-p)\right]\\
 &=& - \Biggl[  \left(1+(1-\alpha)\Gainv{q} \right) (1-r)^2 - (1-\alpha) \Gainv{q}(1-p)(1-r)  \\
 && + \Gainv{q} (1-r)(1-p)  - \Gainv{q} (1-p)    \Biggl]\\
 &=& - \left[  \left(1+(1-\alpha)\Gainv{q} \right) (1-r)^2  + \alpha \Gainv{q}(1-p)(1-r)   - \Gainv{q} (1-p)    \right]\\
 &=& - \left[  q^{\alpha-1} (1-r)^2  + \alpha \Gainv{q}(1-p)(1-r)   - \Gainv{q} (1-p)    \right].
\eearn
Let 
\bearn 
\Delta_{\alpha,q}(p) &=& \left(\alpha \Gainv{q}(1-p) \right)^2 + 4 q^{\alpha-1}\Gainv{q} (1-p).
\eearn
The above is positive and hence the quadratic $\varphi(r)$ admits two roots given by 
\bearn 
r_1 &=&   1+ \frac{\alpha \Gainv{q} (1-p)  + \sqrt{\Delta_{\alpha,q}(p)} }{2 q^{\alpha-1} }, \\
r_2 &=& 1+ \frac{\alpha \Gainv{q} (1-p)  - \sqrt{\Delta_{\alpha,q}(p)}}{2 q^{\alpha-1} }.
\eearn

It is clear that $r_1 \ge 1$. We next establish that $r_2$ belongs to $\left[\rlone, p\right].$ 
\bearn
p-r_2 &=&  p - 1-  \frac{\alpha \Gainv{q} (1-p)  - \sqrt{\Delta_{\alpha,q}(p)}}{2 q^{\alpha-1} }\\
&=&   \frac{ -( 2 q^{\alpha-1} +  \alpha \Gainv{q}) (1-p)   + \sqrt{\Delta_{\alpha,q}(p)}}{2 q^{\alpha-1} }\\
&=&  \frac{ - \left (4q^{2(\alpha-1)} + \left( \alpha \Gainv{q}  \right)^2 +  4 q^{\alpha-1}  \alpha \Gainv{q} \right) (1-p)^2    + \left(\alpha \Gainv{q}(1-p) \right)^2 + 4 q^{\alpha-1}\Gainv{q} (1-p) }{2 q^{\alpha-1} ((2 q^{\alpha-1} + \alpha \Gainv{q}) (1-p)  + \sqrt{\Delta_{\alpha,q}(p)}) }\\
&=&  4 (1-p)q^{\alpha-1} \frac{ - \left (q^{(\alpha-1)}  +     \alpha \Gainv{q} \right) (1-p)     +  \Gainv{q}  }{2 q^{\alpha-1} ((2 q^{\alpha-1} + \alpha \Gainv{q}) (1-p)  + \sqrt{\Delta_{\alpha,q}(p)}) }\\
&=&  4 (1-p)q^{\alpha-1} \frac{ -   (1+ \Gainv{q}) (1-p)     +  \Gainv{q}  }{2 q^{\alpha-1} ((2 q^{\alpha-1} + \alpha \Gainv{q}) (1-p)  + \sqrt{\Delta_{\alpha,q}(p)}) }\\
&=&  4 (1-p)q^{\alpha-1}\Gainv{q} \frac{p/\rlone - 1  }{2 q^{\alpha-1} ((2 q^{\alpha-1} + \alpha \Gainv{q}) (1-p)  + \sqrt{\Delta_{\alpha,q}(p)}) }\\
&\ge& 0,
\eearn
where the last inequality follows since $p \ge \rlone$. 

Now, we also have 
\bearn
 r_2 - \rlone 
&=& 1+  \frac{\alpha \Gainv{q} (1-p)  - \sqrt{\Delta_{\alpha,q}(p)}}{2 q^{\alpha-1}}  - \frac{1}{1+\Gainv{q}}\\
&=&   \frac{\alpha \Gainv{q} (1-p)  - \sqrt{\Delta_{\alpha,q}(p)}}{2 q^{\alpha-1}}  + \Gainv{q} \rlone\\
&=&   \frac{\alpha \Gainv{q}  (1-p) + 2 q^{\alpha-1}\Gainv{q}\rlone   -  \sqrt{\Delta_{\alpha,q}(p)}}{2 q^{\alpha-1}}  \\
&=&   \frac{(\alpha \Gainv{q})^2 (1-p)^2 + 4 q^{2(\alpha-1)}(\Gainv{q})^2 (\rlone)^2 +  4 q^{\alpha-1}(\Gainv{q})^2\alpha  \rlone  (1-p)   }{2 q^{\alpha-1} (\alpha \Gainv{q}  (1-p) + 2 q^{\alpha-1}\Gainv{q}\rlone   + \sqrt{\Delta_{\alpha,q}(p)})}  \\
&& - \frac{     \left(\alpha \Gainv{q}(1-p) \right)^2 + 4 q^{\alpha-1}\Gainv{q} (1-p) }{2 q^{\alpha-1} (\alpha \Gainv{q}  (1-p) + 2 q^{\alpha-1}\Gainv{q}\rlone   + \sqrt{\Delta_{\alpha,q}(p)})}\\
&=&  4 \Gainv{q}q^{\alpha-1} \frac{   q^{\alpha-1}\Gainv{q}(\rlone)^2 +   \alpha \Gainv{q}  \rlone (1-p)  -   (1-p) }{2 q^{\alpha-1} (\alpha \Gainv{q}  (1-p) + 2 q^{\alpha-1}\Gainv{q}\rlone   + \sqrt{\Delta_{\alpha,q}(p)})}  \\
&=&  4 \Gainv{q}q^{\alpha-1} \frac{   q^{\alpha-1} \rlone (p - \rlone) }{2 q^{\alpha-1} (\alpha \Gainv{q}  (1-p) + 2 q^{\alpha-1}\Gainv{q}\rlone   + \sqrt{\Delta_{\alpha,q}(p)})}  \\
&\ge& 0,
\eearn
where the last inequality follows since $p \ge \rlone$. Hence we have established that $r_2$ belongs to $[\rlone,p)$, and $r_1 \ge 1 \ge p$.

Now, note that   the sign of $\varphi$ is non-negative on $[r_2, r_1]$ and non-positive on $[0,r_2]$. We deduce that   the function $\Phi$ is non increasing on $[\rlone, r_2]$ and non decreasing on $[r_2, p)$, thus, on  $[\rlone, p)$, $\Phi$ achieves its minimum at $r_2= \mu_{\alpha, q}\left(p\right)$. In other words we have established that for any $p$ in $[0,1],$
\bear  \label{eq:det_mech_inter_3} 
\inf_{r \in [\rlone, p)}  \frac{p \bF_{\alpha}\left( p  |r, (1, q)\right)}{r} =  \frac{p  \bF_{\alpha}\left( p  |\mu_{\alpha, q}\left(p\right), (1, q)\right)}{\mu_{\alpha, q}\left(p\right)} = \frac{p  \bF_{\alpha}\left( p  |\mu_{\alpha, q}\left(p\right), (1, q)\right)}{\opt(F_{\alpha}\left( \cdot  |\mu_{\alpha, q}\left(p\right), (1, q)\right))}.
\eear

Combining equations \eqref{eq:det_mech_inter_1}, \eqref{eq:det_mech_inter_2}  and \eqref{eq:det_mech_inter_3} yields the result.
\end{proof}

\begin{proof}[\underline{\textbf{Proof of \Cref{ratio_alpha_properties}}}] 
We first start by studying the function $p \to \mu_{\alpha, q}\left(p\right)$. We have
\bearn
\mu_{\alpha, q}\left(p\right) &=& 1- \frac{\sqrt{\left(\alpha \Gainv{q} \left(1-p\right)\right)^2 + 4 \Gainv{q} \left(1-p\right) q^{\alpha-1}}-\alpha \Gainv{q} \left(1-p\right)}{2q^{\alpha-1}} \\
&=& 1 + \alpha \frac{\Gainv{q} \left(1-p\right)}{2 q^{\alpha-1}} - \sqrt{\left(\alpha \frac{\Gainv{q} \left(1-p\right)}{2 q^{\alpha-1}}\right)^2 + 2 \frac{\Gainv{q} \left(1-p\right)}{2 q^{\alpha-1}}} \\
&=& \frac{1-2\left(1-\alpha\right)\frac{\Gainv{q} \left(1-p\right)}{2 q^{\alpha-1}}}{1 + \alpha \frac{\Gainv{q} \left(1-p\right)}{2 q^{\alpha-1}} + \sqrt{\left(\alpha \frac{\Gainv{q} \left(1-p\right)}{2 q^{\alpha-1}}\right)^2 + 2 \frac{\Gainv{q} \left(1-p\right)}{2 q^{\alpha-1}}}}.
\eearn
The numerator of the above ratio $p \to 1-2\left(1-\alpha\right)\frac{\Gainv{q} \left(1-p\right)}{2 q^{\alpha-1}}$ is clearly non-decreasing, and the denominator $p\to 1 + \alpha \frac{\Gainv{q} \left(1-p\right)}{2 q^{\alpha-1}} + \sqrt{\left(\alpha \frac{\Gainv{q} \left(1-p\right)}{2 q^{\alpha-1}}\right)^2 + 2 \frac{\Gainv{q} \left(1-p\right)}{2 q^{\alpha-1}}}$ is clearly non-increasing. Therefore, by composition, $p \to \mu_{\alpha, q}\left(p\right)$ is non-decreasing and $ \mu_{\alpha, q}\left(p\right) \mbox{ in }\left[\mu_{\alpha, q}\left(\frac{1}{1+\Gainv{q}}\right),  \mu_{\alpha, q}\left(1\right)\right]$ with:
\bearn
\mu_{\alpha, q}\left(\frac{1}{1+\Gainv{q}}\right) &=& 1 + \alpha \frac{\Gainv{q}^2}{2 q^{\alpha-1} \left(1+\Gainv{q}\right)}-\sqrt{\left(\alpha \frac{\Gainv{q}^2}{2 q^{\alpha-1} \left(1+\Gainv{q}\right)}\right)^2 + 2 \frac{\Gainv{q}^2}{2 q^{\alpha-1} \left(1+\Gainv{q}\right)}} \\
&=& 1 + \alpha \frac{\Gainv{q}^2}{2 q^{\alpha-1} \left(1+\Gainv{q}\right)}-\frac{\Gainv{q}}{2 q^{\alpha-1} \left(1+\Gainv{q}\right)}\sqrt{\alpha^2 \Gainv{q}^2 + 4 q^{\alpha-1} \left(1+\Gainv{q}\right)}\\
&=& 1 + \frac{\Gainv{q}}{2 q^{\alpha-1} \left(1+\Gainv{q}\right)} \left(\alpha \Gainv{q}-\sqrt{\alpha^2 \Gainv{q}^2 + 4 \left(1+\left(1-\alpha\right)\Gainv{q}\right) \left(1+\Gainv{q}\right)} \right) \\
&=& 1 + \frac{\Gainv{q}}{2 q^{\alpha-1} \left(1+\Gainv{q}\right)} \left(\alpha \Gainv{q}-\sqrt{\left(2+\left(2-\alpha\right)\Gainv{q}\right)^2} \right) \\
&=& 1 - \frac{\Gainv{q}}{2 q^{\alpha-1} \left(1+\Gainv{q}\right)} 2\left(1+ \left(1-\alpha\right)\Gainv{q}\right) = \frac{1}{1+\Gainv{q}}. \\
\mu_{\alpha, q}\left(1\right) &=& 1 + \alpha \frac{\Gainv{q} \left(1-1\right)}{2 q^{\alpha-1}} - \sqrt{\left(\alpha \frac{\Gainv{q} \left(1-1\right)}{2 q^{\alpha-1}}\right)^2 + 2 \frac{\Gainv{q} \left(1-1\right)}{2 q^{\alpha-1}}} = 1.
\eearn
Therefore $\mu_{\alpha, q}\left(p\right) \mbox{ in }\left[\frac{1}{1+\Gainv{q}}, 1\right]$ and $p \to \mu_{\alpha, q}(p)$ is an increasing function in $\left[\frac{1}{1+\Gainv{q}}, 1\right]$ and, for any $p \mbox{ in }\left[\frac{1}{1+\Gainv{q}}, 1\right]$, its inverse is given by
\bearn
\mu_{\alpha, q}^{-1}\left(p\right) := 1-\frac{q^{\alpha-1} \left(1-p\right)^2}{\Gainv{q} \left(1-\alpha \left(1-p\right)\right)}.
\eearn
 Next, we will show each point separately. 

\paragraph{First point}
If $q\ge \underline{q}_\alpha $, we have 
\bearn
\frac{R_{3, \alpha}\left(p, q\right)}{R_{1, \alpha}\left(p, q\right)} &=& \frac{\Gainv{q}}{\Tilde{v}_\alpha}  \mu_{\alpha, q}\left(p\right) \frac{\Gad{\Gainv{q} p}}{\Gad{\Gainv{q} \frac{p-\mu_{\alpha, q}\left(p\right)}{1-\mu_{\alpha, q}\left(p\right)}}} \\
&\overset{\left(a\right)}{=}& \frac{\Gainv{q}}{\Tilde{v}_\alpha}  \mu_{\alpha, q}\left(p\right) \Gad{\Gainv{q} \frac{p-\frac{p-\mu_{\alpha, q}\left(p\right)}{1-\mu_{\alpha, q}\left(p\right)}}{1+\left(1-\alpha\right)\frac{p-\mu_{\alpha, q}\left(p\right)}{1-\mu_{\alpha, q}\left(p\right)} \Gainv{q}}},
\eearn
where in (a), we used the identity $\frac{\Gad{x}}{\Gad{y}} = \Gad{\frac{x-y}{1+\left(1-\alpha\right)y}}$. Let us now focus on simplifying the term (A) inside $\Gad{\cdot}$. We have
\bearn
(A) &=&  \frac{\Gainv{q} \mu_{\alpha, q}\left(p\right) \left(1-p\right)}{1-\mu_{\alpha, q}\left(p\right) + \Gainv{q}\left(1-\alpha\right)\left(p-\mu_{\alpha, q}\left(p\right)\right)}\\
 &\overset{\left(b\right)}{=}&  \frac{\Gainv{q} \mu_{\alpha, q}\left(p\right) \frac{q^{\alpha-1} \left(1-\mu_{\alpha, q}\left(p\right)\right)^2}{\Gainv{q} \left(1-\alpha \left(1-\mu_{\alpha, q}\left(p\right)\right)\right)}}{1-\mu_{\alpha, q}\left(p\right) + \Gainv{q}\left(1-\alpha\right)\left(1-\mu_{\alpha, q}\left(p\right)-\frac{q^{\alpha-1} \left(1-\mu_{\alpha, q}\left(p\right)\right)^2}{\Gainv{q} \left(1-\alpha \left(1-\mu_{\alpha, q}\left(p\right)\right)\right)}\right)} \\
 &=&  \\
 && \frac{q^{\alpha-1}\mu_{\alpha, q}\left(p\right) \left(1-\mu_{\alpha, q}\left(p\right)\right)}{ 1-\alpha\left(1-\mu_{\alpha, q}\left(p\right)\right)+\left(1-\alpha\right)\left(\Gainv{q}-\alpha \Gainv{q} \left(1-\mu_{\alpha, q}\left(p\right)\right)-q^{\alpha-1}\left(1-\mu_{\alpha, q}\left(p\right)\right)\right)} \\
  &=&  \frac{q^{\alpha-1}\mu_{\alpha, q}\left(p\right) \left(1-\mu_{\alpha, q}\left(p\right)\right)}{ 1+\left(1-\alpha\right)\Gainv{q} - \left(1-\mu_{\alpha, q}\left(p\right)\right)\left(\alpha + \alpha \left(1-\alpha\right)\Gainv{q} -\left(1-\alpha\right)q^{\alpha-1}\right)} \\
  &=&  \frac{q^{\alpha-1}\mu_{\alpha, q}\left(p\right) \left(1-\mu_{\alpha, q}\left(p\right)\right)}{ q^{\alpha-1} - \left(1-\mu_{\alpha, q}\left(p\right)\right)q^{\alpha-1}} =  1-\mu_{\alpha, q}\left(p\right),
\eearn
 where in (b), we used that
\bearn
p = \mu_{\alpha, q}^{-1}\left(\mu_{\alpha, q}\left(p\right)\right) = 1-\frac{q^{\alpha-1} \left(1-\mu_{\alpha, q}\left(p\right)\right)^2}{\Gainv{q} \left(1-\alpha \left(1-\mu_{\alpha, q}\left(p\right)\right)\right)}.
\eearn
Therefore we obtain that:
\bearn
\frac{R_{3, \alpha}\left(p, q\right)}{R_{1, \alpha}\left(p, q\right)} &=& \frac{\Gainv{q}}{\Tilde{v}_\alpha} \mu_{\alpha, q}\left(p\right) \Gad{1-\mu_{\alpha, q}\left(p\right)}.
\eearn

Since $p \to \mu_{\alpha, q}\left(p\right)$ is non-decreasing and $\Gad{\cdot}$ is non-increasing, then by composition, we have $p \mapsto R_{13, \alpha, q}\left(p\right) = \frac{R_{3, \alpha}\left(p, q\right)}{R_{1, \alpha}\left(p, q\right)}$ is non-decreasing in $\left[\rlone, 1\right]$ and $\frac{R_{3, \alpha}\left(p, q\right)}{R_{1, \alpha}\left(p, q\right)} \mbox{ in }\left[R_{13, \alpha, q}\left(\rlone\right), R_{13, \alpha, q}\left(1\right)\right]$, we have 
\bearn
R_{13, \alpha, q}\left(\rlone\right) &=& \frac{R_{3, \alpha}\left(\frac{1}{1+\Gainv{q}}, q\right)}{R_{1, \alpha}\left(\frac{1}{1+\Gainv{q}}, q\right)} = \frac{\frac{\Gainv{q}}{1+\Gainv{q}} \Gad{\frac{\Gainv{q}}{1+\Gainv{q}}}}{\Tilde{v}_\alpha} \overset{\left(a\right)}{\leq} 1,
\eearn
where in (a), we used the fact that the revenue function $x \to x \Gad{x}$ is maximized at $x = \frac{1}{\alpha}$ (with the convention that for $\alpha = 0$, $1/\alpha = \infty$) and the maximum value achieved is $\Tilde{v}_\alpha$. Furthermore, we have
\bearn
R_{13, \alpha, q}\left(1\right) &=& \frac{R_{3, \alpha}\left(1, q\right)}{R_{1, \alpha}\left(1, q\right)} = \frac{\Gainv{q}}{\Tilde{v}_\alpha}.
\eearn
Note that $R_{13, \alpha, q}\left(1\right) \geq 1$ iff  $q \le \Gad{\Tilde{v}_\alpha}$. For $q \mbox{ in }(\rlone, \Gad{\Tilde{v}_\alpha})$, we define $p_{13, \alpha, q} = R_{13, \alpha, q}^{-1}\left(1\right)$.

Therefore, we have that when  $q \mbox{ in }\left[\underline{q}_\alpha, \Gad{\Tilde{v}_\alpha}\right)$, then $R_{3, \alpha}\left(p, q\right) \leq R_{1, \alpha}\left(p, q\right)$  if $p \leq p_{13, \alpha, q}$ and $R_{3, \alpha}\left(p, q\right) \geq R_{1, \alpha}\left(p, q\right)$  if $p \geq p_{13, \alpha, q}$. And if $q \mbox{ in }\left[\Gad{\Tilde{v}_\alpha}, 1\right]$ then $R_{1, \alpha}\left(p, q\right) \geq R_{3, \alpha}\left(p, q\right) $ for all $ p \mbox{ in }\left[\rlone,1\right]$. 

\paragraph{Second point}
We have
\bearn
\frac{R_{1, \alpha}\left(p, q\right)}{R_{2, \alpha}\left(p, q\right)} &=& \frac{1}{\mu_{\alpha, q}\left(p\right)} \Gad{\Gainv{q} \frac{p-\mu_{\alpha, q}\left(p\right)}{1-\mu_{\alpha, q}\left(p\right)}} = \frac{1}{\mu_{\alpha, q}\left(p\right)} \Gad{\Gainv{q} \left(1-\frac{1-p}{1-\mu_{\alpha, q}\left(p\right)}\right)} \\
&\overset{\left(a\right)}{=}& \frac{1}{\mu_{\alpha, q}\left(p\right)} \Gad{\Gainv{q} \left(1-\frac{q^{\alpha-1} \left(1-\mu_{\alpha, q}\left(p\right)\right)}{\Gainv{q} \left(1-\alpha \left(1-\mu_{\alpha, q}\left(p\right)\right)\right)}\right)} \\
&=& \frac{1}{\mu_{\alpha, q}\left(p\right)} \Gad{\Gainv{q}-\frac{q^{\alpha-1}}{\left(\frac{1}{1-\mu_{\alpha, q}\left(p\right)}-\alpha\right)}} \\
&=& \frac{1}{\mu_{\alpha, q}\left(p\right)} \Gad{\Gainv{q}-\frac{q^{\alpha-1}}{\left(\frac{1}{1-\mu_{\alpha, q}\left(p\right)}-\alpha\right)}},
\eearn
where in (a), we used:
\bearn
p = \mu_{\alpha, q}^{-1}\left(\mu_{\alpha, q}\left(p\right)\right) = 1-\frac{q^{\alpha-1} \left(1-\mu_{\alpha, q}\left(p\right)\right)^2}{\Gainv{q} \left(1-\alpha \left(1-\mu_{\alpha, q}\left(p\right)\right)\right)}.
\eearn
Therefore, by composition, we have $p \to R_{12, \alpha, q}\left(p\right) = \frac{R_{1, \alpha}\left(p, q\right)}{R_{2, \alpha}\left(p, q\right)}$ is non-increasing in $\left[\rlone, 1\right]$ and $\frac{R_{1, \alpha}\left(p, q\right)}{R_{2, \alpha}\left(p, q\right)} \mbox{ in }\left[R_{12, \alpha, q}\left(1\right), R_{12, \alpha, q}\left(\rlone\right)\right]$, we have 
\bearn
R_{12, \alpha, q}\left(\rlone \right) &=& \frac{1}{\rln} \Gad{\Gainv{q}-\frac{q^{\alpha-1}}{\left(\frac{1}{1-\rln}-\alpha\right)}}\\
&=& \left(1+\Gainv{q}\right) \Gad{\Gainv{q}-\frac{q^{\alpha-1}}{\frac{1+\left(1-\alpha\right)\Gainv{q}}{\Gainv{q}}}} \\
&=& \left(1+\Gainv{q}\right) \geq 1,\\
\lim_{p \to 1} R_{12, \alpha, q}\left(p\right) &=& \lim_{\mu \to 1} \frac{1}{\mu} \Gad{\Gainv{q}-\frac{q^{\alpha-1}}{\left(\frac{1}{1-\mu}-\alpha\right)}} = 0.
\eearn
We define $p_{12, \alpha, q} = R_{12, \alpha, q}^{-1}\left(1\right)$. Therefore, $R_{1, \alpha}\left(p, q\right) \leq R_{2, \alpha}\left(p, q\right)$ if $p \leq p_{12, \alpha, q}$ and $R_{1, \alpha}\left(p, q\right) \geq R_{2, \alpha}\left(p, q\right)$ if $p \geq p_{12, \alpha, q}$.

\paragraph{Third point}
If $q$ belongs to $[\underline{q}_\alpha, 1]$, we have
\bearn
R_{3, \alpha}\left(p, q\right) &=& \frac{\Gainv{q}}{\Tilde{v}_\alpha}  \Gad{\Gainv{q} p} \\
\frac{\partial R_{3, \alpha}\left(p, q\right)}{\partial p} &=& \frac{\Gainv{q}}{\Tilde{v}_\alpha}  \left(\Gad{\Gainv{q} p}-\Gainv{q} p \Gad{\Gainv{q} p}^{2-\alpha}\right) \\ 
&=& \frac{\Gainv{q}}{\Tilde{v}_\alpha}  \Gad{\Gainv{q} p} \left( 1- \frac{\Gainv{q} p}{1+\left(1-\alpha\right)\Gainv{q} p}\right)
\\ 
&=& \begin{cases}
    \frac{\Gainv{q}^2}{\Tilde{v}_\alpha}  \Gad{\Gainv{q} p} \left(\frac{\frac{1}{\Gainv{q}}}{1+\left(1-\alpha\right)\Gainv{q} p}\right) &\quad \mbox{if } \alpha = 0,\\
    \alpha \frac{\Gainv{q}^2}{\Tilde{v}_\alpha}  \Gad{\Gainv{q} p} \left(\frac{\rhone - p}{1+\left(1-\alpha\right)\Gainv{q} p}\right) &\quad \mbox{if } \alpha \in (0,1].\\
\end{cases}
\eearn
Therefore, if $q \mbox{ in }[\underline{q}_\alpha, 1]$, then $\rhone \geq 1$ and therefore $R_{3, \alpha}\left(\cdot, q\right)$ is non-decreasing in $\left[\rln, 1\right]$.

\paragraph{Fourth point}
\subparagraph{Case 1: Regular $\alpha=0$}
In this case, the first function is expressed as follows:
\bearn
R_{1, 0}\left(p, q\right) &=& \frac{p}{\mu_{0, q}\left(p\right) \left(1+ \left(\frac{1}{q}-1\right) \frac{p-\mu_{0, q}\left(p\right)}{1-\mu_{0, q}\left(p\right)}\right)} = \frac{p q \left(\mu_{0, q}\left(p\right) - 1\right)}{\mu_{0, q}\left(p\right) \left(p \left(q - 1\right) - q + \mu_{0, q}\left(p\right)\right)}\\
\mbox{with } \mu_{0, q}\left(p\right) &=& 1- \frac{\sqrt{4 \left(q^{-1}-1\right) \left(1-p\right) q^{-1}}}{2q^{-1}} =  1-\sqrt{\left(1-p\right)\left(1-q\right)}.
\eearn
Therefore 
\bearn
R_{1, 0}\left(p, q\right) &=& \frac{p}{\left(1-\sqrt{\left(1-p\right)\left(1-q\right)}\right) \left(1+ \left(\frac{1}{q}-1\right) \frac{p-1+\sqrt{\left(1-p\right)\left(1-q\right)}}{1-1+\sqrt{\left(1-p\right)\left(1-q\right)}}\right)} \\
&=& \frac{p}{\left(1-\sqrt{\left(1-p\right)\left(1-q\right)}\right) \left(1+ \frac{1-q}{q} \frac{p-1+\sqrt{\left(1-p\right)\left(1-q\right)}}{\sqrt{\left(1-p\right)\left(1-q\right)}}\right)} \\
&=& \frac{p}{\left(1-\sqrt{\left(1-p\right)\left(1-q\right)}\right) \left(1+ \frac{1-q}{q} \left(1-\frac{\sqrt{1-p}}{\sqrt{1-q}}\right) \right)} \\
&=& \frac{pq}{\left(1-\sqrt{\left(1-p\right)\left(1-q\right)}\right) \left(q +  1-q-\sqrt{\left(1-q\right)\left(1-p\right)} \right)} \\
&=& \frac{pq}{\left(1-\sqrt{\left(1-p\right)\left(1-q\right)}\right)^2}.
\eearn
We have, for all $p \mbox{ in }\left[q, 1\right]$:
\bearn
\frac{\partial R_{1, 0}}{\partial p}\left(p, q\right) = -\frac{q\sqrt{1-p}\left(\sqrt{1-q}-\sqrt{1-p}\right)}{\left(1-p\right)\left(1-\sqrt{\left(1-p\right)\left(1-q\right)}\right)^3} \leq 0 \quad \forall p \mbox{ in }\left[q,1\right],
\eearn
and it is easy to see that function $p \to R_2\left(p,q\right)$ is non-decreasing.

\subparagraph{Case 2: mhr case $\alpha=1$}

\noindent For $\alpha = 1$, the first function is expressed as follows:
\bearn
R_{1, 1}\left(p, q\right) &=& \frac{p}{\mu_{1, q}\left(p\right)} e^{-\log\left(q^{-1}\right)\frac{p-\mu_{1, q}\left(p\right)}{1-\mu_{1, q}\left(p\right)}} \\
\mbox{with } \mu_{1, q}\left(p\right) &=& 1- \frac{\sqrt{\log\left(q^{-1}\right) \left(1-p\right))^2 + 4 \log\left(q^{-1}\right) \left(1-p\right) }-\log\left(q^{-1}\right) \left(1-p\right)}{2} \\
\mbox{and }p &=& \mu_{1, q}^{-1}\left(\mu_{1, q}\left(p\right)\right) = 1-\frac{ \left(1-\mu_{1, q}\left(p\right)\right)^2}{\log\left(q^{-1}\right) \mu_{1, q}\left(p\right)}.
\eearn
We therefore have
\bearn
R_{1, 1}\left(p, q\right) &=& \frac{1-\frac{ \left(1-\mu_{1, q}\left(p\right)\right)^2}{\log\left(q^{-1}\right) \mu_{1, q}\left(p\right)}}{\mu_{1, q}\left(p\right)} e^{-\log\left(q^{-1}\right)\frac{1-\mu_{1, q}\left(p\right)-\frac{ \left(1-\mu_{1, q}\left(p\right)\right)^2}{\log\left(q^{-1}\right) \mu_{1, q}\left(p\right)}}{1-\mu_{1, q}\left(p\right)}} \\
&=& \left(\frac{1}{\mu_{1, q}\left(p\right)} - \frac{1}{\log\left(q^{-1}\right)}\left(\frac{1}{\mu_{1, q}\left(p\right)}-1\right)^2\right) e^{-\log\left(q^{-1}\right) + \frac{1}{\mu_{1, q}\left(p\right)} - 1} \\
&=& \frac{q}{e\log\left(q^{-1}\right)} \left(\frac{\log\left(q^{-1}\right)}{\mu_{1, q}\left(p\right)} - \frac{1}{\mu_{1, q}\left(p\right)^2} + \frac{2}{\mu_{1, q}\left(p\right)} -1\right) e^{\frac{1}{\mu_{1, q}\left(p\right)}} \\
&=& -\frac{q}{e\log\left(q^{-1}\right)} \left(\frac{1}{\mu_{1, q}\left(p\right)^2} - \frac{\log\left(q^{-1}\right)+2}{\mu_{1, q}\left(p\right)}  +1\right) e^{\frac{1}{\mu_{1, q}\left(p\right)}} =: \Tilde{R_1}\left(\frac{1}{\mu_{1, q}\left(p\right)}\right),
\eearn
with 
\bearn
\Tilde{R_1}\left(x\right) = -\frac{q}{e \log\left(q^{-1}\right)} \left(x^2 -\left(2+\log\left(q^{-1}\right)\right) x +1\right) e^{x} \quad \mbox{for } x \mbox{ in }\left[1, 1+\log\left(q^{-1}\right)\right].\eearn
On another hand, we have
\bearn
\frac{d \Tilde{R_1}\left(x\right)}{dx} &=& -\frac{q}{e \log\left(q^{-1}\right)} \left(2x - 2-\log\left(q^{-1}\right) + x^2 -\left(2+\log\left(q^{-1}\right)\right)x +1 \right) e^{x} \\
&=& -\frac{q}{e \log\left(q^{-1}\right)} \left(x^2 -\log\left(q^{-1}\right) x - \left(1+\log\left(q^{-1}\right)\right) \right) e^{x} \\
&=& -\frac{q}{e \log\left(q^{-1}\right)} \left(x +1\right) \left(x-\left(1+\log\left(q^{-1}\right)\right)\right) e^{x} \geq 0 \quad \forall x \mbox{ in }\left[1, 1+\log\left(q^{-1}\right)\right].
\eearn
For all $p$ in $[\frac{1}{1+\log\left(q^{-1}\right)}, 1]$, we have $\mu_{1,q}\left(p\right)$ in $[\frac{1}{1+\log\left(q^{-1}\right)}, 1]$, therefore $\frac{1}{\mu_{1,q}\left(p\right)}$ in $[1, 1+\log\left(q^{-1}\right)]$.

Therefore, by composition, $p \to R_{1, 1}\left(p, q\right)$ is non increasing. The function $p \to R_{2, 1}\left(p,q\right) = p$ is non-decreasing.
\end{proof}

\begin{proof}[\underline{\textbf{Proof of \Cref{r3_properties}}}] 
\textbf{Case 1: Regular case $\alpha=0$.}
In this case, for $q \mbox{ in } (\underline{q}_\alpha, \Gad{\Tilde{v}_\alpha}] \overset{\alpha = 0}{=} \left(0, \frac{1}{2}\right]$, $p_{13, 0, q}$ is a solution to the following equation
\bearn
&& \frac{\Gainv{q}}{\Tilde{v}_\alpha} \mu_{\alpha, q}\left(p\right) \Gad{1-\mu_{\alpha, q}\left(p\right)}   = 1 \\
&\text{iff}&  \left(\frac{1}{q}-1\right) \frac{1-\sqrt{\left(1-p\right)\left(1-q\right)}}{1+\sqrt{\left(1-p\right)\left(1-q\right)}} = 1\\
&\text{iff}& \frac{1-2q}{1-q} = \frac{\sqrt{\left(1-p\right)\left(1-q\right)}}{1-q} \\
&\text{iff}& p = p_{13, 0, q} = 1- \frac{\left(1-2q\right)^2}{1-q}.
\eearn
and $p_{12, 0, q}$ is solution to the following equation 
\bearn
&& \frac{1}{\mu_{\alpha, q}\left(p\right)} \Gad{\Gainv{q}-\frac{q^{\alpha-1}}{\left(\frac{1}{1-\mu_{\alpha, q}\left(p\right)}-\alpha\right)}} = 1 \\
&\text{iff}& \frac{q}{\left(1-\sqrt{\left(1-p\right)\left(1-q\right)}\right)^2} = 1\\
&\text{iff}& \sqrt{q} = 1-\sqrt{\left(1-p\right)\left(1-q\right)} \\
&\text{iff}& \sqrt{\left(1-p\right)} = \frac{1-\sqrt{q}}{\sqrt{1-q}} \\
&\text{iff}& p = p_{12, 0, q} = 1 - \frac{\left(1-\sqrt{q}\right)^2}{1-q}.
\eearn
Therefore
\bearn
p_{13, 0, q} \leq p_{12, 0, q} &\text{iff}& 1- \frac{\left(1-2q\right)^2}{1-q}  \leq 1 - \frac{\left(1-\sqrt{q}\right)^2}{1-q} \\
&\text{iff}& \left(1-2q\right)^2  \geq \left(1-\sqrt{q}\right)^2 \\
&\text{iff}& 2q  \leq \sqrt{q} \\
&\text{iff}& q  \leq \frac{1}{4} := \hat{q}_{0}.
\eearn
Furthermore, note that $\hat{q}_{0} \in [\underline{q}_\alpha, \Gad{\Tilde{v}_\alpha}]$, since $\underline{q}_\alpha \overset{\alpha = 0}{=}  0$ and    $\Gad{\Tilde{v}_\alpha} \overset{\alpha = 0}{=} \frac{1}{2}$.

\textbf{Case 2: mhr case $\alpha=1$. }
In this case, for $q \mbox{ in }(\underline{q}_\alpha, \Gad{\Tilde{v}_\alpha}] \overset{\alpha = 1}{=} \left(e^{-1}, e^{-e^{-1}}\right]$, $p_{13, 1, q}$ is solution to the following equation
\bearn
&& \frac{\Gainv{q}}{\Tilde{v}_\alpha} \mu_{\alpha, q}\left(p\right) \Gad{1-\mu_{\alpha, q}\left(p\right)} \overset{\alpha = 1}{=} \log\left(q^{-1}\right) e \mu_{1, q}\left(p\right) e^{\mu_{1, q}\left(p\right)-1} = 1 \\
&\text{iff}& \mu_{1, q}\left(p\right) e^{\mu_{1, q}\left(p\right)-1} = \frac{1}{\log\left(q^{-1}\right)} \\
&\text{iff}& p = p_{13, 1, q} = \mu_{1, q}^{-1}\left(W\left(\frac{1}{\log\left(q^{-1}\right)}\right)\right).
\eearn
And $p_{12, 1, q}$ is solution to the following equation 
\bearn
&& \frac{1}{\mu_{\alpha, q}\left(p\right)} \Gad{\Gainv{q}-\frac{q^{\alpha-1}}{\left(\frac{1}{1-\mu_{\alpha, q}\left(p\right)}-\alpha\right)}} \overset{\alpha = 1}{=} \frac{1}{\mu_{1, q}\left(p\right)} e^{\log\left(q\right)+\frac{1}{\frac{1}{1-\mu_{1, q}\left(p\right)}-1}} = 1 \\
&\text{iff}& \frac{1}{\mu_{1, q}\left(p\right)} e^{\frac{1}{\mu_{1, q}\left(p\right)}} \frac{q}{e} = 1 \\
&\text{iff}& \frac{1}{\mu_{1, q}\left(p\right)} e^{\frac{1}{\mu_{1, q}\left(p\right)}} = \frac{e}{q} \\
&\text{iff}& p = p_{12, 1, q} = \mu_{1, q}^{-1}\left(\frac{1}{W\left(\frac{e}{q}\right)}\right).
\eearn
Therefore
\bearn
p_{13, 1, q} \leq p_{12, 1, q} &\text{iff}& \mu_{1, q}^{-1}\left(W\left(\frac{1}{\log\left(q^{-1}\right)}\right)\right)  \leq \mu_{1, q}^{-1}\left(\frac{1}{W\left(\frac{e}{q}\right)}\right) \\
&\text{iff}&  W\left(\frac{1}{\log\left(q^{-1}\right)}\right) \leq \frac{1}{W\left(\frac{e}{q}\right)} \quad \mbox{as } \mu_{1, q}^{-1}\left(.\right) \mbox{ is increasing} \\
&\text{iff}&  W\left(\frac{1}{\log\left(q^{-1}\right)}\right)W\left(\frac{e}{q}\right) \leq 1.
\eearn
We now study $q \mapsto g\left(q\right):= W\left(\frac{1}{\log\left(q^{-1}\right)}\right) W\left(\frac{e}{q}\right)$ in $\left(e^{-1}, e^{-e^{-1}}\right]$, we have
\bearn
\frac{d g\left(q\right)}{dq} &=& -\frac{W\left(\frac{e}{q}\right) W\left(-\frac{1}{\log \left(q\right)}\right)\left(W\left(\frac{e}{q}\right)+\log \left(q\right) W\left(-\frac{1}{\log \left(q\right)}\right)+\log \left(q\right)+1\right)}{q \log \left(q\right)\left(W\left(\frac{e}{q}\right)+1\right)\left(W\left(-\frac{1}{\log \left(q\right)}\right)+1\right)}.
\eearn
We next analyze the sign of the derivative.
\bearn 
 sign\left(\frac{d g\left(q\right)}{dq}\right) &=& sign\left(W\left(\frac{e}{q}\right)+\log \left(q\right) W\left(-\frac{1}{\log \left(q\right)}\right)+\log \left(q\right)+1\right) = sign\left(h\left(q\right)\right) \\
\mbox{with } h\left(q\right) &:=& W\left(\frac{e}{q}\right)+\log \left(q\right) W\left(-\frac{1}{\log \left(q\right)}\right)+\log \left(q\right)+1 \\
\frac{d h\left(q\right)}{dq} &=& \frac{\left(W\left(\frac{e}{q}\right)+1\right) W\left(-\frac{1}{\log \left(q\right)}\right)^{2}+W\left(-\frac{1}{\log \left(q\right)}\right)+1}{q\left(W\left(\frac{e}{q}\right)+1\right)\left(W\left(-\frac{1}{\log \left(q\right)}\right)+1\right)} \geq 0 \quad \forall q \mbox{ in }\left[e^{-1}, e^{-e^{-1}}\right].
\eearn

Therefore $q \mapsto h\left(q\right)$ is non-decreasing and we have  $h\left(e^{-1}\right) = W\left(e^{2}\right)-W\left(1\right) >0$. Therefore $g$ is increasing in $\left(e^{-1}, e^{-e^{-1}}\right]$. Furthermore, we have $g\left(e^{-1}\right) = W\left(1\right) W\left(e^{2}\right) < 1$ and $g\left(e^{-e^{-1}}\right) = W\left(e^{1+1 / e}\right) > 1$. 
Therefore there exists a unique $\hat{q}_{1}$ solution in $\left(e^{-1},e^{-e^{-1}}\right]$ to the equation $W\left(\frac{1}{\log\left(q^{-1}\right)}\right) W\left(\frac{e}{q}\right) = 1$.

Therefore, we have, for $q \mbox{ in }\left(e^{-1}, e^{-e^{-1}}\right]$
\bearn
p_{13, 1, q} \leq p_{12, 1, q} &\text{iff}& g\left(q\right) \leq 1 \quad \text{iff} \quad q \leq \hat{q}_{1}.
\eearn

\end{proof}

\setcounter{equation}{0}
\setcounter{proposition}{0}
\setcounter{lemma}{0}

\section{Proofs and auxiliary results for  \Cref{sec:rand}}\label{apx:mechanismreduction}\label{apx:LP_q}

\subsection{Proofs and auxiliary results for  \Cref{sec:rand-opt}} 

\begin{proof}[\textbf{Proof of \Cref{thm:mechanismreduction}.}] 
Fix $\alpha \mbox{ in }[0,1]$,  $\Psi \mbox{ in }\MechSet$, $q \mbox{ in }(0,1)$, $N > 1$, and a finite sequence of increasing reals $\IN= \{a_i\}_{i=1}^{N}$ such that $0 < a_1 \leq \pin \leq a_N$. The proof uses two building blocks associated with  uniformly bounding the losses stemming from truncating a mechanism and the losses stemming from local transfers of mass in a mechanism. 


Define the two mechanisms $\Psi_{a_N}$ in $\MechSet$  and $\Psi_{\IN} \mbox{ in }\MechSet_{\IN}$ as follows
\bearn
\Psi_{a_N}(v) &=& \begin{cases}
			\Psi(v) &\quad \mbox{if } v \in [0,a_N),\\
			1 & \quad \mbox{if } v \geq a_N.
		\end{cases}\\
\Psi_{\IN}(x) &=& \begin{cases}
            0 &\quad \mbox{if } x \in [0, a_{1}).\\
			\Psi(a_{i+1}) &\quad \mbox{if } x \in [a_i, a_{i+1}), \quad \mbox{for } 1 \leq i \leq N-1 ,\\
			1 &\quad \mbox{if } x \in [a_{N}, \infty).
		\end{cases}
\eearn
$\Psi_{a_N}$ is a truncated version of $\Psi$ at $a_N$ and $\Psi_{\IN}$ is a discretized and truncated version of $\Psi$. 

Let $F \mbox{ in }\aDistSetqa{q}$. Next we analyze $R(\Psi,F) - R(\Psi_{\IN},F)$ by decomposing it as follows.
\bear \label{eq:decomp}
R(\Psi,F) - R(\Psi_{\IN},F) &=& R(\Psi,F) - R(\Psi_{a_N},F) + R(\Psi_{a_N},F) - R(\Psi_{\IN},F).
\eear

To uniformly bound the maximal losses stemming from truncation $R(\Psi,F) - R(\Psi_{a_N},F)$, we establish the following result, whose proof is deferred to \Cref{app:aux-rand}. 
    \begin{lemma}[Truncation]\label{lem:bsupport}
Fix a mechanism $\Psi \mbox{ in }\MechSet$, $b\ge \pin$, $q \mbox{ in }(0,1)$ and  let
\bearn
\Psi_b(v) &=& \begin{cases}
			\Psi(v) &\quad \mbox{if } v \in [0,b),\\
			1 & \quad \mbox{if } v \geq b.
		\end{cases}
\eearn
Then  for any distribution $F \mbox{ in }\aDistSetqa{q}$,
\bearn
 R (\Psi_b, F) \geq   R (\Psi, F) - \frac{1}{q\left(1+(q^{-1}-1)b/\pin\right)}\mathbf{1}\{b \leq \rh\}.
\eearn
\end{lemma}
In particular, the result upper bounds the maximal performance  losses that can stem from truncating a pricing mechanism at $b$. 

To uniformly bound the the impact of discretization $R(\Psi_{a_N},F) - R(\Psi_{\IN},F)$,  we first establish a result (whose proof is deferred to \Cref{app:aux-rand}) that bounds the performance losses stemming from transferring mass locally in a mechanism. 
    \begin{lemma}[Local transfer of mass]\label{lem:epsupport}
Fix a mechanism $\Psi \mbox{ in } \MechSet$, $0 < \epsilon < v$, and let
\bearn
\Psi_{\epsilon, v}(x) &=& \begin{cases}
			\Psi(x) &\quad \mbox{if } x \in [0,v-\epsilon),\\
			\Psi(v) &\quad \mbox{if } x \in [v-\epsilon, v),\\
			\Psi(x) & \quad \mbox{if } x \geq v.
		\end{cases}
\eearn
Then, for any distribution $F \mbox{ in }\aDistSetqa{q}$
\bearn
R (\Psi_{\epsilon, v}, F) \geq   R (\Psi, F) - \frac{\epsilon}{v}\left(\Psi(v)-\Psi(v-\epsilon)\right).
\eearn
\end{lemma}





Applying \cref{lem:epsupport} on $(v,\epsilon) = (a_i, a_i-a_{i-1})$, $N-1$ times consecutively for $2 \leq i \leq N$ on the mechanism $\Psi_{a_N}$, we obtain
\bearn
R(\Psi_{a_N}, F) - R(\Psi_{\IN}, F) 
&\le&   \sum_{i = 2}^{N} \frac{a_i-a_{i-1}}{a_i} (\Psi(a_i) - \Psi(a_{i-1}))\\
&\stackrel{(a)}{\le}& \frac{\Delta(\IN)}{a_1} \sum_{i = 2}^{N} (\Psi(a_i) - \Psi(a_{i-1}))
= \frac{\Delta(\IN)}{a_1} (\Psi(a_{N})-\Psi(a_{1}))
\le \frac{\Delta(\IN)}{a_1}.
\eearn
where $(a)$ follows from $a_i-a_{i-1} \leq \sup_i (a_i - a_{i-1}) = \sigma(\IN)$ and $a_i \geq a_1 > 0$. 
Using \Cref{lem:bsupport}, we have
\bearn
R (\Psi_b, F) \geq   R (\Psi, F) - \frac{1}{q\left(1+(q^{-1}-1)a_N/\pin\right)}\mathbf{1}\{a_N \leq \rh\}.
\eearn
Returning to the decomposition in \eqref{eq:decomp}, we have established
\bearn
R(\Psi,F) - R(\Psi_{\IN},F) &\le& \frac{\Delta(\IN)}{a_1}+  \frac{1}{q(1+\left(q^{-1}-1)a_N/\pin\right)}\mathbf{1}\{a_N \leq \rh\}.
\eearn 
Noting that the inequality above applies for any $F$ in $\aDistSetqa{q}$ and that the mechanism $\Psi_{\IN}$ does not depend on $F$, the result follows. 
\end{proof}

\setcounter{equation}{0}
\setcounter{theorem}{0}
\setcounter{lemma}{0}
\setcounter{proposition}{0}

\noindent
\underline{\textbf{Proof of \Cref{thm:LB_LP}.}} This result is a special case of \Cref{thm:LB_LP_robust}.

\subsubsection{Proofs of auxiliary results}\label{app:aux-rand}

\begin{proof}[\textbf{Proof of \Cref{lem:bsupport}.}]
Let $Rev(q) = q F^{-1}(1-q)$ denote the revenue curve of associated with $F$ in the quantity space.  Let $r_F$ denote the optimal oracle price, $q_F$ the corresponding quantity,  and recall, from \Cref{lem:decreasingqs} that $r_F \le \rh$. By definition, we have
\bearn
R(\Psi_b, F) &=& R(\Psi, F) + \int_b^{+\infty} \frac{Rev(q_b)-Rev(q_x)}{\opt(F)} d\Psi(x).
\eearn

\textbf{Case 1.} Suppose first $b > \rh$. In this case,  then $r_F \leq b$. Given that $F$ is regular, the revenue curve is monotone for $q \le q_b$, and we have $Rev(q_b)-Rev(q_x)$ for $x\ge b$. We then have 
\bearn
 R (\Psi_b, F) \geq  R (\Psi, F). 
\eearn

\textbf{Case 2.} Suppose now that $b \leq \rh$. In this case, we divide the analysis into two subcases.

\textbf{Case a).} Suppose first that $r_F \leq b$. 
We have for any $x \geq b \geq r$, by monotonicity of the revenue curve, $Rev(q_b)-Rev(q_x) \ge 0$, and therefore
\bearn
 R (\Psi_b, F) \geq  R (\Psi, F). 
\eearn
\textbf{Case b).} $r_F > b$
We have:
\bear
R(\Psi_b, F) &\geq& R(\Psi, F) + \int_b^{+\infty} \frac{Rev(q_b)-Rev(q_x)}{\opt(F)} d\Psi(x) \nonumber\\
            &\geq& R(\Psi, F) + \int_b^{+\infty} \left(\frac{Rev(q_b)}{\opt(F)}-1\right) d\Psi(x) \nonumber \\
            &\ge& R(\Psi, F) +  \left(\frac{Rev(q_b)}{\opt(F)}-1\right). \label{eq:qb}
\eear

Recall that by assumption $b \ge \pin$ and hence $q_b \le q$. Using  concavity of the revenue curve in the quantity space (which follows from regularity of $F$), we have 
\bearn
Rev(q_b) &\ge& Rev(q_F) + \frac{Rev(q)-Rev(q_F)}{q-q_F} (q_b - q_F).
\eearn
This implies that
\bearn
\frac{Rev(q_b)}{Rev(q_F)} &\ge& 1 + \left(\frac{Rev(q)}{Rev(q_F)} -1\right) \frac{q_b - q_F}{q-q_F} 
 \:\ge\: \frac{q - q_b}{q-q_F} + \frac{Rev(q)}{Rev(q_F)} \frac{q_b - q_F}{q-q_F} 
\:\ge\: \frac{q - q_b}{q-q_F}.
\eearn
Noting that $F$ is regular and using \Cref{lemma:singlecross}, we have 
\bearn
q_b \leq \Gamma_{0}\left( \Gamma_{0}^{-1}\left(q\right) \frac{b}{\pin}   \right) =  \frac{1}{1+(q^{-1}-1)b/\pin}.
\eearn
Therefore
\bearn
\frac{Rev(q_b)}{\opt(F)} &\geq& \frac{q}{q-q_F}\left(1-\frac{1}{q\left(1+\left(q^{-1}-1\right)b/\pin \right)}\right) \geq 1-\frac{1}{q\left(1+\left(q^{-1}-1\right)b/\pin\right)}.
\eearn 
Returning to \eqref{eq:qb}, we deduce 
\bearn
 R (\Psi_b, F) \geq   R (\Psi, F) - \frac{1}{q(1+(q^{-1}-1)b/\pin)}.
\eearn
Combining both cases, the result follows.
\end{proof}


\begin{proof}[\textbf{Proof of  \Cref{lem:epsupport}.}]
Let $r_F$ denote the optimal oracle price, $q_F$ the corresponding quantity. We have
\bearn
R(\Psi_{\epsilon, v}, F) &=& R(\Psi, F) + \int_{v-\epsilon}^{v} \frac{Rev(q_{v-\epsilon})-Rev(q_x)}{\opt(F)} d\Psi(x).
\eearn

\textbf{Case 1.} Suppose  $r_F \leq v-\epsilon$. In this case, using the regularity of $F$ and the unimodality  of the revenue curve, we have for any $x \geq v-\epsilon \geq r_F$,  $Rev(q_{v-\epsilon})-Rev(q_x) \geq 0$, and
\bearn
 R (\Psi_{\epsilon, v}, F) \geq   R (\Psi, F).  
\eearn

\textbf{Case 2.} Suppose now $v-\epsilon < r_F \leq v$. In this case, we have
\bearn
R(\Psi_{\epsilon, v}, F) &=& R(\Psi, F) + \int_{v-\epsilon}^{v} \frac{Rev(q_{v-\epsilon})-Rev(q_x)}{opt(F)} d\Psi(x) \\
            &\geq& R(\Psi, F) + \int_{v-\epsilon}^{v} \left(\frac{Rev(q_{v-\epsilon})}{\opt(F)}-1\right) d\Psi(x) \\
            &=& R(\Psi, F) + \left(\frac{Rev(q_{v-\epsilon})}{\opt(F)}-1\right)(\Psi(v)-\Psi(v-\epsilon)).
\eearn

In this case, we have $\opt(F) = r_F q_F \leq v q_{v-\epsilon}$. Therefore
\bearn
R(\Psi_{\epsilon, v}, F) &\geq& R(\Psi, F) + \left(\frac{v-\epsilon}{v}-1\right)(\Psi(v)-\Psi(v-\epsilon)) \\
             &\geq& R(\Psi, F) -\frac{\epsilon}{v}(\Psi(v)-\Psi(v-\epsilon)).
\eearn

\textbf{Case 3.} Suppose now $v < r_F$. In this case, 
for any $v-\epsilon \leq x \leq v < r_F$, by monotonicity of the revenue curve,  $Rev(q_{v-\epsilon})-Rev(q_x) \leq 0$, and furthermore, $Rev(q_x) \le Rev(q_v) \le \opt(F) $. In turn, we have
\bearn
R(\Psi_{\epsilon, v}, F) &=& R(\Psi, F) + \int_{v-\epsilon}^{v} \frac{Rev(q_{v-\epsilon})-Rev(q_x)}{\opt(F)} d\Psi(x) \\
            &\geq& R(\Psi, F) + \int_{v-\epsilon}^{v} \left(\frac{Rev(q_{v-\epsilon})}{Rev(q_v)}-\frac{Rev(q_{x})}{Rev(q_v)}\right) d\Psi(x) \\
            &\geq& R(\Psi, F) + \int_{v-\epsilon}^{v} \left(\frac{Rev(q_{v-\epsilon})}{Rev(q_v)}-1\right) d\Psi(x) \\
            &=& R(\Psi, F) + \left(\frac{Rev(q_{v-\epsilon})}{Rev(q_v)}-1\right)(\Psi(v)-\Psi(v-\epsilon)) \\
            &=& R(\Psi, F) + \left((1-\frac{\epsilon}{v})\frac{q_{v-\epsilon}}{q_v}-1\right)(\Psi(v)-\Psi(v-\epsilon))\\
            &\ge& R(\Psi, F) + \left((1-\frac{\epsilon}{v})-1\right)(\Psi(v)-\Psi(v-\epsilon))\\
            &=& R(\Psi, F)  -\frac{\epsilon}{v}(\Psi(v)-\Psi(v-\epsilon)).
\eearn
Combining the three cases yields the result. 
\end{proof}

\setcounter{equation}{0}
\setcounter{theorem}{0}
\setcounter{lemma}{0}
\setcounter{proposition}{0}

\setcounter{equation}{0}
\setcounter{proposition}{0}
\setcounter{lemma}{0}

\subsection{Proofs for  \Cref{sec:rand-perf}}\label{apx:qzerbounds}

\begin{proof}[\textbf{Proof of \Cref{lem:qzerbounds}.}]

The proof is divided into two steps. In the first step, we will show the lower bound by analyzing the performance of a specific mechanism. Then in a second step, we will derive the upper through the analysis of a family of hard cases when $q$ is close to $0$.

\paragraph{Step 1: Lower bound}
Let  us define the following measure:
\bearn
d \Psi(u) = \begin{cases}
0 &\quad \mbox{if } u < \pin q, \\
\frac{1}{u \log(\frac{1}{q})} &\quad \mbox{if } u \mbox{ in }[q \pin,\pin)\\
0 &\quad \mbox{if } u \geq \pin.
\end{cases}
\eearn

We have that $\Psi(u) $ is a distribution since
\bearn
\int_{0}^{\infty} d\Psi(u) = \frac{1}{ \log(\frac{1}{q})}\int_{\pin q}^{\pin} \frac{1}{u}du = 1,
\eearn

Using \Cref{thm:LB_1D} and the fact that $\aDistSetqa{q} \subseteq \Fb_0(w,q)$ for any $\alpha \in [0,1]$, we have
\bearn 
\inf _{F \in \aDistSetqa{q}} R(\Psi, F) &\geq& \inf _{F \in \Fb_0(w,q)} R(\Psi, F)  \\
&=& \min \Biggl\{ \inf_{x \in [\underline{r}_{0}(\pin, q), \pin)}  \frac{1}{\opt(F_{0}( \cdot | x, (\pin, q)))}  \int_{0}^{\infty}    u \bF_{0}( u | x, (\pin, q)) d\Psi(u), \\
&& \inf_{x \in [\pin, \overline{r}_0(\pin, q)]} \frac{1}{\opt(F_{0}( \cdot | x, (\pin, q)))} \int_{0}^{\infty}  u \bF_{0}( u | x, (\pin, q)))  d\Psi(u) \Biggr\} \\
&=& \min \Biggl\{ \inf_{x \in [\pin q, \pin)}  \frac{1}{x} \Biggl[ \int_{0}^{x} u d\Psi(u) + \int_{x}^{\pin} u \bG_{0, \pin}(u|(x,1), (\pin, q))  d\Psi(u) \Biggl], \\
&& \inf_{x \in [\pin, \infty)} \frac{1}{x \bG_{0, x}(x|(0,1), (\pin, q))} \int_{0}^{x} u \bG_{0, x}(u|(0,1), (\pin, q))  d\Psi(u) \Biggr\}.
 \eearn

We will analyze each term separately depending whether $x\mbox{ in } [\pin q, \pin)$ or $x\mbox{ in }[\pin, \infty).$

\subparagraph{Case 1: $x\mbox{ in } [\pin q, \pin)$} We have 

\bearn
&&\frac{1}{x} \Biggl[ \int_{0}^{x} u d\Psi(u) + \int_{x}^{\pin} u \bG_{0, \pin}(u|(x,1), (\pin, q))  d\Psi(u) \Biggl] \\
&=& \frac{1}{x\log(\frac{1}{q})} \Biggl[ \int_{\pin q}^{x}  u du + \int_{x}^{\pin}   \bG_{0, \pin}(u|(x,1), (\pin, q)) du \Biggr] \\
&=& \frac{1}{x\log(\frac{1}{q})} \Biggl[ x-\pin q + \int_{x}^{\pin} \frac{1}{1 + (1/q-1)\frac{u-x}{\pin-x}} du \Biggr], \\
&=& \frac{1}{x\log(\frac{1}{q})} \left( x-\pin q + \Biggl[(\pin-x)\frac{\log(1 + (1/q-1)\frac{u-x}{\pin-x})}{(1/q-1)}\Biggl]_{u = x}^{u = \pin}\right), \\
&=& \frac{1}{x\log(\frac{1}{q})} \left( x-\pin q + \log(\frac{1}{q})\frac{\pin-x}{(1/q-1)}\right),
\eearn

Hence we get that 
\bearn
&& \frac{1}{x} \Biggl[ \int_{0}^{x} u d\Psi(u) + \int_{x}^{\pin} u \bG_{0, \pin}(u|(x,1), (\pin, q))  d\Psi(u) \Biggl] \\ &=& \frac{1}{\log(\frac{1}{q})} \left(1 + \frac{\pin q}{x}(\frac{\log(q)}{q-1}-1)- \frac{\log(\frac{1}{q})}{(1/q-1)}\right)\\
&\stackrel{(a)}{\ge}& \frac{1}{\log(\frac{1}{q})} \left(1 + \frac{\pin q}{\pin}(\frac{\log(q)}{q-1}-1)- \frac{\log(\frac{1}{q})}{(1/q-1)}\right) = \frac{1-q}{\log(\frac{1}{q})},
\eearn
where (a) is due to the fact that  $\log(q) \leq q-1 \leq 0$, and $x \le \pin.$

Hence we conclude that 
\bear \label{conv_0_0}
\frac{1}{x} \Biggl[ \int_{0}^{x} u d\Psi(u) + \int_{x}^{\pin} u \bG_{0, \pin}(u|(x,1), (\pin, q))  d\Psi(u) \Biggl] \ge \frac{1-q}{\log(\frac{1}{q})}.
\eear

\subparagraph{Case 2: $x\mbox{ in }[\pin, \infty)$} Let us now analyze the second term, we have 
\bearn
&&  \frac{1}{x \bG_{0, x}(x|(0,1), (\pin, q))} \int_{0}^{x} u \bG_{0, x}(u|(0,1), (\pin, q))  d\Psi(u) \\ &=& \frac{1}{x \bG_{0, x}(x|(0,1), (\pin, q))} \Biggl[ \int_{0}^{\pin}  u \bG_{0, x}(u|(0,1), (\pin, q))  d\Psi(u) +   \int_{\pin}^{x}  u \bG_{0, x}(u|(0,1), (\pin, q))  d\Psi(u)\Biggl]\\
 &=&  \frac{1}{x \bG_{0, x}(x|(0,1), (\pin, q))\log(\frac{1}{q})}\int_{\pin q}^{\pin} \frac{1}{1 + (\frac{1}{q}-1)\frac{u}{\pin}} du = \frac{\Biggl[ \pin \log(1 + (\frac{1}{q}-1)\frac{u}{\pin})\Biggl]_{u = \pin q}^{u = \pin}}{x \bG_{0, x}(x|(0,1), (\pin, q))(\frac{1}{q}-1)\log(\frac{1}{q})} \\
 &=& \frac{\pin}{(\frac{1}{q}-1)\log(\frac{1}{q})}(\log(\frac{1}{q})- \log(2-q)) \frac{1}{x \bG_{0, x}(x|(0,1), (\pin, q))} \\
 &=& \frac{\pin}{(\frac{1}{q}-1)}(1- \frac{\log(2-q)}{\log(\frac{1}{q})}) \frac{1}{x \bG_{0, x}(x|(0,1), (\pin, q))} .
\eearn
The revenue function $x \to x \bG_{0, x}(x|(0,1), (\pin, q))$ is non-decreasing in $[0, +\infty)$, therefore
\bearn
x \bG_{0, x}(x|(0,1), (\pin, q)) \leq \lim_{x \to +\infty} x \bG_{0, x}(x|(0,1), (\pin, q)) = \frac{\pin}{(\frac{1}{q}-1)}.
\eearn
Hence 
\bear \label{conv_0_1}
 \frac{1}{x \bG_{0, x}(x|(0,1), (\pin, q))} \int_{0}^{x} u \bG_{0, x}(u|(0,1), (\pin, q))  d\Psi(u)
 & \ge &  (1- \frac{\log(2-q)}{\log(\frac{1}{q})}),
\eear

By combining \eqref{conv_0_0} and \eqref{conv_0_0}, we get that
\bearn
\inf _{F \in \aDistSetqa{q}} R(\Psi, F) \geq \min\left(\frac{1-q}{\log(\frac{1}{q})}, 1- \frac{\log(2-q)}{\log(\frac{1}{q})} \right).
\eearn

For $q \mbox{ in }[0, 1-\frac{1}{\sqrt{2}}]$, we have
\bearn
\frac{1-q}{\log(\frac{1}{q})} \geq \frac{\frac{1}{\sqrt{2}}}{\log(\frac{1}{q})} \geq \frac{\log(2)}{\log(\frac{1}{q})}
\eearn
and
\bearn
1- \frac{\log(2-q)}{\log(\frac{1}{q})} = \frac{\log(\frac{1}{q(2-q)})}{\log(\frac{1}{q})} = \frac{\log(\frac{1}{1-(1-q)^2})}{\log(\frac{1}{q})} \geq \frac{\log(\frac{1}{1-(\frac{1}{\sqrt{2}})^2})}{\log(\frac{1}{q})} = \frac{\log(2)}{\log(\frac{1}{q})}.
\eearn

Hence we we get that 
\bearn
\inf _{F \in \aDistSetqa{q}} R(\Psi, F) \geq \frac{\log(2)}{\log(\frac{1}{q})}.
\eearn
This conclude the lower bound.

\paragraph{Step 2: Upper bound} 

Let $q \mbox{ in }(0, 1)$ and $K \mbox{ in }\mathbb{N}^*$. Define $\varepsilon = q^{\frac{1}{K}} \mbox{ in } (q, 1)$ and $a_k = w \varepsilon^k \mbox{ in }[q\pin, \pin)$ for $k = 1 \cdots K$. Consider the family of distributions $F_{0}(\cdot| a_k, (\pin, q)) \mbox{ in }\Fb_0 (w,q)$.

 Using Yao's principle \citep{yaoprinciple}, we have
\bear \label{ineq:yaos}
\sup_{\Psi \in \MechSet} \inf _{F \in \Fb_0 (\pin,q)} R(\Psi, F) &\leq& \sup_{p \ge 0} \frac{1}{K}  \sum_{i=1}^K \frac{ p \bar{F}_{0}(p| a_i, (\pin, q))}{\opt(F_{0}(\cdot| a_i, (\pin, q)))} \nonumber \\
&=& \frac{1}{K}  \max_{1 \le k \le K} \sup_{p  \in [a_{k+1}, a_{k})}  \sum_{i=1}^K \frac{ p \bar{F}_{0}(p| a_i, (\pin, q))}{\opt(F_{0}(\cdot| a_i, (\pin, q)))}.
\eear

Now let's analyze the sup on each interval $[a_{k+1}, a_{k}).$ For all $1 \le k \le K$, the revenue curve associated with $\bar{F}_{0}(\cdot| a_i, (\pin, q))$ is monotone non-increasing on $[a_i,\pin)$ as the optimal reserve price of $F_{0}(\cdot| a_i, (\pin, q))$ is $a_i$.

Furthermore, the revenue curve is convex in $[0,\pin)$ as we have:
\bearn
(p \bar{F}_{0}(p| a_i, (\pin, q)))^{'} &=& \begin{cases}
            1 &\quad \text{if } p \leq a_i \\
            \frac{\pin (1-\frac{a_i}{\pin q})}{(w-a_i)(1+(\frac{1}{q}-1)\frac{p-a_i}{\pin-a_i})^2} &\quad \text{if } p < \pin \\
            0 &\quad \text{if } p \geq \pin.
\end{cases} 
\eearn
Therefore the derivative of the revenue function $p \bar{F}_{0}(p| a_i, (\pin, q))$ is non-decreasing because $a_i \geq wq$ and $p \mapsto \frac{1}{(\pin-a_i)(1+(\frac{1}{q}-1)\frac{p-a_i}{\pin-a_i})^2}$ is non-increasing. Hence the function $$p \mapsto \sum_{i=1}^K \frac{p \bar{F}_{0}(p| a_i, (\pin, q))}{\opt(F_{0}(\cdot| a_i, (\pin, q)))}$$ is convex on $[a_{k+1}, a_{k}).$ Thus, the sup  on an interval must be attained at one of the extreme points of the interval.

Therefore by \eqref{ineq:yaos}, we have that
\bear  \label{ineq:yaos_simplified}
\sup_{\Psi \in \MechSet} \inf _{F \in \Fb_0 (w,q)} R(\Psi, F) \leq \sup_{p \ge 0} G(p) &=& \frac{1}{K}  \max_{1 \le k \le K}  \sum_{i=1}^K \frac{ a_k \bar{F}_{0}(a_k| a_i, (\pin, q))}{\opt(F_{0}(\cdot| a_i, (\pin, q)))}.
\eear

Now let us analyze the elementary term, $\frac{ a_k \bar{F}_{0}(a_k| a_i, (\pin, q))}{\opt(F_{0}(\cdot| a_i, (\pin, q)))}$ for any $i,k$. Note that $\opt(F_{0}(\cdot| a_i, (\pin, q)) = a_i$ by \Cref{lem:opt_feas}.

There are two cases of interest  either $ i \le k$ or $i >k$, let us analyze each case separately. 

\subparagraph{Case 1, $ i \le k$:} We have $\bar{F}_{0}(a_k| a_i, (\pin, q)) = 1,$ thus
\bearn
\frac{a_k \bar{F}_{0, a_i, \pin, q}(a_k)}{\opt(F_{0, a_i, \pin, q})} = \frac{a_k}{a_i} = \varepsilon^{k-i},
\eearn
which implies that
\bearn
\sum_{i=1}^k \frac{ a_k \bar{F}_{0}(a_k| a_i, (\pin, q))}{\opt(F_{0}(\cdot| a_i, (\pin, q)))} = \sum_{i=1}^k \varepsilon^{k-i} =  \sum_{i=0}^{k-1}  \varepsilon^{i} = \frac{1-\varepsilon^k}{1-\varepsilon} \le \frac{1}{1-\varepsilon} .
\eearn

Hence we conclude that 
\bear \label{ineq:conv_0_1}
\sum_{i=1}^k \frac{ a_k \bar{F}_{0}(a_k| a_i, (\pin, q))}{\opt(F_{0}(\cdot| a_i, (\pin, q)))}  \le \frac{1}{1-\varepsilon} .
\eear

\subparagraph{Case 2, $ i \ge k$:} We have that 
\bearn
\frac{ a_k \bar{F}_{0}(a_k| a_i, (\pin, q))}{\opt(F_{0}(\cdot| a_i, (\pin, q)))} = \frac{ a_k }{a_i}  \frac{1}{1+(\frac{1}{q}-1)\frac{a_k-a_i}{\pin-a_i}} &=&  \varepsilon^{k-i} \frac{1-\varepsilon^{i}}{1- \varepsilon^{i}+\frac{1}{q}(\varepsilon^{k}-\varepsilon^{i})-\varepsilon^{k}+\varepsilon^{i}}\\
&=&  \varepsilon^{k-i} \frac{1-\varepsilon^{i}}{1-\varepsilon^{k}+\frac{1}{q}(\varepsilon^{k}-\varepsilon^{i})}\\
&=& \frac{\varepsilon^{-i}-1}{(\varepsilon^{-k}-1)+\frac{1}{q}(1-\varepsilon^{i-k})} \\
&\leq&  \frac{\varepsilon^{-i}-1}{(\varepsilon^{-k}-1)+\frac{1}{q}(1-\varepsilon)},
\eearn
where in the last inequality we used the fact that $\varepsilon \le 1.$

Hence we conclude that 
\bearn
\sum_{i=k+1}^{K} \frac{ a_k \bar{F}_{0}(a_k| a_i, (\pin, q))}{\opt(F_{0}(\cdot| a_i, (\pin, q)))} &\leq& \sum_{i=k+1}^{K} \frac{\varepsilon^{-i}-1}{(\varepsilon^{-k}-1)+\frac{1}{q}(1-\varepsilon)}\\
&\stackrel{(a)}{\le} & \frac{1}{\frac{1}{\varepsilon}-1}\frac{\frac{1}{\varepsilon^{K+1}}-\frac{1}{\varepsilon^{k+1}}}{(\varepsilon^{-k}-1)+\frac{1}{q}(1-\varepsilon)} \\
&\leq& \frac{1}{1-\varepsilon}   \frac{\frac{1}{\varepsilon^{K}}-\frac{1}{\varepsilon^{k}}}{(\varepsilon^{-k}-1)+\frac{1}{q}(1-\varepsilon)}  \\
&\stackrel{(b)}{\le}& \frac{1}{1-\varepsilon}  \frac{\frac{1}{q}}{(\varepsilon^{-k}-1)+\frac{1}{q}(1-\varepsilon)},
\eearn
where in (a) we used $\varepsilon^{-i}-1 \leq \varepsilon^{-i}$ and (b) we used the fact that $\varepsilon \ge 0$ and that $\varepsilon^K = q.$ From the last inequality we conclude that
\bear \label{ineq:conv_0_2}
\sum_{i=k+1}^{K} \frac{ a_k \bar{F}_{0}(a_k| a_i, (\pin, q))}{\opt(F_{0}(\cdot| a_i, (\pin, q)))} 
&\leq& \frac{1}{1-\varepsilon}  \frac{1}{q(\varepsilon^{-k}-1)+(1-\varepsilon)} \le  \frac{1}{(1-\varepsilon)^2} .
\eear

By combining the last two cases, in particular \eqref{ineq:yaos_simplified}, \eqref{ineq:conv_0_1} and \eqref{ineq:conv_0_2}, we get that
\bearn
\sup_{\Psi \in \MechSet} \inf _{F \in \Fb_0 (w,q)} R(\Psi, F) \leq \frac{1}{K} \left(\frac{1}{1-\varepsilon}+ \frac{1}{(1-\varepsilon)^2} \right).
\eearn

By choosing $K = \log(1/q)$, thus $\varepsilon = e^{-1}$, we get:
\bearn
\sup_{\Psi \in \MechSet} \inf _{F \in \Fb_0 (w,q)} R(\Psi, F) \leq \frac{c_2}{\log(1/q)} \mbox{ with } c_2 =  \frac{1}{1-e^{-1}} \left( 1 +  \frac{1}{1-e^{-1}} \right)
\eearn

This concludes the proof.
\end{proof}

\begin{proof}[\textbf{Proof of \Cref{lem:qonebounds}.}]

This proof is divided into two steps. In the first step, we will show the lower bound by analyzing the performance of a specific mechanism. Then in a second step, we will derive the upper through the analysis of a family of hard cases when $q$ is close to $1$.

Throughout the proof we will assume that $q \ge 3/4$ since we are interested in the limit when  when $q$ is close to $1$.

\paragraph{Step 1: Lower bound}
Let  us define the following measure parameterized by $a, b \geq 0$:
\bearn
d \Psi(u) = \begin{cases}
a &\quad \mbox{if } u = \pin\\
b \frac{(u \bG_{0, u}(u|(0,1), (\pin, q)))'}{u \bG_{0, u}(u|(0,1), (\pin, q)) } &\quad \mbox{if } u > \pin.
\end{cases}
\eearn

Note that $d \Psi(u) \ge 0$  since  the revenue function $u \to u \bG_{0, u}(u|(0,1), (\pin, q))  d\Psi(u)$ is increasing in $[\pin, \infty).$ Let us determine the condition on  the parameters $a$ and $b$ so that $\Psi $ is a distribution. For that we need the following
\bearn
\int_{0}^{\infty} d\Psi(u) = 1,
\eearn
which implies that 
\bearn
a + b \log\left(\frac{\lim_{u \to \infty} u \bG_{0, u}(u|(0,1), (\pin, q))  }{\pin \bG_{0, \pin}(\pin |(0,1), (\pin, q))  }\right) = 1.
\eearn
Since $\bG_{0, \pin}(\pin |(0,1), (\pin, q))  d\Psi(u) = q $ and $\lim_{u \to \infty} u \bG_{0, u}(u|(0,1), (\pin, q)) = \frac{\pin}{\frac{1}{q}-1}$, we get that 
\bearn
a + b \log\left(\frac{1}{1-q}\right) = 1.
\eearn
Hence the relation between $a$ and $b$ is as follows
\bearn
 b  = \frac{1-a}{\log\left(\frac{1}{1-q}\right)} \quad \mbox{ and } a \mbox{ in }[0,1].
\eearn

Using \Cref{thm:LB_1D}, we have
\bear
\inf _{F \in \aDistSetqa{q}} R(\Psi, F) &\geq& \inf _{F \in \Fb_0(w,q)} R(\Psi, F) \nonumber \\ &=& \min \Biggl\{ \inf_{x \in [\pin q, \pin)}  \frac{1}{x} \Biggl[ \int_{0}^{x} u d\Psi(u) + \int_{x}^{\pin} u \bG_{0, \pin}(u|(x,1), (\pin, q))  d\Psi(u) \Biggl], \nonumber \\
&& \inf_{x \in [\pin, \infty)} \frac{1}{x \bG_{0, x}(x|(0,1), (\pin, q))} \int_{0}^{x} u \bG_{0, x}(u|(0,1), (\pin, q))  d\Psi(u) \Biggr\}.  \label{conv_1_1_main}
 \eear

We will analyze each term separately depending if $x\mbox{ in } [\pin q, \pin)$ or $x\mbox{ in }[\pin, \infty).$

\subparagraph{Case 1: $x\mbox{ in } [\pin q, \pin)$} We have
\bearn
\frac{1}{x} \Biggl[ \int_{0}^{x} u d\Psi(u) + \int_{x}^{\pin} u \bG_{0, \pin}(u|(x,1), (\pin, q))  d\Psi(u) \Biggl] \stackrel{(a)}{=} \frac{aq}{x} \ge aq,
\eearn
where the last inequality is due to the fact that $x \le \pin$ and $(a)$ is due to $d\Psi(u) = 0$ for $u<\pin$, $d\Psi(\pin) = a$ and $\bG_{0, \pin}(\pin|(x,1), (\pin, q))= q$. 

Hence we conclude that 
\bear \label{ineq:con_1_0}
\inf_{x \in [\pin q, \pin)}  \frac{1}{x} \Biggl[ \int_{0}^{x} u d\Psi(u) + \int_{x}^{\pin} u \bG_{0, \pin}(u|(x,1), (\pin, q))  d\Psi(u) \Biggl] \ge aq.
\eear

\subparagraph{Case 2: $x\mbox{ in }[\pin, \infty)$} Let us now analyze the second term, we have 
\bearn
  \int_{0}^{x} u \bG_{0, x}(u|(0,1), (\pin, q))  d\Psi(u)  &=&  \int_{0}^{\pin}  u \bG_{0, x}(u|(0,1), (\pin, q))  d\Psi(u) +   \int_{\pin}^{x}  u \bG_{0, x}(u|(0,1), (\pin, q))  d\Psi(u)\\
 &\stackrel{(a)}{=}& aq + b \int_{\pin}^{x}  u \bG_{0, x}(u|(0,1), (\pin, q)) \frac{(u \bG_{0, u}(u|(0,1), (\pin, q)))'}{u \bG_{0, u}(u|(0,1), (\pin, q))} du\\
 &=& a\pin q + b  \left[u \bG_{0, u}(u|(0,1), (\pin, q))\right]_{\pin}^{x} \\
 & \ge & a\pin q + b \left(x \bG_{0, x}(x|(0,1), (\pin, q)) - \pin q \right),
\eearn
$(a)$ is due to $d\Psi(u) = 0$ for $u<\pin$, $d\Psi(\pin) = a$ and $\pin \bG_{0, \pin}(\pin|(0,1), (\pin, q)) = \pin q$.  Hence we conclude that 
\bearn
\frac{1}{x \bG_{0, x}(x|(0,1), (\pin, q))} \int_{0}^{x} u \bG_{0, x}(u|(0,1), (\pin, q))  d\Psi(u)
 & \ge &  \pin q\frac{a-b}{x \bG_{0, x}(x|(0,1), (\pin, q))}+ b\\
 &\ge& (a-b)(1-q) + b,
\eearn
where the last inequality we used the fact that $x \to x \bG_{0, x}(x|(0,1), (\pin, q))$ is non-decreasing in $[\pin, \infty)$ and that $\lim_{x \to \infty} x \bG_{0, x}(x|(0,1), (\pin, q)) = \frac{\pin q}{1-q}.$

Thus we conclude that 
\bear \label{conv_1_1}
  \inf_{x \mbox{ in }[\pin, +\infty)} \frac{1}{x \bG_{0, x}(x|(0,1), (\pin, q))} \int_{0}^{x} u \bG_{0, x}(u|(0,1), (\pin, q))  d\Psi(u)
 &\ge& (a-b)(1-q) + b.
\eear

By combining (\ref{conv_1_1_main}), \eqref{ineq:con_1_0} and \eqref{conv_1_1} we get that
\bearn
\inf _{F \in \aDistSetqa{q}} R(\Psi, F) \geq \min\{aq, (a-b)(1-q) + b\}.
\eearn

Now let us set \bearn
a = \frac{q}{2(q-\frac{1}{2})\log\left(\frac{1}{1-q}\right) +q}. 
\eearn

Note that $a\mbox{ in }[0,1]$ as $q \mbox{ in }[3/4, 1],$ this also leads to the fact that 
\bearn
(a-b)(1-q) + b = aq = \frac{q^2}{2(q-\frac{1}{2})\log\left(\frac{1}{1-q}\right) +q}\stackrel{(a)}{
\ge } \frac{9}{16} \frac{1}{2(1-\frac{1}{2})\log\left(\frac{1}{1-q}\right) +1} &=& \frac{9}{16( \log\left(\frac{1}{1-q}\right) + 1) }\\ &\stackrel{(b)}{\ge}& \frac{9}{32} \frac{1}{\log\left(\frac{1}{1-q}\right)} ,
\eearn
where  inequality (a) stems from the fact $q \mbox{ in }[3/4, 1],$ and in (b) we have used $\log\left(\frac{1}{1-q}\right) \ge 1.$  
 
 Hence we we get that 
 \bearn
\inf _{F \in \aDistSetqa{q}} R(\Psi, F) \geq \frac{9}{32} \frac{1}{\log\left(\frac{1}{1-q}\right)}.
\eearn
This conclude the lower bound.

\paragraph{Step 2: Upper bound}
To show the upper bound we will introduce a family of ``hard'' cases.  We consider the family of distributions $(F_{0}(\cdot| r, (\pin, q)) \mbox{ in }\Fb_0 (w,q))_{r\geq \pin}$ and the following weight distribution:

\bearn
d\lambda(r) &=& \begin{cases}
            0 &\quad \text{if } r < \pin \\
           \frac{1}{\log\left(\frac{1}{1-q}\right)}\frac{(r \bG_{0, r}(r|(0,1), (\pin, q)))'}{r \bG_{0, r}(r|(0,1), (\pin, q)) } &\quad \text{if } r \geq \pin 
\end{cases} 
\eearn

One can verify that  $\int_0^{\infty} d\lambda(r) =1$ and that  $ d\lambda(r) \ge 0.$ Let us define
\bearn
G(p) &=& \int_{0}^{\infty} \frac{ p \bF_{0}(p| r, (\pin, q))}{\opt(F_{0}(\cdot| r, (\pin, q)))} d\lambda(r).
\eearn

Using Yao's principle \citep{yaoprinciple}, we have
\bear \label{ineq:yao_1}
\sup_{\Psi \in \MechSet} \inf _{F \in \Fb_0 (w,q)} R(\Psi, F) &\leq& \sup_{p \ge 0} G(p)
= \sup_{p \ge 0} \int_{0}^{\infty} \frac{ p \bF_{0}(p| r, (\pin, q))}{\opt(F_{0}(\cdot| r, (\pin, q)))} d\lambda(r).
\eear
Note that
\bearn
\sup_{p \ge 0} \int_{0}^{\infty} \frac{ p \bF_{0}(p| r, (\pin, q))}{\opt(F_{0}(\cdot| r, (\pin, q)))} d\lambda(r) &=& \sup_{p \ge 0} \int_{\pin}^{\infty} \frac{ p \bF_{0}(p| r, (\pin, q))}{\opt(F_{0}(\cdot| r, (\pin, q)))} d\lambda(r) \\
&=& \sup_{p \ge \pin} \int_{\pin}^{\infty} \frac{ p \bF_{0}(p| r, (\pin, q))}{\opt(F_{0}(\cdot| r, (\pin, q)))} d\lambda(r),
\eearn
where the last equality follows from the fact that $p \mapsto p \bF_{0}(p| r, (\pin, q))$ is increasing on $[0,\pin]$ for any $r \ge \pin$. 

Fix $p \ge \pin$ and let us analyze the integral term. We have 
\bearn
& &\int_{\pin}^{\infty} \frac{ p \bF_{0}(p| r, (\pin, q))}{\opt(F_{0}(\cdot| r, (\pin, q)))} d\lambda(r)\\ &=& \int_{p}^{\infty} \frac{ p \bF_{0}(p| r, (\pin, q))}{r \bG_{0, r}(r|(0,1), (\pin, q))} \frac{1}{\log\left(\frac{1}{1-q}\right)}\frac{(r \bG_{0, r}(r|(0,1), (\pin, q)))'}{r \bG_{0, r}(r|(0,1), (\pin, q)) } dr \\
&=& \frac{1}{\log\left(\frac{1}{1-q}\right)}   p \bG_{0, p}(p|(0,1), (\pin, q)) \int_{p}^{\infty} \frac{(r \bG_{0, r}(r|(0,1), (\pin, q)))'}{(r \bG_{0, r}(r|(0,1), (\pin, q)))^2} dr\\
&=& \frac{1}{\log\left(\frac{1}{1-q}\right)}   p \bG_{0, p}(p|(0,1), (\pin, q)) \left(\frac{1}{p \bG_{0, p}(p|(0,1), (\pin, q))}  - \lim_{r \to \infty} \frac{1}{r \bG_{0, r}(r|(0,1), (\pin, q))} \right) \\
&=& \frac{1}{\log\left(\frac{1}{1-q}\right)} \left( 1- \left(\frac{1}{q}-1 \right) \frac{p \bG_{0, p}(p|(0,1), (\pin, q))}{\pin} \right) \leq \frac{1}{\log\left(\frac{1}{1-q}\right)}.
\eearn
By using the last inequality, together with \eqref{ineq:yao_1}, we obtain the result.
\end{proof}
\endproof

\setcounter{equation}{0}
\setcounter{theorem}{0}
\setcounter{lemma}{0}
\setcounter{proposition}{0}

\setcounter{equation}{0}
\setcounter{proposition}{0}
\setcounter{lemma}{0}

\section{Proofs and auxiliary results for \Cref{sec:I}}

\begin{proof}[\underline{\textbf{Proof of \Cref{thm:LB_LP_robust}}}] 
We aim to show that one can approximate the value of the maximin ratio via  lower and upper bounds and we quantify the asymptotic error of this approximation as a function of the grid size $N>0$. 

We will do that in different steps:
\begin{itemize}
\item In a first step, we extend previous results to the interval uncertainty case:
    \begin{itemize}
        \item We will first show in \Cref{thm:LB_1D_uncertainty_1} that in  the interval uncertainty case,  we reduce the family of worst case distributions by generalizing \Cref{thm:LB_1D}.
    \item Under such a reduction, we then show in \Cref{thm:mechanismreduction_uncertainty_1} that we can still approximate the performance of any mechanism by its discrete version by generalizing \Cref{thm:mechanismreduction}.  
    \end{itemize}
    
    \item In a second step, we derive lower  bounds on the maximin ratio in the form of linear programs.
    
    \item In a third step, we show that through an appropriate choice of the support of a discrete mechanism, one can approximate the maximin ratio arbitrarily closely through the lower bound. 
        

\end{itemize}

\textbf{Step 1.} We first reduce the possible set of worst-cases to consider by  extending \Cref{thm:LB_1D}.  For that, let us  define the following subset of distributions
\bear \label{eq:subset-u}
\aDistSetqprime{\ql, \qh} &=& \left\{ F_{\alpha}( \cdot| r, (\pin, \ql)): r \mbox{ in }\left [\rl[\ql], \pin \right) \right\} \cup \left\{ F_{\alpha}( \cdot| r, (\pin, \qh)): r \mbox{ in }\left [\pin, \rh[\qh] \right] \right\}. 
\eear
where we use the convention that whenever $\rh[\qh]<\pin$, $[\pin,\rh[\qh] ]:= \emptyset$. We have the following result, whose proof is deferred to  \Cref{app:aux-uncer}.

\begin{proposition}
\label{thm:LB_1D_uncertainty_1} 
For any $\ql, \qh \mbox{ in } (0, 1)^2$ such that $\ql \le \qh$, and for any subset of mechanisms $\MechSet' \subseteq \MechSet$, 
\bearn
\Rb(\MechSet', \aDistSet{[\ql, \qh]}) = \Rb(\MechSet', \aDistSetqprime{\ql, \qh}).
\eearn
\end{proposition}

In addition, the next proposition generalizes \Cref{thm:mechanismreduction}, and its proof is deferred to  \Cref{app:aux-uncer}.

\begin{proposition}\label{thm:mechanismreduction_uncertainty_1} 
Let $\ql, \qh \mbox{ in } (0, 1)^2$ such that $\ql \le \qh$. Fix a  mechanism $\Psi \mbox{ in }\MechSet$,  $N > 1$, and any finite sequence of increasing reals $\IN= \{a_i\}_{i=0}^{N}$ such that $a_0 = \rl[\ql],  a_N \geq \pin$. Then there exists $\Psi_{\IN} \mbox{ in }\MechSet_{\IN}$ such that 
    \bearn
    \inf_{F \in \aDistSet{[\ql, \qh]}} R(\Psi_{\IN}, F) \geq \inf_{F \in \aDistSet{[\ql, \qh]}} R(\Psi, F) - \frac{\Delta(\IN)}{\rl[\ql]}  -  \frac{\mathbf{1}\{a_N < \rh[\qh]\}}{\qh(1+(\qh^{-1}-1)a_N/\pin)} ,
    \eearn    
    where $\Delta(\IN) = \sup_i \{a_i - a_{i-1}\}$.
\end{proposition}

\textbf{Step 2.}  Fix an arbitrary sequence of increasing reals $\IN= \{a_i\}_{i=0}^{2N+1}$ such that $a_0 = \rl[\ql]$, $a_{N+1} = \pin$ and $a_{2N+1} \le \rh[\qh]$. Set $a_{2N+2}:=\rh[\qh]$. Note that $\rh[\qh] =\infty$ when $\alpha =0$. With some abuse of notation, we will use intervals that include  $\rh[\qh]$. These should be interpreted as open when $\alpha =0$. 

 We next develop a lower bound on the maximin ratio $\Rb (\MechSet_{\IN}, \aDistSet{[\ql, \qh]})$ in the form of a linear program. Fix a mechanism $\Psi \mbox{ in }\MechSet_{\IN}$ and denote by $p_0,...,p_{2N+1}$ the corresponding probabilities. We set $p_{2N+2}:=0$.  Using \cref{thm:LB_1D_uncertainty_1}. Then we have: 
\bearn
&& \inf _{F \in \aDistSet{[\ql, \qh]}} R(\Psi, F)  \\
&=& \min \Biggl\{ \inf_{x \in [\rl[\ql], \pin)}  \frac{1}{\opt(F_{\alpha}( \cdot | x, (\pin, q_l)))}  \int_{0}^{\infty}    u \bF_{\alpha}( u | x, (\pin, q_l)) d\Psi(u), \\
&&\inf_{x \in [\pin, \rh[\qh]]} \frac{1}{\opt(F_{\alpha}( \cdot | x, (\pin, q_h)))} \int_{0}^{\infty}  u \bF_{\alpha}( u | x, (\pin, q_h)))  d\Psi(u) \Biggr\} \\
  &=& \min \Biggl\{ \min_{i = 0, \cdots, N}\inf_{x \in [a_i, a_{i+1})}  \frac{1}{\opt(F_{\alpha}( \cdot | x, (\pin, q_l)))}  \int_{0}^{\infty}    u \bF_{\alpha}( u | x, (\pin, q_l)) d\Psi(u), \\
 && \min_{i = N+1, \cdots, 2N+1}\inf_{x \in [a_i, a_{i+1}]} \frac{1}{\opt(F_{\alpha}( \cdot | x, (\pin, q_h)))} \int_{0}^{\infty}  u \bF_{\alpha}( u | x, (\pin, q_h)))  d\Psi(u)\Biggr\}.
\eearn
Note that, for $x \in [\rl[\ql], \pin)$,  $\bF_{\alpha}( \cdot | x, (\pin, q_l)) $ is non-decreasing in $x$ and that the revenue function $u\mapsto u \bF_{\alpha}( \cdot | x, (\pin, q_l))$ is increasing in $u$ on $[0,x)$ and decreasing on $(x,\pin)$. In addition, note that, for $x \in [\pin, \rh[\qh]]$, the revenue function  $u \mapsto u \bF_{\alpha}( u | x, (\pin, q_h))$ is non-decreasing on $[0,x]$. We let $\opt(F_{\alpha}( \cdot | a_{i+1}^{-}, (\pin, q_l))) = \lim_{x \to a_{i+1}^-} \opt(F_{\alpha}( \cdot | x, (\pin, q_l)))$ for any $i = 0, \cdots, N$. 
 Hence, we have
\bear
 \inf _{F \in \aDistSet{[\ql, \qh]}} R(\Psi, F) 
  &\ge&  \min \Biggl\{ \min_{i = 0, \cdots, N}  \frac{1}{\opt(F_{\alpha}( \cdot | a_{i+1}^{-}, (\pin, q_l)) )} \int_{0 }^{\infty}    u \bF_{\alpha}( u | a_i, (\pin, q_l)) d\Psi(u)  ,\nonumber \\
 && \min_{i = N+1, \cdots, 2N+1} \frac{1}{\opt(F_{\alpha}( \cdot | a_{i+1}^{-}, (\pin, q_h)))} \int_{0 }^{\infty}  u \bF_{\alpha}( u | a_i, (\pin, q_h))  d\Psi(u) \Biggr\} \nonumber\\
&=& \min \Biggl\{ \min_{i = 0, \cdots, N}  \frac{1}{\opt(F_{\alpha}( \cdot | a_{i+1}^{-}, (\pin, q_l)) )}   \sum_{j=0}^{2N+1} a_j \bF_{\alpha}( a_j | a_{i}, (\pin, q_l))  p_j , \nonumber\\
 && \min_{i = N+1, \cdots, 2N+1}\frac{1}{\opt(F_{\alpha}( \cdot | a_{i+1}, (\pin, q_h)))} \sum_{j=0}^{2N+1}   a_j \bF_{\alpha}( a_j | a_{i}, (\pin, q_h)) p_j \Biggr\}, \label{eq:ucp}
\eear
where the equality simply stems from the fact that $\Psi \mbox{ in }\MechSet_{\IN}$.
The problem of maximizing over mechanisms in $\MechSet_{\IN}$ is  clearly lower bounded by the problem of maximizing the RHS above over $p_0,...,p_{2N+1}$. The latter problem admits exactly \ref{eq:small-LP-2} as its epigraph formulation, and hence we have 
\bearn
\Rb (\MechSet_{\IN}, \aDistSet{[\ql, \qh]}) &\geq&  
\underline{\Lb}_{\alpha, \ql, \qh, \IN}.
\eearn

\textbf{Step 3.}  We next establish that with a proper choice of sequence $\IN$, $\underline{\Lb}_{\alpha, \ql, \qh, \IN}$ may be arbitrarily close to the maximin ratio $\Rb (\MechSet, \aDistSet{[\ql, \qh]})$. To do so, we will first develop an upper bound on $\Rb (\MechSet_{\IN}, \aDistSet{[\ql, \qh]})$. Then, we will construct a particular sequence $\IN$ and establish for this sequence, the gap between   $\Rb (\MechSet_{\IN}, \aDistSet{[\ql, \qh]})$ and $\underline{\Lb}_{\alpha, \ql, \qh, \IN}$ is small and that the gap between $\Rb (\MechSet_{\IN}, \aDistSet{[\ql, \qh]})$ and $\Rb (\MechSet, \aDistSet{[\ql, \qh]})$ is also small. This will yield the result. 

Suppose that $a_0>0$. Following the same reasoning as in step 2 above, we may also obtain an upper bound on $\inf _{F \in\aDistSet{[\ql, \qh]}} R(\Psi, F)$. Indeed, we have
 \bearn
 \inf _{F \in \aDistSet{[\ql, \qh]}} R(\Psi, F) 
  &\le& \min \Biggl\{ \min_{i = 0, \cdots, N}  \frac{1}{\opt(F_{\alpha}( \cdot | a_{i}, (\pin, q_l)) )} \int_{0 }^{\infty}    u \bF_{\alpha}( u | a_{i+1}^{-}, (\pin, q_l)) d\Psi(u)  , \\
 && \min_{i = N+1, \cdots, 2N+1} \frac{1}{\opt(F_{\alpha}( \cdot | a_{i}, (\pin, q_h)))} \int_{0 }^{\infty}  u \bF_{\alpha}( u | a_{i+1}^{-}, (\pin, q_h))  d\Psi(u) \Biggr\}\\
&=& \min \Biggl\{ \min_{i = 0, \cdots, N}  \frac{1}{\opt(F_{\alpha}( \cdot | a_{i}, (\pin, q_l)) )}   \sum_{j=0}^{2N+1} a_j \bF_{\alpha}( a_j | a_{i+1}^{-}, (\pin, q_l))  p_j , \\
 && \min_{i = N+1, \cdots, 2N+1}\frac{1}{\opt(F_{\alpha}( \cdot | a_{i}, (\pin, q_h)))} \sum_{j=0}^{2N+1}   a_j \bF_{\alpha}( a_j | a_{i+1}^{-}, (\pin, q_h)) p_j \Biggr\}.
\eearn

With $u \bF_{\alpha}( u | a_{i+1}^{-}, (\pin, q_l)) = \lim_{x \to a_{i+1}^-} u \bF_{\alpha}( u | x, (\pin, q_l))$ for any $u \geq 0$ and $i = 0, \cdots, 2N+1$.
The problem of maximizing over mechanisms in $\MechSet_{\IN}$ is  clearly upper bounded by the problem of maximizing the RHS above over $p_0,...,p_{2N+1}$. The epigraph formulation of the latter problem can be written as 

\begin{align}
\overline{\Lb}_{\alpha, \ql, \qh, \IN} \:\:=\:\:  \max_{ \mathbf{p}, c} &  ~~c  \label{eq:small-LP-UP-1}\tag{\PFLPIU}\\
s.t.    ~~   & \frac{1}{\opt(F_{\alpha}( \cdot | a_{i}, (\pin, q_l)))} \sum_{j=0}^{2N+1}   a_j \bF_{\alpha}( a_j | a_{i+1}^-, (\pin, q_l)) p_j \ge c \quad i=0,...N, \nonumber\\
& \frac{1}{\opt(F_{\alpha}( \cdot | a_{i}, (\pin, q_h)))} \sum_{j=0}^{2N+1}   a_j \bF_{\alpha}( a_j | a_{i+1}^-, (\pin, q_h)) p_j \ge c \quad i=N+1,...2N+1, \nonumber\\
& \sum_{j=0}^{2N+1} p_j \le 1, \quad p_i \ge 0 \quad i=0,...2N+1.\nonumber
\end{align}

Therefore, we have
\bearn
\Rb (\MechSet_{\IN}, \aDistSet{[\ql, \qh]}) 
& \leq& \overline{\Lb}_{\alpha, \ql, \qh, \IN}.
\eearn
Hence, we have established the following.
\bearn
\underline{\Lb}_{\alpha, \ql, \qh, \IN} \leq \Rb (\MechSet_{\IN}, \aDistSet{[\ql, \qh]}) \leq \overline{\Lb}_{\alpha, \ql, \qh, \IN}.
\eearn

 We next quantify the gap $\overline{\Lb}_{\alpha, \ql, \qh, \IN} - \underline{\Lb}_{\alpha, \ql, \qh, \IN}$ as a function the discretization grid size $N$ for a particular sequence. For $N > 1$, $b = \rh[\qh]$ if $\alpha \in (0,1]$, $b > \pin$ if $\alpha = 0$ and $\eta \mbox{ in }(0, \rl[\ql])$, consider the following finite sequence of prices  $\IN= \{a_i\}_{i=0}^{2N+1}$ in $ [\rl[\ql] , \min\{b, \rh[\qh]\}]$:
 \bearn
 a_i &=& \begin{cases}
           \rl[\ql] + \frac{i}{N} \left(  (\pin-\eta)-\rl[\ql] \right) &\quad \mbox{if } 0 \leq i \leq N,\\
		    \pin + \frac{i-(N+1)}{N} \left(\min\{b, \rh[\qh]\}-\pin \right) &\quad \mbox{if } N+1 \leq i \leq 2N+1.\\
		\end{cases}
 \eearn

When fixing the probability weights $\textbf{p}$, let $\underline{c}(\textbf{p})$ denote the maximum value achievable (as a function of $c$) in the inner problem in \eqref{eq:small-LP-2}. In particular, it can be expressed as the minimum in \eqref{eq:ucp}.

Let $\textbf{p}$ correspond be a probability weight vector corresponding to an optimal  solution to the upper bound linear program \eqref{eq:small-LP-UP-1}. 
 We have
\bearn
\overline{\Lb}_{\alpha, \ql, \qh, \IN} - \underline{\Lb}_{\alpha, \ql, \qh, \IN} 
\:\le\: \overline{\Lb}_{\alpha, \ql, \qh, \IN} - \underline{c}(\textbf{p}).
\eearn
We next analyze upper bound the gap $ \overline{\Lb}_{\alpha, \ql, \qh, \IN} - \underline{c}(\textbf{p})$ as a function of the constraints that lead to the minimum value  when solving $\underline{c}(\textbf{p})$.

\textbf{Case 1:} If $\underline{c}(\textbf{p}) =  \frac{1}{\opt(F_{\alpha}( \cdot | a_{i+1}^{-}, (\pin, q_l)))} \sum_{j=0}^{2N+1}  a_j \bF_{\alpha}( a_j | a_{i}, (\pin, q_l)) p_j$ for some $0 \leq i \leq N-1$. Then we have
\bearn
\overline{\Lb}_{\alpha, \ql, \qh, \IN} - \underline{c}(\textbf{p})  &=&\overline{\Lb}_{\alpha, \ql, \qh, \IN} - \frac{1}{\opt(F_{\alpha}( \cdot | a_{i+1}^{-}, (\pin, q_l)))} \sum_{j=0}^{2N+1}   a_j \bF_{\alpha}( a_j | a_{i}, (\pin, q_l)) p_j \\
&\le& \frac{1}{\opt(F_{\alpha}( \cdot | a_{i}, (\pin, q_l)))} \sum_{j=0}^{2N+1}   a_j \bF_{\alpha}( a_j | a_{i+1}^-, (\pin, q_l)) p_j \\
&& - \frac{1}{\opt(F_{\alpha}( \cdot | a_{i+1}^{-}, (\pin, q_l)))} \sum_{j=0}^{2N+1}   a_j \bF_{\alpha}( a_j | a_{i}, (\pin, q_l)) p_j \\
&=& \left(\frac{1}{\opt(F_{\alpha}( \cdot | a_{i}, (\pin, q_l)) )}-\frac{1}{\opt(F_{\alpha}( \cdot | a_{i+1}^-, (\pin, q_l)) )}\right) \sum_{j=0}^{2N}  a_{j} \bF_{\alpha}( a_j | a_{i+1}^-, (\pin, q_l)) p_j  \\
&& + \frac{1}{\opt(F_{\alpha}( \cdot | a_{i+1}^-, (\pin, q_l)) )} \sum_{j=0}^{2N+1}  a_{j} \left[ \bF_{\alpha}( a_j | a_{i+1}^-, (\pin, q_l)) - \bF_{\alpha}( a_j | a_{i}, (\pin, q_l))\right] p_j\\
&=& \left(\frac{a_{i+1}-a_{i}}{a_{i}}\right) \sum_{j=0}^{2N+1}  \frac{a_{j} \bF_{\alpha}( a_j | a_{i+1}, (\pin, q_l))}{{\opt(F_{\alpha}( \cdot | a_{i+1}, (\pin, q_l)) )}} p_j  \\
&& + \frac{1}{\opt(F_{\alpha}( \cdot | a_{i+1}, (\pin, q_l)) )} \sum_{j=i+1}^{N+1}  a_{j} \left[ \bF_{\alpha}( a_j | a_{i+1}, (\pin, q_l)) - \bF_{\alpha}( a_j | a_{i}, (\pin, q_l))\right] p_j,
\eearn
where in the last equality, we have used that $a_{i+1}= \opt(F_{\alpha}(\cdot| a_{i+1}^-, (\pin ,\ql)))$ for $0 \leq i \leq N-1$ (cf. \Cref{lem:rr}), $\bF_{\alpha}( \cdot | a_{i+1}^-, (\pin, q_l)) = \bF_{\alpha}( \cdot | a_{i+1}, (\pin, q_l))$ for $0 \leq i \leq N-1$ and the fact that $\bF_{\alpha}( \cdot | a_{i+1}, (\pin, q_l)) = \bF_{\alpha}( \cdot | a_{i}, (\pin, q_l))$ on $[0,a_i]$ and on $(\pin,+\infty)$.
 We analyze the two terms on the RHS above separately. 
\bearn
\left(\frac{a_{i+1}-a_{i}}{a_{i}}\right) \sum_{j=0}^{2N+1}  \frac{a_{j} \bF_{\alpha}( a_j | a_{i+1}, (\pin, q_l))}{{\opt(F_{\alpha}( \cdot | a_{i+1}, (\pin, q_l)) )}} p_j
&\le& \left(\frac{a_{i+1}-a_{i}}{a_{i}}\right) \sum_{j=0}^{2N+1}   p_j  
\:\le\: \frac{1}{N} \left(\frac{\pin - \eta - \rl[q_l]}{ \rl[q_l]}\right),
\eearn
where the first inequality follows from the definition of $\opt$, and the second from the fact that $\mathbf{p}$ belongs to the simplex, from definition and from lower bounding $a_i$ by $a_0=\rl[q_l]$. 

Now, let for $j = i+1,...,N$, $g_j(x) =  \bF_{\alpha}( a_j | x, (\pin, q_l)) $. Note that $g_j(\cdot)$ is differentiable in $[a_i,a_j]$ with derivative bounded as follows
\bearn
g_j'(x) &=& \left( \Gad{\Gainv{\ql} \frac{a_j-x}{\pin-x}} \right)' \\
           &=&  \Gainv{\ql}  \frac{(\pin-a_j)}{(\pin-x)^2}\left(\Gad{\Gainv{\ql} \frac{a_j-x}{\pin-x}} \right)^{2-\alpha}
           \le  \frac{\Gainv{\ql}}{\pin-x}
           \le  \frac{\Gainv{\ql}}{\pin-\eta}.
\eearn
We deduce that
\bearn
&& \frac{1}{\opt(F_{\alpha}( \cdot | a_{i+1}, (\pin, q_l)) )} \sum_{j=i+1}^{N+1}  a_{j} \left[ \bF_{\alpha}( a_j | a_{i+1}, (\pin, q_l)) - \bF_{\alpha}( a_j | a_{i}, (\pin, q_l))\right] a_j p_j\\
&\le& \frac{1}{a_{i+1}} \sum_{j=i+1}^{N+1}   \frac{\Gainv{\ql}}{\pin-\eta} (a_{i+1}-a_{i}) a_j p_j\\
&\le& \frac{1}{N} \left(\frac{\pin - \eta - \rl[q_l]}{ \rl[q_l]}\right) \pin \frac{\Gainv{\ql}}{\pin - \eta}.
\eearn
Hence, we have, in this case
\bearn
\overline{\Lb}_{\alpha, \ql, \qh, \IN} - \underline{\Lb}_{\alpha, \ql, \qh, \IN} &\le& \frac{1}{N} \left(\frac{\pin - \eta - \rl[q_l]}{ \rl[q_l]}\right) \left[1 +  \pin \frac{\Gainv{\ql}}{\pin-\eta}\right].
\eearn

\textbf{Case 2:} Suppose $\underline{c}(\textbf{p}) =  \frac{1}{\opt(F_{\alpha}( \cdot | a_{i+1}^{-}, (\pin, q_l)))} \sum_{j=0}^{2N+1}   a_j \bF_{\alpha}( a_j | a_{i}, (\pin, q_l)) p_j$ for $i = N$. In this case, we have
\bearn
\underline{c}(\textbf{p}) &=& \frac{1}{\pin} \left(\sum_{j=0}^{N} a_j p_j + \pin q_l p_{N+1} \right),
\eearn
and
\bearn
\overline{\Lb}_{\alpha, \ql, \qh, \IN} - \underline{\Lb}_{\alpha, \ql, \qh, \IN}  &\leq& \frac{1}{\pin-\eta} \left(\sum_{j=0}^{N} a_j p_j + \pin q_l p_{N+1} \right) - \frac{1}{\pin} \left(\sum_{j=0}^{N} a_j p_j + \pin q_l p_{N+1} \right) \\
&\leq& \left( \frac{1}{\pin-\eta}  - \frac{1}{\pin} \right)\left(\sum_{j=0}^{N} a_j p_j + \pin q_l p_{N+1} \right) \\
&\le& \frac{\eta }{\pin-\eta}  \pin \sum_{j=0}^{N+1} p_j \\
&\le& \frac{\eta \pin}{\pin-\eta} .
\eearn

\textbf{Case 3:} Suppose $\underline{c}(\textbf{p}) = \frac{1}{\opt(F_{\alpha}( \cdot | a_{i+1}, (\pin, q_h)))} \sum_{j=0}^{2N+1}   a_j \bF_{\alpha}( a_j | a_{i}, (\pin, q_h)) p_j$ for some $i=N+1,...,2N$.
\bearn
\overline{\Lb}_{\alpha, \ql, \qh, \IN} - \underline{\Lb}_{\alpha, \ql, \qh, \IN}  
&\leq&  \frac{1}{\opt(F_{\alpha}( \cdot | a_{i}, (\pin, q_h)))} \sum_{j=0}^{2N+1}   a_j \bF_{\alpha}( a_j | a_{i+1}^-, (\pin, q_h)) p_j \\
&&  - \frac{1}{\opt(F_{\alpha}( \cdot | a_{i+1}^-, (\pin, q_h)))} \sum_{j=0}^{2N+1}   a_j \bF_{\alpha}( a_j | a_{i}, (\pin, q_h)) p_j\\
&=&  \left( \frac{1}{\opt(F_{\alpha}( \cdot | a_{i}, (\pin, q_h)))}  - \frac{1}{\opt(F_{\alpha}( \cdot | a_{i+1}, (\pin, q_h)))}  \right)\sum_{j=0}^{2N+1}   a_j \bF_{\alpha}( a_j | a_{i+1}^-, (\pin, q_h)) p_j \\
&&  + \frac{1}{\opt(F_{\alpha}( \cdot | a_{i+1}, (\pin, q_h)))} \sum_{j=0}^{2N+1}   a_j \left[\bF_{\alpha}( a_j | a_{i+1}^-, (\pin, q_h))-\bF_{\alpha}( a_j | a_{i}, (\pin, q_h))\right] p_j\\
&=&  \frac{\opt(F_{\alpha}( \cdot | a_{i+1}, (\pin, q_h)))-\opt(F_{\alpha}( \cdot | a_{i}, (\pin, q_h)))}{\opt(F_{\alpha}( \cdot | a_{i}, (\pin, q_h)))} \sum_{j=0}^{2N+1}   \frac{a_j \bF_{\alpha}( a_j | a_{i+1}, (\pin, q_h))}{\opt(F_{\alpha}( \cdot | a_{i+1}, (\pin, q_h)))} p_j,
\eearn
where in the last equality, we have used that $\bF_{\alpha}( \cdot | a_{i+1}^-, (\pin, q_h)) = \bF_{\alpha}( \cdot | a_{i}, (\pin, q_l))$ on $\{a_{0}, \dots, a_{i+1}\}$. We analyze the above term on the RHS.
\bearn
\frac{\opt(F_{\alpha}( \cdot | a_{i+1}, (\pin, q_h)))-\opt(F_{\alpha}( \cdot | a_{i}, (\pin, q_h)))}{\opt(F_{\alpha}( \cdot | a_{i}, (\pin, q_h)))}
 \sum_{j=0}^{2N+1}   \frac{a_j \bF_{\alpha}( a_j | a_{i+1}, (\pin, q_h))}{\opt(F_{\alpha}( \cdot | a_{i+1}, (\pin, q_h)))} p_j \\ \leq \left( \frac{a_{i+1} \bG_{\alpha, a_{i+1}}(a_{i+1} | (0,1), (\pin ,\qh))- a_{i} \bG_{\alpha, a_{i}}(a_{i} | (0,1), (\pin ,\qh))}{a_{i} \bG_{\alpha, a_{i}}(a_{i} | (0,1), (\pin ,\qh))}  \right) \sum_{j=0}^{2N+1} p_j \\
\leq \left( \frac{a_{i+1} \bG_{\alpha, a_{i+1}}(a_{i+1} | (0,1), (\pin ,\qh))- a_{i} \bG_{\alpha, a_{i}}(a_{i} | (0,1), (\pin ,\qh))}{a_{i} \bG_{\alpha, a_{i}}(a_{i} | (0,1), (\pin ,\qh))}  \right),
\eearn
where the first inequality follows from the definition of $\opt$ and $\opt(F_{\alpha}( \cdot | a_{i+1}, (\pin, q_h)))$, and the second from the fact that $\mathbf{p}$ belongs to the simplex.

Now, let $g_{\alpha, \qh}(x) =  x \bG_{\alpha, x}(x | (0,1), (\pin ,\qh)) $. Note that $g_{\alpha, \qh}(\cdot)$ is differentiable in $[\pin, \rh[\qh])$ with derivative bounded as follows
\bearn
g_{\alpha, \qh}'(x) &=& \left( x \Gad{\Gainv{\qh} \frac{x}{\pin}} \right)' \\
           &=&  \Gad{\Gainv{\qh} \frac{x}{\pin}} - \frac{\Gainv{q_h}x}{\pin} \Gad{\Gainv{\qh} \frac{x}{\pin}}^{2-\alpha} \\
           &=&  \Gad{\Gainv{\qh} \frac{x}{\pin}} \left(1- \frac{\Gainv{q_h}x}{\pin} \Gad{\Gainv{\qh} \frac{x}{\pin}}^{1-\alpha} \right)\\
           &=&  \Gad{\Gainv{\qh} \frac{x}{\pin}} \left(1- \frac{\Gainv{q_h}x}{\pin\left(1 + (1-\alpha) \Gainv{\qh} \frac{x}{\pin}\right)} \right) \\
           &=&  \Gad{\Gainv{\qh} \frac{x}{\pin}} \left(1- \frac{\Gainv{q_h}\frac{x}{\pin}}{1 + (1-\alpha) \Gainv{\qh} \frac{x}{\pin}} \right) \\
           &=&  \Gad{\Gainv{\qh} \frac{x}{\pin}} 
           \frac{1-\alpha \Gainv{q_h}\frac{x}{\pin}}{1 + (1-\alpha) \Gainv{\qh} \frac{x}{\pin}}.  \\
\eearn
Therefore, since $x \leq \rh[\qh] := \frac{w}{\alpha \Gainv{\qh}}$, we have that 
\bearn
|g_{\alpha, \qh}'(x)| = \left|\Gad{\Gainv{\qh} \frac{x}{\pin}} 
           \frac{1-\alpha \Gainv{q_h}\frac{x}{\pin}}{1 + (1-\alpha) \Gainv{\qh} \frac{x}{\pin}} \right| 
        =  \Gad{\Gainv{\qh} \frac{x}{\pin}} 
           \frac{1-\alpha \Gainv{q_h}\frac{x}{\pin}}{1 + (1-\alpha) \Gainv{\qh} \frac{x}{\pin}} \leq 1.
\eearn
We deduce that
\bearn
\left( \frac{a_{i+1} \bG_{\alpha, a_{i+1}}(a_{i+1} | (0,1), (\pin ,\qh))- a_{i} \bG_{\alpha, a_{i}}(a_{i} | (0,1), (\pin ,\qh))}{a_{i} \bG_{\alpha, a_{i}}(a_{i} | (0,1), (\pin ,\qh))}  \right) &\le& \frac{a_{i+1}-a_{i}}{g_{\alpha, \qh}(a_i)} \\
&\le& \frac{\min\{b, \rh[\qh]\}-\pin}{N g_{\alpha, \qh}(\pin)} \\
&=& \frac{\min\{b, \rh[\qh]\}-\pin}{N \pin q}.
\eearn
Hence, we have, in this case
\bearn
\overline{\Lb}_{\alpha, \ql, \qh, \IN} - \underline{\Lb}_{\alpha, \ql, \qh, \IN} &\le& \frac{\min\{b, \rh[\qh]\}-\pin}{N \pin q}.
\eearn

\textbf{Case 4:} Suppose $\underline{c}(\textbf{p}) = \frac{1}{\opt(F_{\alpha}( \cdot | a_{i+1}, (\pin, q_h)))} \sum_{j=0}^{2N+1}   a_j \bF_{\alpha}( a_j | a_{i}, (\pin, q_h)) p_j$ for $i=2N+1$.
\bearn
&& \overline{\Lb}_{\alpha, \ql, \qh, \IN} - \underline{\Lb}_{\alpha, \ql, \qh, \IN}  \\
&\leq&  \frac{1}{\opt(F_{\alpha}( \cdot | a_{2N+1}, (\pin, q_h)))} \sum_{j=0}^{2N+1}   a_j \bF_{\alpha}( a_j | a_{2N+1}^-, (\pin, q_h)) p_j \\
&&  - \frac{1}{\opt(F_{\alpha}( \cdot | a_{2N+2}^-, (\pin, q_h)))} \sum_{j=0}^{2N+1}   a_j \bF_{\alpha}( a_j | a_{2N}, (\pin, q_h)) p_j\\
&=&  \left( \frac{1}{\opt(F_{\alpha}( \cdot | a_{2N+1}, (\pin, q_h)))}  - \frac{1}{\opt(F_{\alpha}( \cdot | a_{2N+2}^-, (\pin, q_h)))}  \right)\sum_{j=0}^{2N+1}   a_j \bF_{\alpha}( a_j | a_{2N+2}^-, (\pin, q_h)) p_j \\
&&  + \frac{1}{\opt(F_{\alpha}( \cdot | a_{2N+2}^-, (\pin, q_h)))} \sum_{j=0}^{2N+1}   a_j \left[\bF_{\alpha}( a_j | a_{2N+2}^-, (\pin, q_h))-\bF_{\alpha}( a_j | a_{2N+1}, (\pin, q_h))\right] p_j\\
&=&  \frac{\opt(F_{\alpha}( \cdot | a_{2N+2}^-, (\pin, q_h)))-\opt(F_{\alpha}( \cdot | a_{2N+1}, (\pin, q_h)))}{\opt(F_{\alpha}( \cdot | a_{2N+1}, (\pin, q_h)))} \sum_{j=0}^{2N+1}   \frac{a_j \bF_{\alpha}( a_j | a_{2N+2}^-, (\pin, q_h))}{\opt(F_{\alpha}( \cdot | a_{2N+2}^-, (\pin, q_h)))} p_j,
\eearn

where in the last equality, we have used the fact that $\bF_{\alpha}( \cdot | a_{2N+2}^-, (\pin, q_h)) = \bF_{\alpha}( \cdot | a_{2N+1}, (\pin, q_l))$ on $\{a_{0}, \cdots, a_{2N+1}\}$. 

We analyze the above term on the RHS in two separate cases $\alpha \in (0,1]$ and $\alpha = 0$. 

 In the case where $\alpha \in (0,1]$, we have
\bearn
&&\frac{\opt(F_{\alpha}( \cdot | a_{2N+2}^-, (\pin, q_h)))-\opt(F_{\alpha}( \cdot | a_{2N+1}, (\pin, q_h)))}{\opt(F_{\alpha}( \cdot | a_{2N+1}, (\pin, q_h)))} \sum_{j=0}^{2N+1}   \frac{a_j \bF_{\alpha}( a_j | a_{2N+2}^-, (\pin, q_h))}{\opt(F_{\alpha}( \cdot | a_{2N+2}^-, (\pin, q_h)))} p_j \\ &\leq& \left( \frac{\rh[\qh] \bG_{\alpha, \rh[\qh]}(\rh[\qh] | (0,1), (\pin ,\qh))- a_{2N+1} \bG_{\alpha, a_{2N+1}}(a_{2N+1} | (0,1), (\pin ,\qh))}{a_{2N+1} \bG_{\alpha, a_{2N+1}}(a_{2N+1} | (0,1), (\pin ,\qh))}  \right) \sum_{j=0}^{2N+1} p_j \\
&\leq& \left( \frac{\rh[\qh] \bG_{\alpha, \rh[\qh]}(\rh[\qh] | (0,1), (\pin ,\qh))- a_{2N+1} \bG_{\alpha, a_{2N+1}}(a_{2N+1} | (0,1), (\pin ,\qh))}{a_{2N+1} \bG_{\alpha, a_{i}}(a_{i} | (0,1), (\pin ,\qh))}  \right) \\
&\leq& \frac{g_{\alpha, \qh}(\rh[\qh]) - g_{\alpha, \qh}(a_{2N+1})}{g_{\alpha, \qh}(a_{2N+1})} \leq \frac{\rh[\qh]-a_{2N+1}}{g_{\alpha, \qh}(\pin)} = \frac{\rh[\qh]-\pin}{N \pin q},
\eearn
where the first inequality follows from the definition of $\opt$, the second from the fact that $\mathbb{p}$ belongs to the simplex, and the fourth from the fact that the derivative $g'_{\alpha, q_h}(\cdot)$ is bounded (established in the previous case).

 In the case where $\alpha = 0$, we have
\bearn
&&\frac{\opt(F_{\alpha}( \cdot | a_{2N+2}^-, (\pin, q_h)))-\opt(F_{\alpha}( \cdot | a_{2N+1}, (\pin, q_h)))}{\opt(F_{\alpha}( \cdot | a_{2N+1}, (\pin, q_h)))} \sum_{j=0}^{2N+1}   \frac{a_j \bF_{\alpha}( a_j | a_{2N+2}^-, (\pin, q_h))}{\opt(F_{\alpha}( \cdot | a_{2N+2}^-, (\pin, q_h)))} p_j \\ &\leq& \frac{\lim_{x \to \infty}\opt(F_{0}( \cdot | x, (\pin, q_h)))-\opt(F_{0}( \cdot | b, (\pin, q_h)))}{\opt(F_{0}( \cdot | b, (\pin, q_h)))} \sum_{j=0}^{2N+1} p_j \\
&\leq&  \frac{\frac{1}{\frac{1}{\qh}-1}-\frac{b}{1+\left(\frac{1}{\qh}-1\right)b}}{\frac{b}{1+\left(\frac{1}{\qh}-1\right)b}}  = \frac{1+\left(\frac{1}{\qh}-1\right)b}{\left(\frac{1}{\qh}-1\right)b} - 1 = \frac{1}{\left(\frac{1}{\qh}-1\right)b},
\eearn
where the first inequality follows from the definition of $\opt$, the second from the fact that $\mathbf{p}$ belongs to the simplex, and the definition of $\opt(F_{\alpha}( \cdot | a_{2N+1}^-, (\pin, q_h)))$.
Hence, we have, in this case
\bearn
\overline{\Lb}_{\alpha, \ql, \qh, \IN} - \underline{\Lb}_{\alpha, \ql, \qh, \IN} &\le& \frac{\qh}{(1-\qh)b}.
\eearn

We are now in a position to combine all cases and conclude. 

If $\alpha \in (0,1]$, we have established that
\bearn
\overline{\Lb}_{\alpha, \ql, \qh, \IN} - \underline{\Lb}_{\alpha, \ql, \qh, \IN}  &\leq& \max \left\{\frac{1}{N} \left(\frac{\pin - \eta - \rl[q_l]}{ \rl[q_l]}\right) \left[1 +  \pin \frac{\Gainv{\ql}}{\pin-\eta}\right], \frac{\eta \pin}{\pin-\eta}, 
 \frac{\rh[\qh]-\pin}{N \pin q} \right\}. 
\eearn
Recall that \Cref{thm:mechanismreduction_uncertainty_1} implies that
\bearn
\Rb (\MechSet_{\IN}, \aDistSet{[\ql, \qh]}) &\leq&   \Rb (\MechSet, \aDistSet{[\ql, \qh]})  \: \leq\: \Rb (\MechSet_{\IN}, \aDistSet{[\ql, \qh]}) + \frac{\Delta(\IN)}{\rl[\ql] }.
\eearn
Noting that  $\Delta(\IN) = \max \{ \frac{\pin-\eta - \rl[\ql]}{N}, \eta, \frac{\rh[\qh]-\pin}{N} \}$, we have
\bearn
\underline{\Lb}_{\alpha, \ql, \qh, \IN} &\leq& \Rb (\MechSet, \aDistSet{[\ql, \qh]}) \\ 
&\le& \Rb (\MechSet_{\IN}, \aDistSet{[\ql, \qh]}) + \frac{\Delta(\IN)}{\rl[\ql] } \\
&\leq& \underline{\Lb}_{\alpha, \ql, \qh, \IN} + \left(\overline{\Lb}_{\alpha, \ql, \qh, \IN}-\underline{\Lb}_{\alpha, \ql, \qh, \IN}\right) + \max \left\{ \frac{\pin-\eta - \rl[\ql]}{N\rl[\ql]}, \eta, \frac{\rh[\qh]-\pin}{N\rl[\ql]} \right\} \\
&\leq& \underline{\Lb}_{\alpha, \ql, \qh, \IN} + \max \left\{\frac{1}{N} \left(\frac{\pin - \eta - \rl[q_l]}{ \rl[q_l]}\right) \left[1 +  \pin \frac{\Gainv{\ql}}{\pin-\eta}\right], \frac{\eta \pin}{\pin-\eta}, \frac{\rh[\qh]-\pin}{N \pin q} \right \}\\
&&  +  \max \left\{ \frac{\pin-\eta - \rl[\ql]}{N\rl[\ql]}, \eta, \frac{\rh[\qh]-\pin}{N\rl[\ql]} \right\}.
\eearn
By choosing $\eta = \frac{\pin}{\sqrt{N}}$, we obtain
\bearn
 \underline{\Lb}_{\alpha, \ql, \qh, \IN} \leq \Rb (\MechSet, \aDistSet{[\ql, \qh]}) \leq \underline{\Lb}_{\alpha, \ql, \qh, \IN} +  \mathcal{O}(\frac{1}{\sqrt{N}}).
\eearn

Suppose now $\alpha = 0$. In this case.
\bearn
\overline{\Lb}_{\alpha, \ql, \qh, \IN} - \underline{\Lb}_{\alpha, \ql, \qh, \IN}  &\leq& \max \left\{\frac{1}{N} \left(\frac{\pin - \eta - \rl[q_l]}{ \rl[q_l]}\right) \left[1 +  \pin \frac{\Gainv{\ql}}{\pin-\eta}\right], \frac{\eta \pin}{\pin-\eta},  \frac{b-\pin}{N \pin q}, \frac{\qh}{(1-\qh)b} \right\}. 
\eearn
Using again \Cref{thm:mechanismreduction_uncertainty_1} and the fact that $\Delta(\IN) = \max \{ \frac{\pin-\eta - \rl[\ql]}{N}, \eta, \frac{b-\pin}{N} \}$, we have 
\bearn
\underline{\Lb}_{\alpha, \ql, \qh, \IN} &\leq& \Rb (\MechSet, \aDistSet{[\ql, \qh]}) \\ 
&\le& \Rb (\MechSet_{\IN}, \aDistSet{[\ql, \qh]}) + \frac{\Delta(\IN)}{\rl[\ql] } \\
&\leq& \underline{\Lb}_{\alpha, \ql, \qh, \IN} + \left(\overline{\Lb}_{\alpha, \ql, \qh, \IN}-\underline{\Lb}_{\alpha, \ql, \qh, \IN}\right) + \max \left\{ \frac{\pin-\eta - \rl[\ql]}{N\rl[\ql]}, \eta, \frac{b-\pin}{N\rl[\ql]} \right\} \\
&\leq& \underline{\Lb}_{\alpha, \ql, \qh, \IN} + \max \left\{\frac{1}{N} \left(\frac{\pin - \eta - \rl[q_l]}{ \rl[q_l]}\right) \left[1 +  \pin \frac{\Gainv{\ql}}{\pin-\eta}\right], \frac{\eta \pin}{\pin-\eta},  \frac{b-\pin}{N \pin q}, \frac{\qh}{(1-\qh)b}\right\} \\
&&+  \max \left\{ \frac{\pin-\eta - \rl[\ql]}{N\rl[\ql]}, \eta, \frac{b-\pin}{N\rl[\ql]} \right\}.
\eearn
By choosing $\eta = \frac{\pin}{\sqrt{N}}$ and $b = \pin \sqrt{N}$, we obatin
\bearn
 \underline{\Lb}_{\alpha, \ql, \qh, \IN} \leq \Rb (\MechSet, \aDistSet{[\ql, \qh]}) \leq \underline{\Lb}_{\alpha, \ql, \qh, \IN} +  \mathcal{O}(\frac{1}{\sqrt{N}}).
\eearn
This concludes the proof. 
\end{proof}

\subsection{Proofs of auxiliary results} \label{app:aux-uncer}

\begin{proof}[\textbf{Proof of \Cref{thm:LB_1D_uncertainty_1}.}]  First we show that 
\bearn
\inf _{F \in \aDistSet{[\ql, \qh]}} R(\Psi, F) = \inf_{q \in [\ql, \qh]} \inf _{F \mbox{ in }\aDistSetqa{q}} R(\Psi, F)
\eearn
Let $q \in[\ql, \qh]$ and $F \in\aDistSetqa{q}$, we have:
\bearn
R(\Psi, F) &\geq& \inf _{F \in\aDistSet{[\ql, \qh]}} R(\Psi, F) \\
\implies \inf _{F \in\aDistSetqa{q}} R(\Psi, F) &\geq& \inf _{F \in\aDistSet{[\ql, \qh]}} R(\Psi, F) \\
\implies \inf_{q \in[\ql, \qh]} \inf _{F \in\aDistSetqa{q}} R(\Psi, F) &\geq& \inf _{F \in\aDistSet{[\ql, \qh]}} R(\Psi, F).
\eearn
Let $\epsilon > 0$ and $F_{\epsilon} \in\Fa(w,[\ql, \qh]) $ such that:
\bearn
\inf _{F \in\aDistSet{[\ql, \qh]}} R(\Psi, F) &\geq&  R(\Psi, F_{\epsilon}) - \epsilon \\
\implies \inf _{F \in\aDistSet{[\ql, \qh]}} R(\Psi, F) &\geq&  \inf _{F \in\aDistSetq{\bF_{\epsilon}(w)}} R(\Psi, F) - \epsilon \\
\implies \inf _{F \in\aDistSet{[\ql, \qh]}} R(\Psi, F) &\geq&  \inf_{q \in[\ql, \qh]} \inf _{F \in\aDistSetqa{q}} R(\Psi, F) - \epsilon,
\eearn
by taking $\epsilon \rightarrow 0$, we obtain the desired result.

Let $F \in\Fa(w,[\ql, \qh])$ and $q = \bF(w) \in[\ql, \qh]$, by \Cref{thm:LB_1D}, we have:
\bearn
\inf _{F \in\aDistSetqa{q}} R(\Psi, F)  &=& \min \Biggl\{ \inf_{x \in [\rl[q], \pin)}  \frac{1}{\opt(F_{\alpha}( \cdot | x, (\pin, q)))}  \int_{0}^{\infty}    u \bF_{\alpha}( u | x, (\pin, q)) d\Psi(u), \\
&& \inf_{x \in [\pin, \rh[q]]} \frac{1}{\opt(F_{\alpha}( \cdot | x, (\pin, q)))} \int_{0}^{\infty}  u \bF_{\alpha}( u | x, (\pin, q)))  d\Psi(u) \Biggr\} \\
&=& \min \Biggl\{ \inf_{x \in [\rl[q], \pin)}  \frac{1}{x} \Biggl[ \int_{[0, x)} u d\Psi(u) + \int_{[x, \pin]} u \bG_{\alpha, \pin}(u|(x,1), (\pin, q))  d\Psi(u) \Biggl], \\
&& \inf_{x \in [\pin, \rh[q]]} \frac{1}{x \bG_{\alpha, x}(x|(0,1), (\pin, q))} \int_{[0, x]} u \bG_{\alpha, x}(u|(0,1), (\pin, q))  d\Psi(u) \Biggr\}. 
\eearn
 Using the non-decreasing monotonicity of the functions $q \to \rl = \frac{\pin}{\Gainv{q}+1}, q \to \rh = \frac{\pin}{\alpha \Gainv{q}}$, we have:
\bearn
\inf _{F \in\aDistSetqa{q}} R(\Psi, F)   &\geq& \min \Biggl\{ \inf_{x \in [\rl[\ql], \pin)}  \frac{1}{x} \Biggl[ \int_{[0, x)} u d\Psi(u) + \int_{[x,\pin]} u \bG_{\alpha, \pin}(u|(x,1), (\pin, q))  d\Psi(u) \Biggl], \\
&& \inf_{x \in [\pin, \rh[\qh]]} \frac{1}{x \bG_{\alpha, x}(x|(0,1), (\pin, q))} \int_{[0, x]} u \bG_{\alpha, x}(u|(0,1), (\pin, q))  d\Psi(u) \Biggr\}.
\eearn 
We have, for fixed $(u, x)$ such that $x \in[\rl[\ql], \pin), u \in [x, \pin)$, the following function is clearly non-decreasing
\bearn
q \to \bG_{\alpha, \pin}(u|(x,1), (\pin, q)) = \Gad{\Gainv{q}\frac{u-x}{\pin-x}}.
\eearn 
We have, for fixed $(u, x)$ such that $x \in[\pin, \rh[\qh]], u \in [\pin, x]$, the following function is non-increasing
\bearn
q \to \frac{u \bG_{\alpha, x}(u|(0,1), (\pin, q))}{x \bG_{\alpha, x}(x|(0,1), (\pin, q))}  &=& \frac{u \Gad{\Gainv{q}\frac{u}{\pin}}}{x \Gad{\Gainv{q}\frac{x}{\pin}}} = 
    \begin{cases}
        \frac{u}{x} (\frac{u}{x} + \frac{x-u}{x(1+(1-\alpha)\Gainv{q}\frac{x}{\pin})})^{\frac{1}{\alpha-1}}  &\quad \mbox{if } \alpha \in[0,1)\\
	    \frac{u}{x} q^{\frac{u-x}{\pin}}   &\quad \mbox{if } \alpha = 1.
	\end{cases}
\eearn 
Using the monotonicity of the above functions we get :
\bearn
\inf _{F \mbox{ in }\aDistSetqa{q}} R(\Psi, F)  &\geq& \min \Biggl\{ \inf_{x \in [\rl[\ql], \pin)}  \frac{1}{x} \Biggl[ \int_{[0, x)} u d\Psi(u) + \int_{[x, \pin]} u \bG_{\alpha, \pin}(u|(x,1), (\pin, \ql))  d\Psi(u) \Biggl], \\
&& \inf_{x \in [\pin, \rh[\qh]]} \frac{1}{x \bG_{\alpha, x}(x|(0,1), (\pin, q))} \int_{[0, x]} u \bG_{\alpha, x}(u|(0,1), (\pin, \qh))  d\Psi(u) \Biggr\}.
\eearn 
 Since the right hand-side does not depend on $q$, we take the minimum on $q$
 \bearn
\inf _{F \in\aDistSet{[\ql, \qh]}} R(\Psi, F)  &\geq& \min \Biggl\{ \inf_{x \in [\rl[\ql], \pin)}  \frac{1}{x} \Biggl[ \int_{[0, x)]} u d\Psi(u) + \int_{[x, \pin]} u \bG_{\alpha, \pin}(u|(x,1), (\pin, \ql))  d\Psi(u) \Biggl], \\
&& \inf_{x \in [\pin, \rh[\qh]]} \frac{1}{x \bG_{\alpha, x}(x|(0,1), (\pin, q))} \int_{[0,x]} u \bG_{\alpha, x}(u|(0,1), (\pin, \qh))  d\Psi(u) \Biggr\}  \\ 
 &=& \min \Biggl\{ \inf_{x \in [\rl[\ql], \pin)}  \frac{1}{\opt(F_{\alpha}( \cdot | x, (\pin, \ql)))}  \int_{0}^{\infty}    u \bF_{\alpha}( u | x, (\pin, \ql)) d\Psi(u), \\
&& \inf_{x \in [\pin, \rh[\qh]]} \frac{1}{\opt(F_{\alpha}( \cdot | x, (\pin, \qh)))} \int_{0}^{\infty}  u \bF_{\alpha}( u | x, (\pin, \qh)))  d\Psi(u) \Biggr\}.
\eearn 
This concludes the proof.
\end{proof}

\begin{proof}[\textbf{Proof of \Cref{thm:mechanismreduction_uncertainty_1}.}] 
Fix $q \mbox{ in }[\ql, \qh]$. Then, using \cref{thm:mechanismreduction}, there exists $\Psi_{\IN} \mbox{ in }\MechSet_{\IN}$ such that 
\bearn
\inf_{F \in \aDistSetqa{q}} R(\Psi_{\IN}, F) \geq   \inf_{F \in \aDistSetqa{q}} R(\Psi, F) - \frac{\Delta(\IN)}{a_1}  -  \frac{1}{q(1+(q^{-1}-1)a_N/\pin)} \mathbf{1}\{a_N < \rh\},
\eearn
where $\Delta(\IN) = \sup_i \{a_i - a_{i-1}\}$. We have the following function $$q \to -\frac{1}{q(1+(q^{-1}-1)a_N/\pin)} = -\frac{1}{\frac{a_N}{\pin} + q(1-\frac{a_N}{\pin})},$$ is non-increasing because $a_N > \pin$ therefore 
$$-\frac{1}{q(1+(q^{-1}-1)a_N/\pin)} \geq -  \frac{1}{\qh(1+(\qh^{-1}-1)a_N/\pin)},$$ 
moreover we have $q \to \rh = \frac{\pin}{\alpha \Gainv{q}}$ is non-decreasing, therefore $$\mathbf{1}\{a_N < \rh\} \leq \mathbf{1}\{a_N < \rh[\qh]\}.$$ Hence, since $-\frac{1}{q(1+(q^{-1}-1)a_N/\pin)} <0$, we have
$$
-  \frac{\mathbf{1}\{a_N < \rh\}}{q(1+(q^{-1}-1)a_N/\pin)} \geq -  \frac{\mathbf{1}\{a_N < \rh[\qh]\}}{\qh(1+(\qh^{-1}-1)a_N/\pin)}.
$$
Using these lower bounds, we obtain 
\bearn
\inf_{F \mbox{ in }\aDistSetqa{q}} R(\Psi_{\IN}, F) \geq   \inf_{F \mbox{ in }\aDistSetqa{q}} R(\Psi, F) - \frac{\Delta(\IN)}{\rl[\ql]} -  \frac{\mathbf{1}\{a_N < \rh[\qh]\}}{\qh(1+(\qh^{-1}-1)a_N/\pin)} ,
\eearn 
taking the infinimum over $q$ from both sides concludes the proof.
\end{proof}


\setcounter{equation}{0}
\setcounter{theorem}{0}
\setcounter{lemma}{0}
\setcounter{proposition}{0}

\setcounter{equation}{0}
\setcounter{proposition}{0}
\setcounter{lemma}{0}

\section{Upper bound linear program and implementation parameters}\label{apx:upb_params}

In this section, we show that one can obtain an upper bound on the maxmin ratio $\Rb (\MechSet, \aDistSet{[\ql, \qh]})$ by solving a linear program. Fix an arbitrary sequence of increasing reals $\IN= \{a_i\}_{i=0}^{2N}$ such that $a_0 = 0, a_1 = \rl[\ql]$, $a_{N+1} = \pin$ and $a_{2N} \le \rh[\qh]$. Set $a_{2N+1}:=\rh[\qh]$.

Fix a mechanism $\Psi \mbox{ in }\MechSet$ and denote by $p_{j+1} = \int_{I_{j}} d \Psi(u)$ where we define the intervals $(I_j)_{j = 0, \cdots, 2N}$ as follows:
\bearn
I_j = \begin{cases}
[a_j, a_{j+1}) &\quad \mbox{if } 0 \leq j < 2N, \\
[a_{2N}, a_{2N+1}] &\quad \mbox{if } j = 2N.
\end{cases}
\eearn
 Using \cref{thm:LB_1D_uncertainty_1}, we have
\bearn
&& \inf _{F \in \aDistSet{[\ql, \qh]}} R(\Psi, F)  \\
&=& \min \Biggl\{ \inf_{x \in [\rl[\ql], \pin)}  \frac{1}{\opt(F_{\alpha}( \cdot | x, (\pin, q_l)))}  \int_{0}^{\infty}    u \bF_{\alpha}( u | x, (\pin, q_l)) d\Psi(u), \\
&&\inf_{x \in [\pin, \rh[\qh]]} \frac{1}{\opt(F_{\alpha}( \cdot | x, (\pin, q_h)))} \int_{0}^{\infty}  u \bF_{\alpha}( u | x, (\pin, q_h)))  d\Psi(u) \Biggr\} \\
  &=& \min \Biggl\{ \min_{i = 1, \cdots, N}\inf_{x \in [a_i, a_{i+1})}  \frac{1}{\opt(F_{\alpha}( \cdot | x, (\pin, q_l)))}  \int_{0}^{\infty}    u \bF_{\alpha}( u | x, (\pin, q_l)) d\Psi(u), \\
 && \min_{i = N+1, \cdots, 2N}\inf_{x \in [a_i, a_{i+1}]} \frac{1}{\opt(F_{\alpha}( \cdot | x, (\pin, q_h)))} \int_{0}^{\infty}  u \bF_{\alpha}( u | x, (\pin, q_h)))  d\Psi(u)\Biggr\}.
\eearn

Following the same reasoning as in the proof of \Cref{thm:LB_LP_robust}, we may also obtain an upper bound on $\inf _{F \in\aDistSet{[\ql, \qh]}} R(\Psi, F)$. Indeed, we have 
 \bearn
 \inf _{F \in \aDistSet{[\ql, \qh]}} R(\Psi, F) 
  &\le& \min \Biggl\{ \min_{i = 1, \cdots, N}  \frac{1}{\opt(F_{\alpha}( \cdot | a_{i}, (\pin, q_l)) )} \int_{0 }^{\infty}    u \bF_{\alpha}( u | a_{i+1}^{-}, (\pin, q_l)) d\Psi(u)  , \\
 && \min_{i = N+1, \cdots, 2N} \frac{1}{\opt(F_{\alpha}( \cdot | a_{i}, (\pin, q_h)))} \int_{0 }^{\infty}  u \bF_{\alpha}( u | a_{i+1}^{-}, (\pin, q_h))  d\Psi(u) \Biggr\}\\
&=& \min \Biggl\{ \min_{i = 1, \cdots, N}  \frac{1}{\opt(F_{\alpha}( \cdot | a_{i}, (\pin, q_l)) )}   \sum_{j=0}^{2N} \int_{I_j} u \bF_{\alpha}( u | a_{i+1}^{-}, (\pin, q_l))  d \Psi(u) , \\
 && \min_{i = N+1, \cdots, 2N}\frac{1}{\opt(F_{\alpha}( \cdot | a_{i}, (\pin, q_h)))} \sum_{j=0}^{2N} \int_{I_j}  u \bF_{\alpha}( u | a_{i+1}^{-}, (\pin, q_h)) d \Psi(u) \Biggr\}.
\eearn

For $x \in [\rl[\ql], \pin)$, the revenue function $u\mapsto u \bF_{\alpha}( \cdot | x, (\pin, q_l))$ is increasing in $u$ on $[0,x)$ and decreasing on $(x,\pin)$. In addition, note that, for $x \in [\pin, \rh[\qh]]$, the revenue function  $u \mapsto u \bF_{\alpha}( u | x, (\pin, q_h))$ is non-decreasing on $[0,x]$ and $u \bF_{\alpha}( u | x, (\pin, q_h)) = 0$ for $u > x$. Hence, we have
\bearn
 \inf _{F \in \aDistSet{[\ql, \qh]}} R(\Psi, F) 
  &\le& \min \Biggl\{ \min_{i = 1, \cdots, N}  \frac{1}{\opt(F_{\alpha}( \cdot | a_{i}, (\pin, q_l)) )}   \sum_{j=0}^{2N} \int_{I_j} u \bF_{\alpha}( u | a_{i+1}^{-}, (\pin, q_l))  d \Psi(u) , \\
 && \min_{i = N+1, \cdots, 2N}\frac{1}{\opt(F_{\alpha}( \cdot | a_{i}, (\pin, q_h)))} \sum_{j=0}^{2N} \int_{I_j}  u \bF_{\alpha}( u | a_{i+1}^{-}, (\pin, q_h)) d \Psi(u) \Biggr\} \\
 &\le& \min \Biggl\{ \min_{i = 1, \cdots, N}  \frac{1}{\opt(F_{\alpha}( \cdot | a_{i}, (\pin, q_l)) )}  \Biggl[ \sum_{j=0}^{i} a_{j+1} \bF_{\alpha}( a_{j+1} | a_{i+1}^{-}, (\pin, q_l)) \int_{I_j}   d \Psi(u) + \\
 && \sum_{j=i+1}^{2N} a_{j} \bF_{\alpha}( a_{j} | a_{i+1}^{-}, (\pin, q_l)) \int_{I_j}    d \Psi(u) \Biggl], \\
 && \min_{i = N+1, \cdots, 2N}\frac{1}{\opt(F_{\alpha}( \cdot | a_{i}, (\pin, q_h)))} \sum_{j=0}^{2N} a_{j+1} \bF_{\alpha}( a_{j+1} | a_{i+1}^{-}, (\pin, q_h))  \int_{I_j} d \Psi(u) \Biggr\} \\
 &=& \min \Biggl\{ \min_{i = 1, \cdots, N}  \frac{1}{\opt(F_{\alpha}( \cdot | a_{i}, (\pin, q_l)) )}  \Biggl[ \sum_{j=1}^{i+1} a_{j} \bF_{\alpha}( a_{j} | a_{i+1}^{-}, (\pin, q_l)) p_j + \\
 && \sum_{j=i+2}^{2N+1} a_{j-1} \bF_{\alpha}( a_{j-1} | a_{i+1}^{-}, (\pin, q_l)) p_j \Biggl], \\
 && \min_{i = N+1, \cdots, 2N}\frac{1}{\opt(F_{\alpha}( \cdot | a_{i}, (\pin, q_h)))} \sum_{j=1}^{2N+1} a_{j} \bF_{\alpha}( a_{j} | a_{i+1}^{-}, (\pin, q_h)) p_j \Biggr\}. \\
\eearn
The problem of maximizing over mechanisms in $\MechSet$ is  clearly upper bounded by the problem of maximizing the RHS above over $p_1,...,p_{2N+1}$. The epigraph formulation of the latter problem can be written as 
\begin{align}
\overline{\Lb_U}_{\alpha, \ql, \qh, \IN} \:\:=\:\:  \max_{ \mathbf{p}, c} &  ~~c  \\
s.t.    ~~   & \frac{1}{\opt(F_{\alpha}( \cdot | a_{i}, (\pin, q_l)) )}  \Biggl[ \sum_{j=1}^{i+1} a_{j} \bF_{\alpha}( a_{j} | a_{i+1}^{-}, (\pin, q_l)) p_j + \sum_{j=i+2}^{2N+1} a_{j-1} \bF_{\alpha}( a_{j-1} | a_{i+1}^{-}, (\pin, q_l)) p_j \Biggl] \ge c \nonumber\\
& i=1,...N, \nonumber\\
& \frac{1}{\opt(F_{\alpha}( \cdot | a_{i}, (\pin, q_h)))} \sum_{j=1}^{2N+1} a_{j} \bF_{\alpha}( a_{j} | a_{i+1}^{-}, (\pin, q_h)) p_j \ge c \quad i=N+1,...2N, \nonumber\\
& \sum_{j=1}^{2N+1} p_j \le 1, \quad p_i \ge 0 \quad i=1,...2N+1.\nonumber
\end{align}

Therefore we obtain that:
\bearn
\Rb (\MechSet, \aDistSet{[\ql, \qh]}) \leq \overline{\Lb_U}_{\alpha, \ql, \qh, \IN}.
\eearn

\textbf{Implementation parameters:} 
For all reported values in the main text, we use the following sequence in the Linear Programs
 \bearn
 a_i &=& \begin{cases}
           \rlone[\ql] + \frac{i}{N} \left(  1-\eta-\rlone[\ql] \right) &\quad \mbox{if } 0 \leq i \leq N.\\
		    \pin + \frac{i-N-1}{N} \left( \min(b, \rhone[\qh])-1 \right) &\quad \mbox{if } N+1 \leq i \leq 2N+1,\\
		\end{cases}
 \eearn
 with $N = 2500, \eta = 10^{-5}, b = 250$.


\setcounter{equation}{0}
\setcounter{proposition}{0}
\setcounter{lemma}{0}

\section{Additional Illustrations of near optimal mechanisms for Section \ref{sec:rand}}\label{apx:optmechq}

\begin{figure}[!ht]
\centering
\begin{minipage}{.5\textwidth}
\begin{center}
\begin{tikzpicture}
	\begin{axis}[ xmin=0.001, xmax=250,
	ymin=0, ymax=1,
	restrict y to domain=0:1.0,
	grid=both,
	xmode=log,
	minor tick num=1,
	axis line style={->},,
	xlabel={ $p$},
	ylabel={ $\Psi(p)$},
	width=7cm, legend pos=north west,
	height=7cm, cycle list name=black white]
	]
	\addplot [blue, very thick ]  table [ col sep=comma ] {optmech_001.csv};
	\addlegendentry{$q=0.01$};
	\addplot [red, very thick]  table [ col sep=comma ] {optmech_025.csv};
	\addlegendentry{$q=0.25$};
	\addplot [black!60!green, very thick ]  table [ col sep=comma ] {optmech_050.csv};
	\addlegendentry{$q=0.50$};
	\addplot [black!100, very thick ]  table [ col sep=comma ] {optmech_075.csv};
	\addlegendentry{$q=0.75$};
	\end{axis}
	\end{tikzpicture}
	\caption*{\textbf{regular distributions}}
\end{center}
\end{minipage}%
\begin{minipage}{.5\textwidth}
  \begin{center}
\begin{tikzpicture}
	\begin{axis}[ xmin=0, xmax=4,
	ymin=0, ymax=1,
	restrict y to domain=0:1.0,
	grid=both,
	minor tick num=1,
	axis line style={->},
	xlabel={ $p$},
	ylabel={ $\Psi(p)$},
	width=7cm, legend pos=south east,
	height=7cm, cycle list name=black white]
	]
	\addplot [blue, very thick ]  table [ col sep=comma ] {optmech_101.csv};
	\addlegendentry{$q=0.01$};
	\addplot [red, very thick ]  table [ col sep=comma ] {optmech_125.csv};
	\addlegendentry{$q=0.25$};
	\addplot [black!60!green, very thick ]  table [ col sep=comma ] {optmech_150.csv};
	\addlegendentry{$q=0.50$};
	\addplot [black!100, very thick ]  table [ col sep=comma ] {optmech_175.csv};
	\addlegendentry{$q=0.75$};
	\end{axis}
	\end{tikzpicture}
	\caption*{\textbf{mhr distributions}}
\end{center}
\end{minipage}
\caption{\textbf{Illustration of near optimal mechanisms.} The figure depicts near optimal pricing distributions for $\pin = 1$, $q \mbox{ in }\{0.01, 0.25, 0.5, 0.75\}$. The left panel corresponds to regular distributions (plotted using a log scale) and the right panel to  mhr distributions (on a regular scale).} \label{fig:rand-mech-apx}
\end{figure}


\end{document}